\documentclass[%
 reprint,
 amsmath,amssymb,
 aps,
floatfix,
]{revtex4-2}

\usepackage{graphicx}
\usepackage{dcolumn}
\usepackage{bm}
\usepackage{amsmath}
\usepackage{enumitem}
\usepackage{hyperref}
\usepackage[mathlines]{lineno}

\begin{document}

\preprint{APS/123-QED}

\title{Strong decays analysis of excited nonstrange charmed mesons:\\Implications for spectroscopy}

\author{Keval Gandhi}
 \altaffiliation[]{keval.physics@yahoo.com}
\author{Ajay Kumar Rai}%
 \email{raiajayk@gmail.com}
\affiliation{%
Department of Applied Physics, Sardar Vallabhbhai National Institute of Technology, Surat 395007, Gujarat, India.\\
}%




\date{\today}

\begin{abstract}
The strong decays of $D_1(2420)^0$, $D_2^*(2460)^0$, $D_2^*(2460)^+$, $D_2^*(2460)^-$, $D(2550)^0$, $D_J^*(2600)^0$, $D(2740)^0$, $D_3^*(2750)^0$, $D_3^*(2750)^+$, $D_3^*(2750)^-$, $D_J(3000)^0$, $D{_{J}^*}(3000)^0$ and $D_2^*(3000)^0$ resonance states are analyzed in the heavy quark mass limit of Heavy Quark Effective Theory (HQET). The individual decay rates and the branching ratios among the strong decays determine their spin and parity. From such states the Regge trajectories are constructed in $(J, M^2)$ and $(n_r, M^2)$ planes and further predict the masses of higher excited states (1$^1D_2$, 1$^3D_3$, 3$^1S_0$, 3$^3S_1$, 1$^1F_3$, 1$^3F_4$, 2$^3D_3$, 3$^3P_2$ and 2$^3F_4$) lying on Regge lines by fixing their slopes and intercepts. Moreover, the strong decay rates and the branching ratios of these higher excited states are also examined, which can help the experimentalists to search these states into their respective decay modes.         
\end{abstract}

\maketitle

\section{Introduction}
\label{sec1}

Remarkable progress has been made in the field of charmed meson spectroscopy recently by experimental observations as well as theoretical computations. Different experimental facilities have provided new informations in this sector like masses, decay widths, branching ratios, isospin mass splittings, spin, parity, polarization amplitude etc.. At latest, the LHCb Collaboration has studied the amplitude contribution in $B^- \rightarrow D^+ \pi^- \pi^-$ decay using the Dalitz plot analysis technique \cite{Aaij2016}. They found that the main contributions are coming from the $D_2^*(2460)^0$, $D_1^*(2680)^0$, $D_3^*(2760)^0$ and $D_2^*(3000)^0$ resonances which are decaying into $S$-wave $D^+ \pi^-$. Their masses and decay widths are measured precisely (with statistical and systematic uncertainties) and make a spin parity assignment of $D_2^*(3000)^0$ as $2^+$ first time. The LHCb group in their earlier analysis of decay $B^0 \rightarrow \bar{D}^0 \pi^+ \pi^-$ has measured $D_0^*(2400)^-$ and $D_0^*(2460)^-$ mesons and identified the $D_3^*(2760)^-$ with a spin parity $3^-$ in the squared invariant mass region of $\bar{D}^0 \pi^-$ \cite{Aaij2015}.

\begin{table*}
\caption{\label{tab1}
The experimental results (masses and decay widths) from LHCb(2016) \cite{Aaij2016}, LHCb(2015) \cite{Aaij2015}, LHCb(2013) \cite{Aaij2013} and $BABAR$(2010) \cite{del2010} of nonstrange charmed mesons (in MeV).}
\begin{ruledtabular}
\begin{tabular}{cccccccc}
Meson & LHCb(2016) \cite{Aaij2016} & LHCb(2015) \cite{Aaij2015} & LHCb(2013) \cite{Aaij2013} & $BABAR$(2010) \cite{del2010} & Decay mode \\
\noalign{\smallskip}\hline\noalign{\smallskip}
$D_1(2420)^0$ & & & 2419.6 $\pm$ 0.1 $\pm$ 0.7 & 2420.1 $\pm$ 0.1 $\pm$ 0.8 & $D^{*+} \pi^-$ &\\
& & & 35.2 $\pm$ 0.4 $\pm$ 0.9  & 31.4 $\pm$ 0.5 $\pm$ 1.3 \\
& & & $1^+$ & Unnatural \\
\noalign{\smallskip}\noalign{\smallskip}
$D{_{2}^*}(2460)^0$ & 2463.7 $\pm$ 0.4 $\pm$ 0.4 $\pm$ 0.6 &  & 2460.4 $\pm$ 0.1 $\pm$ 0.1 & 2462.2 $\pm$ 0.1 $\pm$ 0.8 & $D^{+} \pi^-$  \\
& 47.0 $\pm$ 0.8 $\pm$ 0.9 $\pm$ 0.3 & & 45.6 $\pm$ 0.4 $\pm$ 1.1 & 50.5 $\pm$ 0.6 $\pm$ 0.7 \\
& $2^+$ & & $2^+$ & Natural \\
\noalign{\smallskip}\noalign{\smallskip}
$D{_{2}^*}(2460)^{+}$ & && 2463.1 $\pm$ 0.2 $\pm$ 0.6 && $D^{0} \pi^+$ \\
&&& 48.6 $\pm$ 1.3 $\pm$ 1.9 \\
&&& $2^+$\\
\noalign{\smallskip}\noalign{\smallskip}
$D{_{2}^*}(2460)^{-}$ & & 2468.6 $\pm$ 0.6 $\pm$ 0.3 &&& $D^{0} \pi^-$ \\
&& 47.3 $\pm$ 1.5 $\pm$ 0.7 & \\
&& $2^+$ &\\
\noalign{\smallskip}\noalign{\smallskip}
$D(2550)^0$ & & & 2579.5 $\pm$ 3.4 $\pm$ 5.5 & 2539.4 $\pm$ 4.5 $\pm$ 6.8 & $D^{*+} \pi^-$ \\
& & & 177.5 $\pm$ 17.8 $\pm$ 46.0 & 130 $\pm$ 12 $\pm$ 13\\
& & & Unnatural & $0^-$ \\
\noalign{\smallskip}\noalign{\smallskip}
$D{_{J}^*}(2600)^0$ & 2681.1 $\pm$ 5.6 $\pm$ 4.9 $\pm$ 13.1 &  & & 2608.7 $\pm$ 2.4 $\pm$ 2.5 & $D^{+} \pi^-$ \\
& 186.7 $\pm$ 8.5 $\pm$ 8.6 $\pm$ 8.2 & & & 93 $\pm$ 6 $\pm$ 13 & \\
& $1^-$ & & & Natural \\
&&& 2649.2 $\pm$ 3.5 $\pm$ 3.5 && $D^{*+} \pi^-$ \\
&&& 140.2 $\pm$ 17.1 $\pm$ 18.6 & \\
&&& Natural \\
\noalign{\smallskip}\noalign{\smallskip}
$D(2740)^0$ & & & 2737.0 $\pm$ 3.5 $\pm$ 11.2 && $D^{*+} \pi^-$ \\
&&& 73.2 $\pm$ 13.4 $\pm$ 25.0\\
&&& Unnatural\\
\noalign{\smallskip}\noalign{\smallskip}
$D{_{3}^*}(2750)^0$ & 2775.5 $\pm$ 4.5 $\pm$ 4.5 $\pm$ 4.7 &  & 2760.1 $\pm$ 1.1 $\pm$ 3.7 & 2763.3 $\pm$ 2.3 $\pm$ 2.3 & $D^{+} \pi^-$\\
& 95.3 $\pm$ 9.6 $\pm$ 7.9 $\pm$ 33.1 & & 74.4 $\pm$ 3.4 $\pm$ 19.1 & 60.9 $\pm$ 5.1 $\pm$ 3.6 \\
& $3^-$ & & Natural & Natural \\
&&& 2761.1 $\pm$ 5.1 $\pm$ 6.5 & 2752.4 $\pm$ 1.7 $\pm$ 2.7 & $D^{*+} \pi^-$    \\
&&& 74.4 $\pm$ 3.4 $\pm$ 37.0 & 71 $\pm$ 6 $\pm$ 11  \\
&&& Natural & Natural\\
\noalign{\smallskip}\noalign{\smallskip}
$D{_{3}^*}(2750)^+$ &&& 2771.7 $\pm$ 1.7 $\pm$ 3.8 & 2769.7 $\pm$ 3.8 $\pm$ 1.5 & $D^{0} \pi^+$ \\
&&& 66.7 $\pm$ 6.6 $\pm$ 10.5 & 60.9 \\
&&& $3^-$ & Natural \\
\noalign{\smallskip}\noalign{\smallskip}
$D{_{3}^*}(2750)^-$ &&2798 $\pm$ 7 $\pm$ 1 $\pm$ 7 &&& $D^{0} \pi^-$ \\
&&105 $\pm$ 18 $\pm$ 6 $\pm$ 23 && \\
&&$3^-$ &&\\
\noalign{\smallskip}\noalign{\smallskip}
$D_J(3000)^0$ &  & &2971.8 $\pm$ 8.7 && $D^{*+} \pi^-$\\
& & & 188.1 $\pm$ 44.8 \\
& & & Unnatural\\
\noalign{\smallskip}\noalign{\smallskip}
$D{_{J}^*}(3000)^0$ & & & 3008.1 $\pm$ 4.0 && $D^{+} \pi^-$ \\
& & & 110.5 $\pm$ 11.5 \\
&& & Natural\\
\noalign{\smallskip}\noalign{\smallskip}
$D_2^*(3000)^0$ & 3214 $\pm$ 29 $\pm$ 33 $\pm$ 36 &&&& $D^{+} \pi^-$ \\
&186 $\pm$ 38 $\pm$ 34 $\pm$ 63 \\
& $2^+$\\
\end{tabular}
\end{ruledtabular}
\end{table*}

\begin{table*}
\caption{\label{tab2}
Spectra of nonstrange charmed mesons obtained from different models (in MeV).}
\begin{ruledtabular}
\begin{tabular}{ccccccccccccc}
$\cal{N}$$^{2S+1}L_J$ & $J^P$ & Ref. \cite{Chen2018} & Ref. \cite{Kher2017} & Ref. \cite{Godfrey2016} & Ref. \cite{Sun2013} & Ref. \cite{Li2011} & Ref. \cite{Ebert2010} & Ref. \cite{DiPierro2001} & Ref. \cite{Lahde2000} & Ref. \cite{Cichy2016} \\
\noalign{\smallskip}\hline\noalign{\smallskip}

$1^1S_0$ & $0^-$ & 1869 & 1884 & 1877 & 1874 & 1867 & 1871 & 1868 & 1874 & 1865 \\
$1^3S_1$ & $1^-$ & 2002 & 2010 & 2041 & 2038 & 2010 & 2010 & 2005 & 2006 & 2027 \\
$2^1S_0$ & $0^-$ & 2562 & 2582 & 2581 & 2583 & 2555 & 2581 & 2589 & 2540 \\
$2^3S_1$ & $1^-$ & 2616 & 2655 & 2643 & 2645 & 2636 & 2632 & 2692 & 2601 \\
$3^1S_0$ & $0^-$ & 2970 & 3186 & 3110 & 3068 & & 3062 & 3141 & 2904 \\
$3^3S_1$ & $1^-$ & 3004 & 3239 & 3068 & 3111 & & 3096 & 3226 & 2947 \\

$1^3P_0$ & $0^+$ & 2319 & 2357 & 2399 & 2398 & 2252 & 2406 & 2377 & 2341 & 2325\\
$1^1P_1$ & $1^+$ & 2411 & 2425 & 2456 & 2457 & 2402 & 2426 & 2417 & 2389 & 2468 \\
$1^3P_1$ & $1^+$ & 2427 & 2447 & 2467 & 2465 & 2417 & 2469 & 2490 & 2407 & 2631 \\
$1^3P_2$ & $2^+$ & 2456 & 2461 & 2502 & 2501 & 2466 & 2460 & 2460 & 2477 & 2743\\

$2^3P_0$ & $0^+$ & & 2976 & 2931 & 2932 & 2752 & 2919 & 2949 & 2758\\
$2^1P_1$ & $1^+$ & & 3016 & 2924 & 2933 & 2866 & 2932 & 2995 & 2792 \\
$2^3P_1$ & $1^+$ & & 3034 & 2961 & 2952 & 2926 & 3021 & 3045 & 2802\\
$2^3P_2$ & $2^+$ & 2893 & 3039 & 2957 & 2957 & 2971 & 3012 & 3035 & 2860 \\

$3^3P_0$ & $0^+$ & & 3536 & 3343 &&& 3346 && 3050 \\
$3^1P_1$ & $1^+$ & & 3567 & 3328 &&& 3365 && 3082 \\
$3^3P_1$ & $1^+$ & & 3582 & 3360 &&& 3461 && 3085\\
$3^3P_2$ & $2^+$ & 3214 & 3584 & 3353 &&& 3407 && 3142\\

$1^3D_1$ & $1^-$ & 2775 & 2755 & 2817 & 2816 & 2740 & 2788 & 2795 & 2750 \\
$1^1D_2$ & $2^-$ & 2789 & 2754 & 2816 & 2827 & 2693 & 2806 & 2775 & 2639 \\
$1^3D_2$ & $2^-$ & 2737 & 2783 & 2845 & 2834 & 2789 & 2850 & 2833 & 2727\\
$1^3D_3$ & $3^-$ & 2796 & 2788 & 2833 & 2833 & 2719 & 2863 & 2799 & 2633\\

$2^3D_1$ & $1^-$ && 3315 & 3231 & 3231 & 3.168 & 3228 && 3052\\
$2^1D_2$ & $2^-$ && 3318 & 3212 & 3225 & 3.145 & 3259 && 2997\\
$2^3D_2$ & $2^-$ && 3341 & 3248 & 3235 & 3.215 & 3307 && 3029\\
$2^3D_3$ & $3^-$ && 3355 & 3226 & 3226 & 3.170 & 3335 && 2999\\

$1^3F_2$ & $2^+$ & 3105 & & 3132 & 3132 && 3090 & 3091\\
$1^1F_3$ & $3^+$ & 3087 & & 3108 & 3123 && 3129 & 3074 \\
$1^3F_3$ & $3^+$ & 2998 & & 3143 & 3129 && 3145 & 3123 \\
$1^3F_4$ & $4^+$ & 3073 & & 3132 & 3113 && 3187 & 3101\\
\end{tabular}
\end{ruledtabular}
\end{table*}

In 2013, the LHCb detector found $D^+ \pi^-$, $D^0 \pi^+$ and $D^{*+} \pi^-$ final state mass spectra at the centre-of-mass energy 7 TeV of $pp$ collision \cite{Aaij2013}. They have observed the rich spectrum of nonstrange charmed mesons, $D_J^*(2580)^0$ and $D_J^0(2740)^0$ with unnatural parity ($0^-, 1^+, 2^-,...$) in the $D^{*+} \pi^-$ decay mode. The mass spectra analysis of $D^+ \pi^-$, $D^0 \pi^+$ and $D^{*+} \pi^-$ reconstruct the masses and the decay widths of $D_2^*(2460)$. The $D_J^*(2650)^0$ and $D_J^*(2760)^0$ are found with the natural parity ($0^+, 1^-, 2^+,...$) in the $D^{*+} \pi^-$ mass spectra. Along with these they have also got the resonant structures in a region around 3 GeV. The $D_{J}(3000)^0$ was observed in $D^{*+} \pi^-$ decay mode with unnatural parity and the $D{_{J}^*}(3000)^0$ in $D^+ \pi^-$ with natural parity \cite{Aaij2013}.

Earlier, the $BABAR$ experiment had collected the data sample of excited $D$ mesons resonances corresponding to an integrated luminosity 454 fb$^{-1}$ of $e^+ e^-$ collision at the center-of-mass energy 10.58 GeV \cite{del2010}. The masses and decay widths of the observed $D$ mesons ($D(2550)^0$, $D^*(2600)^{0/+}$, $D(2750)^0$ and $D^*(2760)^{0/+}$), are reconstructed from $D^+\pi^-$, $D^0\pi^+$ and $D^{*+}\pi^-$ decay resonances. Moreover, the helicity distribution analysis identified $D(2550)^0$ and $D^*(2600)^0$ as a $2S$ doublet of spin-parity $0^-$ and $1^-$ respectively; and the states $D(2750)^0$ and $D^*(2760)^{0/+}$ belong to $L=2$ ($L$ is the orbital angular momentum). The masses, decay widths, spin-parity observed by the experimental groups LHCb \cite{Aaij2016,Aaij2015,Aaij2013} and the $BABAR$ \cite{del2010} are presented in Table \ref{tab1} with their respective observed decay modes.

Experimentally, the Dalitz plot model in the $B$ decay production determines the spin-parity and the prompt production analysis differentiate the hadrons with natural  and unnatural parity. Moreover, the ratio of the branching fractions measurement of strong decay modes can help to classify the decaying mesons. It is very crucial to assign the spin-parity of hadrons which facilitate the determination of experimental properties. According to the latest Review of Particle Physics (RPP) by Particle Data Group (PDG), the $J^P$ ($J$ is the total spin and $P$ is parity) values of $D_1(2420)^{\pm}$, $D(2550)^0$, $D_J^*(2600)$, $D^*(2640)^{\pm}$, $D(2740)^0$ and $D(3000)^0$ mesons are not yet confirmed from the known experimental measurements \cite{Tanabashi2018-19}. Many theoretical groups have computed the excited state masses of charmed mesons with the help of various potential models. Recently, Jiao-Kai Chen obtained the radial and orbital Regge trajectories by applying the Bohr-Sommerfeld quantization approach \cite{Chen2018}. Other variants include semi-relativistic approach \cite{Kher2017}, Godfrey-Isgur (GI) relativized quark model \cite{Godfrey2016}, relativistic quark model \cite{Sun2013}, Lakhina and Swanson proposed nonrelativistic constituent quark model \cite{Li2011}, the Quantum Chromodynamics (QCD) ­motivated relativistic quark model based on the quasipotential approach \cite{Ebert2010}, relativistic quark model including the leading order corrections in $1/m$ \cite{DiPierro2001}, the Blankenbecler-Sugar equation in the framework of heavy-light interaction models \cite{Lahde2000}, the lattice QCD \cite{Cichy2016} etc..

We summarize the predicted mass spectra in Table \ref{tab2} (the symbol $\cal{N}$$^{2S+1}L_J$ is used to represent the meson quantum state; where $\cal{N}$, $L$ and $S$ denote the radial, orbital and the intrinsic spin quantum number respectively). Here, we take their comparison with experimental data and make following conclusions,

\begin{enumerate}[label=\roman*.]
\item Two $1S$ states ($D$ and $D^*$) and the four $1P$ states ($D{_{0}^*}(2300)$, $D_1(2420)$, $D_1(2430)$ and $D{_{2}^*}(2460)$) are well established with their respective $J^P$ values.
\item $D(2550)^0$ was observed by experimental groups LHCb \cite{Aaij2013} and $BABAR$ \cite{del2010}. They both suggested that such a state has unnatural parity (but the PDG-2018 \cite{Tanabashi2018-19} need more confirmation). The theoretical studies identified its quantum state $2^1S_0$.
\item $D{_{J}^*}(2600)^0$ and $D^*(2640)^0$ are probably the same state. From LHCb \cite{Aaij2013} and $BABAR$ \cite{del2010} its $J^P$ value is consistent with natural parity and it can be a candidate of $2^3S_1$.
\item $D(2740)^0$ was observed in a single experiment LHCb \cite{Aaij2013} with unnatural parity and it can be a candidate of $1^1D_2$ or $1^3D_2$ state.
\item $D{_{3}^*}(2750)$ belongs to $1^3D_3$ quantum state. Experimentally, the LHCb \cite{Aaij2015} determined its $J^P$ value $3^-$. Yet the state $D{_{1}^*}(2750)$ is not observed experimentally.
\item So far the nature of $D_J(3000)^0$, $D{_{J}^*}(3000)^0$ and $D{_{2}^*}(3000)^0$ mesons are unsolved theoretically. According to LHCb \cite{Aaij2013} the $D_J(3000)^0$ has unnatural parity. So it can be a candidate of $3^1S_0$ and $2^3P_1$ states. $D{_{J}^*}(3000)^0$ has natural parity and may belongs to $3^3S_1$, $2^3P_2$, $1^3F_2$ and $1^3F_4$ quantum states. The LHCb \cite{Aaij2016} measured the spin parity of $D{_{2}^*}(3000)^0$ as $2^+$ and can belongs to quantum states $3^3P_2$ and $1^3F_2$.
\end{enumerate}

In Ref. \cite{Godfrey2016}, S. Godfrey and K. Moats are used $^3P_0$ quark-pair-creation (QPC) model, and identified $D_J(2550)^0$, $D_J^*(2600)^0$, $D_1^*(2760)^0$, $D_3^*(2760)^-$, $D_J(2750)^0$, $D_J(3000)^0$, and $D_J^*(3000)^0$ states as $2^1S_0$, $2^3S_1$, $1^3D_1$, $1^3D_3$, $1D_2$, $3^1S_0$, and $1^3F_4$ respectively; through their strong decays analysis. Y. Sun \textit{et al.}  \cite{Sun2013} calculate the strong decays of $3S$, $2P$, $2D$, and $1F$ states of $D$ mesons in the $^3P_0$ QPC model. They assigned $D_J(3000)^0$ and $D_J^*(3000)^0$ as $2P(J^P = 1^+)$ and $2^3P_0$ respectively. Also, the Ref.  \cite{Li2011} used the same model and examined: $D(2550)$ as $2^1S_0$, $D(2750)$ (or $D(2760)$) as $1^3D_3$ state, and the state $D(2600)$ identify as the low-mass mixing of $1^3D_1 - 2^3S_1$ states. Refs. \cite{Zhong2010,Xiao2014} are determined the strong decay rates of excited heavy-light mesons in the chiral constituent quark model. They predict $D(2760)$ as $1^3D_3$ and $D_J^*(3000)^0$ as $1^3F_4$, and $D(2600)$, $D(2750)$ and $D_J(3000)^0$ are found to be a mixed state of $1^3D_1 - 2^3S_1$, $1^1D_2 - 1^3D_1$ and $2^1P_1 - 2^3P_1$ states respectively.         

In this work, we analyze the strong decays of excited nonstrange charmed mesons observed by LHCb \cite{Aaij2016,Aaij2015,Aaij2013} and $BABAR$ \cite{del2010} Collaborations using the Heavy Quark Effective Theory (HQET) in the leading order approximations. On the basis of the strong decay widths and the branching fractions predictions of $D_1(2420)$, $D(2550)$, $D(2740)$, $D_J(3000)$, and $D{_{2}^*}(3000)^0$ and their spin partners $D^*(2640)$, $D{_{J}^*}(2600)^0$, $D_3^*(2750)$, and $D{_{J}^*}(3000)$ we have assigned their spin and parity. Also, the strong coupling constants are determined by comparing the computed strong decay widths with experimental measurements. Similar kind of studies have been done by \cite{Colangelo2006_1,Colangelo2006_2,Colangelo2008,Colangelo2010,Colangelo2012,Wang2011,Wang2012,Wang2013,Batra2015,Gupta2018} to identify the higher charmed mesonic states. The spectroscopy of a system containing one light (up $(u)$ or down $(d)$ or strange $(s)$) and one heavy (charm $(c)$ or bottom $(b)$) quark provides an excellent base to study the Quantum Chromodynamics (QCD) in the low energy regime. Additionally, our tentative spin-parity assignment of nonstrange charmed mesons allow to construct the Regge trajectory in $(J, M^2)$ and $(n_r, M^2)$ planes, where $J$ is the total spin, $n_r$ is the radial principal quantum number and $M^2$ is the square of the meson mass. They estimate the masses of 1$^1D_2$, 1$^3D_3$, 3$^1S_0$, 3$^3S_1$, 1$^1F_3$, 1$^3F_4$, 2$^3D_3$, 3$^3P_2$ and 2$^3F_4$ states. Their strong decay rates and branching fraction studies can guide to the experimentalists for searching them in a respective decay channels.      

This paper is arranged as follows: after the introduction, section \ref{sec2} is a brief description of HQET used to study the strong decays. Section \ref{sec3} presents results and discussion, where we attempt to identify the spin and parity of experimentally known excited nonstrange charmed mesons. In section \ref{sec4} we plot the Regge trajectories in $(J, M^2)$ and $(n_r, M^2)$ planes using the masses from PDG-2018 \cite{Tanabashi2018-19}. Further, we analyzed the strong decay rates and the branching fractions of 1$^1D_2$, 1$^3D_3$, 3$^1S_0$, 3$^3S_1$, 1$^1F_3$, 1$^3F_4$, 2$^3D_3$, 3$^3P_2$ and 2$^3F_4$ states lying on the Regge lines. Finally, the conclusions are presented in section \ref{sec5}.

\section{Theoretical Framework}
\label{sec2}

In the framework of heavy quark effective theory (HQET) the properties of heavy-light mesons can be determined systematically by considering infinite mass of one heavy quark, i.e. $m_Q \rightarrow \infty$ \cite{Neubert1994}. The heavy quark spin and the flavor symmetry arising from the QCD are demolished in this heavy quark (HQ) mass limit and classify the heavy-light mesons according to the total angular momentum of the light antiquark $\vec{s}_l$, $\vec{s}_l =  \vec{s}_{\bar{q}} + \vec{l}$, where $\vec{s}_{\bar{q}}$ and $\vec{l}$ are the spin and the orbital angular momentum of the light antiquark respectively \cite{Neubert1994}. 

\begin{table*}
\caption{\label{tab3}
The strong decay widths of nonstrange charmed mesons with possible quantum state assignments (in MeV).}
\begin{ruledtabular}
\begin{tabular}{ccccccccccccc}
Meson & $\cal{N}$$^{2S+1}L_J$ & Decay mode & LHCb(2016) \cite{Aaij2016} & LHCb(2015) \cite{Aaij2015} & LHCb(2013) \cite{Aaij2013} & $BABAR$(2010) \cite{del2010}\\
\noalign{\smallskip}\hline\noalign{\smallskip} 
$D_1(2420)^0$ & 1$^1P_1$ & $D^{*+} \pi^-$ &&& 56.2711$h_T^2$ & 56.6228$h_T^2$\\
&& $D^{*0} \pi^0$ &&& 29.3228$h_T^2$ & 29.5040$h_T^2$ \\
&& $D{_{s}^{*+}} K^-$ &&& $-$ & $-$ \\
&& $D^{*0} \eta$ &&& $-$ & $-$\\
&& Total &&& 85.5939$h_T^2$ & 86.1268$h_T^2$ \\
&& $h_T$ &&& 0.641 & 0.604 \\
\noalign{\smallskip}\hline\noalign{\smallskip} 
$D{_{2}^*}(2460)^0$ & 1$^3P_2$ & $D^+ \pi^-$ & 127.978$h_T^2$ && 124.786$h_T^2$ & 126.52$h_T^2$ \\
&& $D^0 \pi^0$ & 66.8656$h_T^2$ && 65.2218$h_T^2$ & 66.1147$h_T^2$ \\
&& $D^+_s K^-$ & $\approx$ 0 && $\approx$ 0 & $\approx$ 0  \\
&& $D^0 \eta$ & $-$ &&  $-$ &  $-$   \\
&& $D^{*+} \pi^-$ & 56.3891$h_T^2$ && 54.3938$h_T^2$ & 55.4757$h_T^2$ \\
&& $D^{*0} \pi^0$ & 29.7173$h_T^2$ && 28.6838$h_T^2$ & 29.2442$h_T^2$\\
&& $D{_{s}^{*+}} K^-$ & $-$ && $-$ & $-$  \\
&& $D^{*0} \eta$ & $-$ && $-$ &  $-$ \\
&& Total & 280.95$h_T^2$ && 273.085$h_T^2$ & 277.355$h_T^2$ \\
&& $h_T$ & 0.409 && 0.409 & 0.427 \\
\noalign{\smallskip}\hline\noalign{\smallskip} 
$D{_{2}^*}(2460)^+$ & 1$^3P_2$ & $D^0 \pi^+$ &&& 131.875$h_T^2$   \\
&& $D^+ \pi^0$ &&& 63.6968$h_T^2$ \\
&& $D^+_s K^0$ &&& $\approx$ 0 \\
&& $D^{+} \eta$ &&& $-$ \\
&& $D^{*0} \pi^+$ &&& 58.0702$h_T^2$\\
&& $D^{*+} \pi^0$ &&& 28.494$h_T^2$ \\
&& $D{_{s}^{*+}} K^0$ &&& $-$  \\
&& $D^{*+} \eta$ &&& $-$ \\
&& Total &&& 282.136$h_T^2$ \\
&& $h_T$ &&& 0.415 \\
\noalign{\smallskip}\hline\noalign{\smallskip} 
$D{_{2}^*}(2460)^-$ & 1$^3P_2$ & $D^0 \pi^-$ && 137.440$h_T^2$ \\
&& $D^- \pi^0$ && 66.4136$h_T^2$ \\
&& $D^-_s K^0$ && $\approx$ 0 \\
&& $D^{-} \eta$ && $-$ \\
&& $D^{*0} \pi^-$ && 61.5865$h_T^2$ \\
&& $D^{*-} \pi^0$ && 30.2238$h_T^2$ \\
&& $D{_{s}^{*-}} K^0$ && $-$ \\
&& $D^{*-} \eta$ && $-$ \\
&& Total && 295.664$h_T^2$ \\
&& $h_T$ && 0.400 \\
\noalign{\smallskip}\hline\noalign{\smallskip}
$D(2550)^0$ & 2$^1S_0$ & $D^{*+} \pi^-$ &&& 864.734$g_H^{\dag2}$ & 709.405$g_H^{\dag2}$ \\
&& $D^{*0} \pi^0$ &&& 441.692$g_H^{\dag2}$ & 363.314$g_H^{\dag2}$ \\
&& $D{_{s}^{*+}} K^-$ &&& $-$ & $-$ \\
&& $D^{*0} \eta$ &&& 3.87486$g_H^{\dag2}$ & $-$ \\
&& Total &&& 1310.30$g_H^{\dag2}$ & 1072.72$g_H^{\dag2}$ \\
&& $g_H^{\dag}$ &&& 0.368 & 0.348 \\
\noalign{\smallskip}
\end{tabular}
\end{ruledtabular}
{continued...}
\end{table*}

\begin{table*}
\addtocounter{table}{-1}
\caption{\label{tab3}
The strong decay widths of nonstrange charmed mesons with possible quantum state assignments (in MeV).}
\begin{ruledtabular}
\begin{tabular}{ccccccccccccc}
Meson & $\cal{N}$$^{2S+1}L_J$ & Decay mode & LHCb(2016) \cite{Aaij2016} & LHCb(2015) \cite{Aaij2015} & LHCb(2013) \cite{Aaij2013} & $BABAR$(2010) \cite{del2010}\\
\noalign{\smallskip}\hline\noalign{\smallskip} 
$D{_{J}^*}(2600)^0$ & 2$^3S_1$ & $D^+ \pi^-$ & 680.382$g_H^{\dag2}$ &&& 541.421$g_H^{\dag2}$ \\
&& $D^0 \pi^0$ & 345.515$g_H^{\dag2}$ &&& 274.992$g_H^{\dag2}$ \\
&& $D^+_s K^-$ & 199.173$g_H^{\dag2}$ &&& 104.757$g_H^{\dag2}$ \\
&& $D^0 \eta$ & 47.9086$g_H^{\dag2}$ &&& 29.1069$g_H^{\dag2}$ \\
&& $D^{*+} \pi^-$ & 886.679$g_H^{\dag2}$ &&& 656.589$g_H^{\dag2}$  \\
&& $D^{*0} \pi^0$ & 450.689$g_H^{\dag2}$ &&& 334.839$g_H^{\dag2}$ \\
&& $D{_{s}^{*+}} K^-$ & 78.3291$g_H^{\dag2}$ &&& $\approx$ 0\\
&& $D^{*0} \eta$ & 31.0273$g_H^{\dag2}$ &&& 8.24532$g_H^{\dag2}$ \\
&& Total & 2719.70$g_H^{\dag2}$ &&& 1949.95$g_H^{\dag2}$ \\
&& $g_H^{\dag}$ & 0.262 &&& 0.218 \\
\noalign{\smallskip}\hline\noalign{\smallskip} 
$D{_{J}^*}(2600)^0$ & 2$^3S_1$ & $D^{*+} \pi^-$ &&& 781.919$g_H^{\dag2}$ \\
&& $D^{*0} \pi^0$ &&& 397.965$g_H^{\dag2}$\\
&& $D{_{s}^{*+}} K^-$ &&& 33.6215$g_H^{\dag2}$ \\
&& $D^{*0} \eta$ &&& 19.8058$g_H^{\dag2}$ \\
&& $D^+ \pi^-$ &&& 617.246$g_H^{\dag2}$ \\
&& $D^0 \pi^0$ &&& 313.774 $g_H^{\dag2}$\\
&& $D^+_s K^-$ &&& 155.109$g_H^{\dag2}$ \\
&& $D^0 \eta$ &&& 39.2693$g_H^{\dag2}$ \\
&& Total &&& 2358.71$g_H^{\dag2}$ \\
&& $g_H^{\dag}$ &&& 0.244 \\
\noalign{\smallskip}\hline\noalign{\smallskip} 
$D(2740)^0$ & 1$^3D_2$ & $D^{*+} \pi^-$ &&& 126.986$k_Y^2$ \\
&& $D^{*0} \pi^0$ &&& 65.8248$k_Y^2$\\
&& $D{_{s}^{*+}} K^-$ &&& 1.92685$k_Y^2$ \\
&& $D^{*0} \eta$ &&& 1.30793$k_Y^2$ \\
&& Total &&& 196.046$k_Y^2$ \\
&& $k_Y$ &&& 0.611 \\
\noalign{\smallskip}\hline\noalign{\smallskip} 
$D{_{3}^*}(2750)^0$ & 1$^3D_3$ & $D^+ \pi^-$ & 190.520$k_Y^2$ && 172.087$k_Y^2$ & 175.794$k_Y^2$\\
&& $D^0 \pi^0$ & 98.5331$k_Y^2$ && 89.0767$k_Y^2$ & 90.979$k_Y^2$ \\
&& $D^+_s K^-$ & 20.954$k_Y^2$ && 17.2091$k_Y^2$ & 17.9416$k_Y^2$ \\
&& $D^0 \eta$ & 7.03403$k_Y^2$ && 5.94594$k_Y^2$ & 6.16072$k_Y^2$\\
&& $D^{*+} \pi^-$ & 99.8604$k_Y^2$ && 88.0932$k_Y^2$ & 90.4411$k_Y^2$ \\
&& $D^{*0} \pi^0$ & 51.6265$k_Y^2$ && 45.5895$k_Y^2$ & 46.7945$k_Y^2$ \\
&& $D{_{s}^{*+}} K^-$ & 2.88624$k_Y^2$ && 2.01803$k_Y^2$ & 2.17967$k_Y^2$ \\
&& $D^{*0} \eta$ & 1.53565$k_Y^2$ && 1.16923$k_Y^2$ & 1.2393$k_Y^2$ \\
&& Total & 472.95$k_Y^2$ && 421.189$k_Y^2$ & 431.53$k_Y^2$ \\
&& $k_Y$ & 0.449 && 0.420 & 0.376 \\
\noalign{\smallskip}\hline\noalign{\smallskip} 
$D{_{3}^*}(2750)^0$ & 1$^3D_3$ & $D^{*+} \pi^-$ &&& 88.8216$k_Y^2$ & 82.6448$k_Y^2$ \\
&& $D^{*0} \pi^0$ &&& 45.9633$k_Y^2$ & 42.7926$k_Y^2$ \\
&& $D{_{s}^{*+}} K^-$ &&& 2.06754$k_Y^2$ & 1.66585$k_Y^2$ \\
&& $D^{*0} \eta$ &&& 1.1908$k_Y^2$ & 1.0128$k_Y^2$ \\
&& $D^+ \pi^-$ &&& 173.239$k_Y^2$ & 163.424$k_Y^2$\\
&& $D^0 \pi^0$ &&& 89.6677$k_Y^2$ & 84.6299$k_Y^2$\\
&& $D^+_s K^-$ &&& 17.4355$k_Y^2$ & 15.5399$k_Y^2$ \\
&& $D^0 \eta$ &&& 6.01244$k_Y^2$ & 5.45238$k_Y^2$  \\
&& Total &&& 424.398$k_Y^2$ & 397.162$k_Y^2$  \\
&& $k_Y$ &&& 0.419 & 0.423 \\
\noalign{\smallskip}
\end{tabular}
\end{ruledtabular}
{continued...}
\end{table*}

\begin{table*}
\addtocounter{table}{-1}
\caption{\label{tab3}
The strong decay widths of nonstrange charmed mesons with possible quantum state assignments (in MeV).}
\begin{ruledtabular}
\begin{tabular}{ccccccccccccc}
Meson & $\cal{N}$$^{2S+1}L_J$ & Decay mode & LHCb(2016) \cite{Aaij2016} & LHCb(2015) \cite{Aaij2015} & LHCb(2013) \cite{Aaij2013} & $BABAR$(2010) \cite{del2010}\\
\noalign{\smallskip}\hline\noalign{\smallskip} 
$D{_{3}^*}(2750)^+$ & 1$^3D_3$ & $D^0 \pi^+$ &&& 191.164$k_Y^2$ & 188.68$k_Y^2$ \\
&& $D^+ \pi^0$ &&& 93.4520$k_Y^2$ & 92.2321$k_Y^2$ \\
&& $D^+_s K^0$ &&& 19.4484$k_Y^2$ & 18.8051$k_Y^2$ \\
&& $D^+ \eta$ &&& 6.43052$k_Y^2$ & 6.291$k_Y^2$ \\
&&$D^{*0} \pi^+$ &&& 99.3632$k_Y^2$ & 97.7744$k_Y^2$ \\
&& $D^{*+} \pi^0$ &&& 48.8147$k_Y^2$ & 48.0319$k_Y^2$ \\
&& $D{_{s}^{*+}} K^0$ &&& 2.46429$k_Y^2$ & 2.35189$k_Y^2$ \\
&& $D^{*+} \eta$ &&& 1.35745$k_Y^2$ & 1.31015$k_Y^2$ \\
&& Total &&& 462.494$k_Y^2$ & 455.476$k_Y^2$ \\
&& $k_Y$ &&& 0.380 & 0.366 \\
\noalign{\smallskip}\hline\noalign{\smallskip} 
$D{_{3}^*}(2750)^-$ & 1$^3D_3$ & $D^0 \pi^-$ &&226.341$k_Y^2$& \\
&& $D^- \pi^0$ &&110.734$k_Y^2$ &\\
&& $D^-_s K^0$ &&26.6515$k_Y^2$& \\
&& $D^- \eta$ &&8.49284$k_Y^2$ & \\
&&$D^{*0} \pi^-$ &&122.268$k_Y^2$\\
&& $D^{*-} \pi^0$ &&60.1023$k_Y^2$&\\
&& $D{_{s}^{*-}} K^0$ &&4.34870$k_Y^2$ &\\
&& $D^{*-} \eta$ &&2.10648$k_Y^2$&\\
&& Total &&561.045$k_Y^2$&\\
&& $k_Y$ &&0.433&\\
\noalign{\smallskip}\hline\noalign{\smallskip} 
$D_J(3000)^0$ & 3$^1S_0$ & $D^{*+} \pi^-$ &&& 3216.82$g_H^{\ddag2}$ \\
&& $D^{*0} \pi^0$ &&& 1623.35$g_H^{\ddag2}$ \\
&& $D{_{s}^{*+}} K^-$ &&& 1434.74$g_H^{\ddag2}$ \\
&& $D^{*0} \eta$ &&& 305.64$g_H^{\ddag2}$ \\
&& Total &&& 6580.55$g_H^{\ddag2}$ \\
&& $g_H^{\ddag}$ &&& 0.169 \\
\noalign{\smallskip}\hline\noalign{\smallskip} 
$D_J(3000)^0$ & 2$^3P_1$ & $D^{*+} \pi^-$ &&& 3315.44$h_S^{\dag2}$ \\
&& $D^{*0} \pi^0$ &&& 1669.56$h_S^{\dag2}$ \\
&& $D{_{s}^{*+}} K^-$ &&& 2409.03$h_S^{\dag2}$ \\
&& $D^{*0} \eta$ &&& 515.393$h_S^{\dag2}$ \\
&& Total &&& 7909.42$h_S^{\dag2}$ \\
&& $h_S^{\dag2}$ &&& 0.154 \\
\noalign{\smallskip}\hline\noalign{\smallskip} 
$D_J^*(3000)^0$& 3$^3S_1$ & $D^+ \pi^-$ &&& 1493.41$g_H^{\ddag2}$\\
&& $D^0 \pi^0$ &&& 753.344$g_H^{\ddag2}$ \\
&& $D^+_s K^-$ &&& 867.203$g_H^{\ddag2}$ \\
&& $D^0 \eta$ &&& 170.321$g_H^{\ddag2}$ \\
&& $D^{*+} \pi^-$ &&& 2338.80$g_H^{\ddag2}$\\
&& $D^{*0} \pi^0$ &&& 1179.62$g_H^{\ddag2}$ \\
&& $D{_{s}^{*+}} K^-$ &&& 1116.37$g_H^{\ddag2}$ \\
&& $D^{*0} \eta$ &&& 233.034$g_H^{\ddag2}$ \\
&& Total &&& 8152.1$g_H^{\ddag2}$ \\
&& $g_H^{\ddag}$ &&& 0.116 \\
\noalign{\smallskip}\hline\noalign{\smallskip} 
$D_J^*(3000)^0$& 2$^3P_2$ & $D^+ \pi^-$ &&& 2003.50$h_T^{\dag2}$ \\
&& $D^0 \pi^0$ &&& 1018.38$h_T^{\dag2}$ \\
&& $D^+_s K^-$ &&& 782.29$h_T^{\dag2}$ \\
&& $D^0 \eta$ &&& 177.739$h_T^{\dag2}$ \\
&& $D^{*+} \pi^-$ &&& 1904.84$h_T^{\dag2}$\\
&& $D^{*0} \pi^0$ &&& 967.421$h_T^{\dag2}$\\
&& $D{_{s}^{*+}} K^-$ &&& 537.317$h_T^{\dag2}$ \\
&& $D^{*0} \eta$ &&& 134.843$h_T^{\dag2}$ \\
&& Total &&& 7526.33$h_T^{\dag2}$ \\
&& $h_T^{\dag}$ &&& 0.121 \\
\end{tabular}
\end{ruledtabular}
{continued...}
\end{table*}

\begin{figure}
  \includegraphics[width=0.49\textwidth]{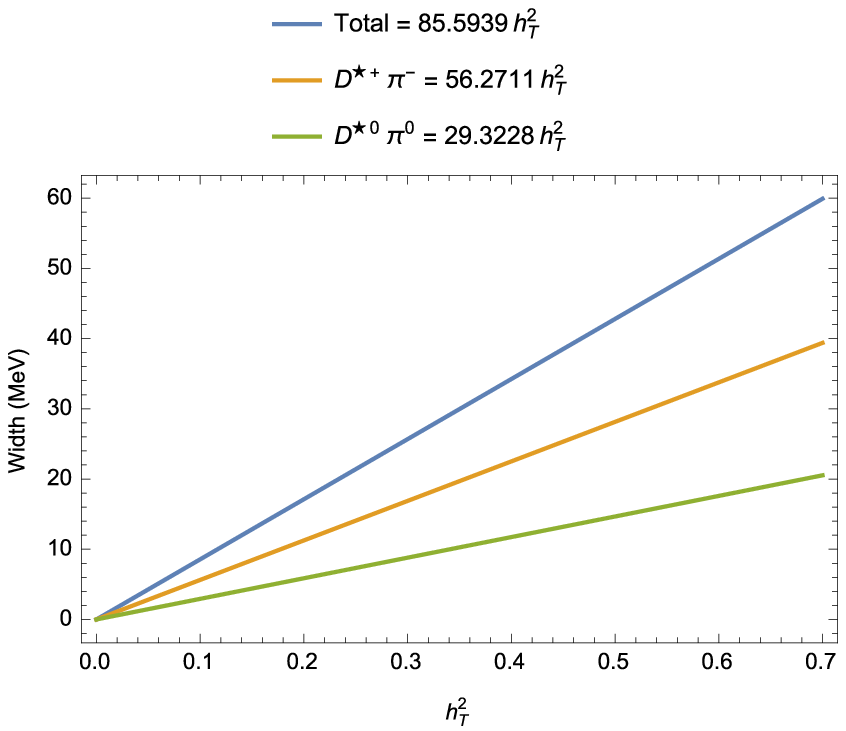}
  \includegraphics[width=0.49\textwidth]{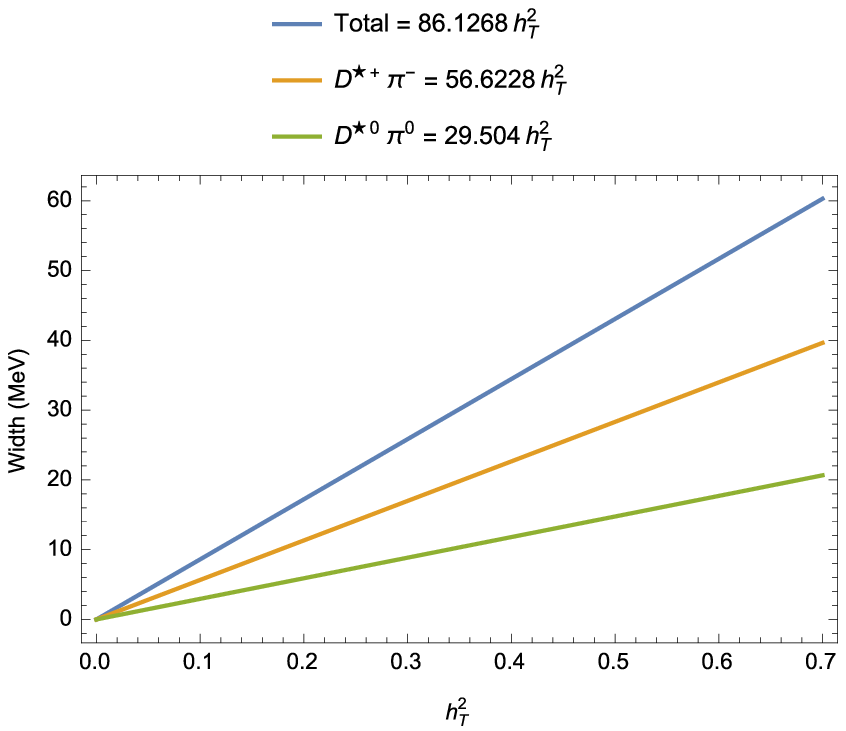}
\caption{Strong decay widths of $D_1(2420)^0$ (in MeV) changing with the square of the coupling $h_T^2$ in HQET. The masses of $D_1(2420)^0$ observed (in the decay mode $D^{*+}\pi^-$) by LHCb(2013) \cite{Aaij2013} (upper) and $BABAR$(2010) \cite{del2010} (lower) are used.}
\label{fig1}       
\end{figure}

\begin{figure}
  \includegraphics[width=0.49\textwidth]{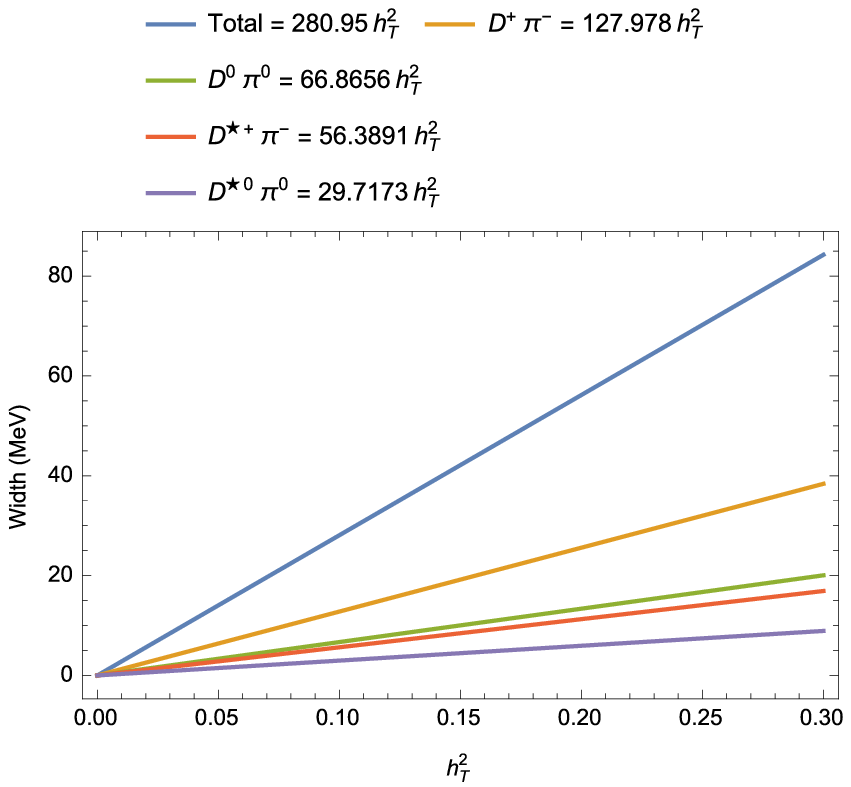}
      \includegraphics[width=0.49\textwidth]{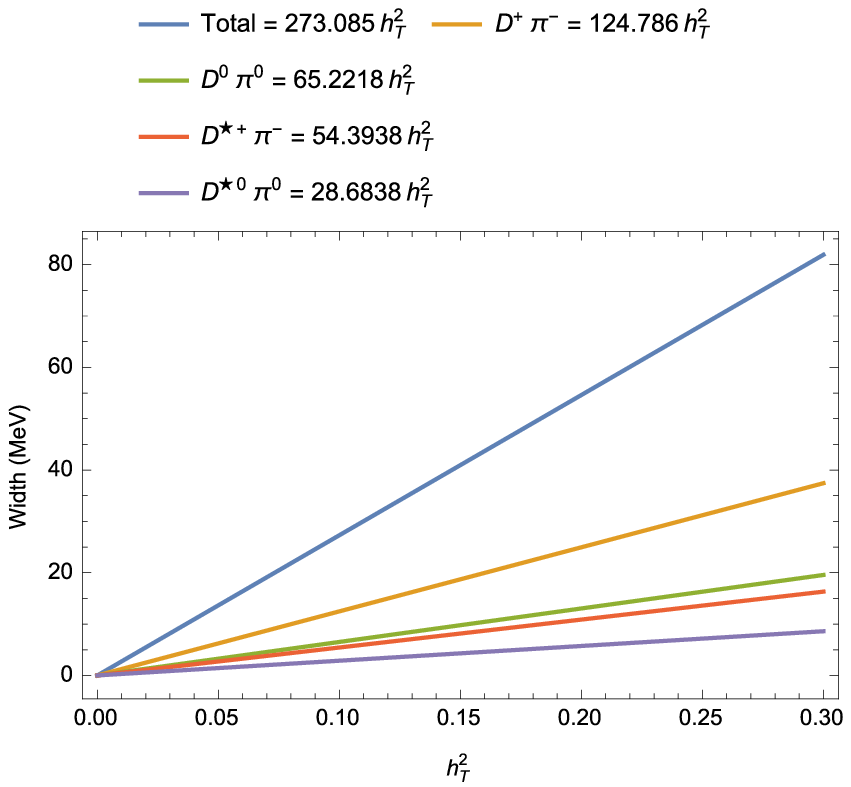}
        \includegraphics[width=0.49\textwidth]{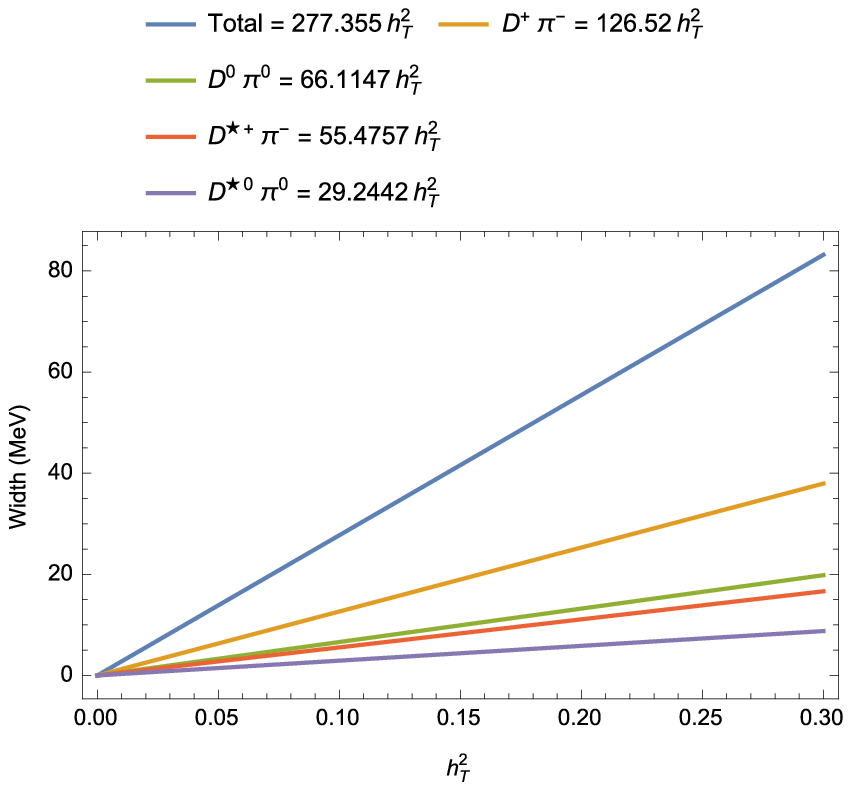}
\caption{Strong decay widths of $D_2^*(2460)^0$ (in MeV) changing with the square of the coupling $h_T^2$ in HQET. The masses of $D_2^*(2460)^0$ observed (in the decay mode $D^+\pi^-$) by LHCb(2016) \cite{Aaij2016} (upper), LHCb(2013) \cite{Aaij2013} (middle) and $BABAR$(2010) \cite{del2010} (lower) are used.}
\label{fig2}       
\end{figure}

\begin{figure}
  \includegraphics[width=0.49\textwidth]{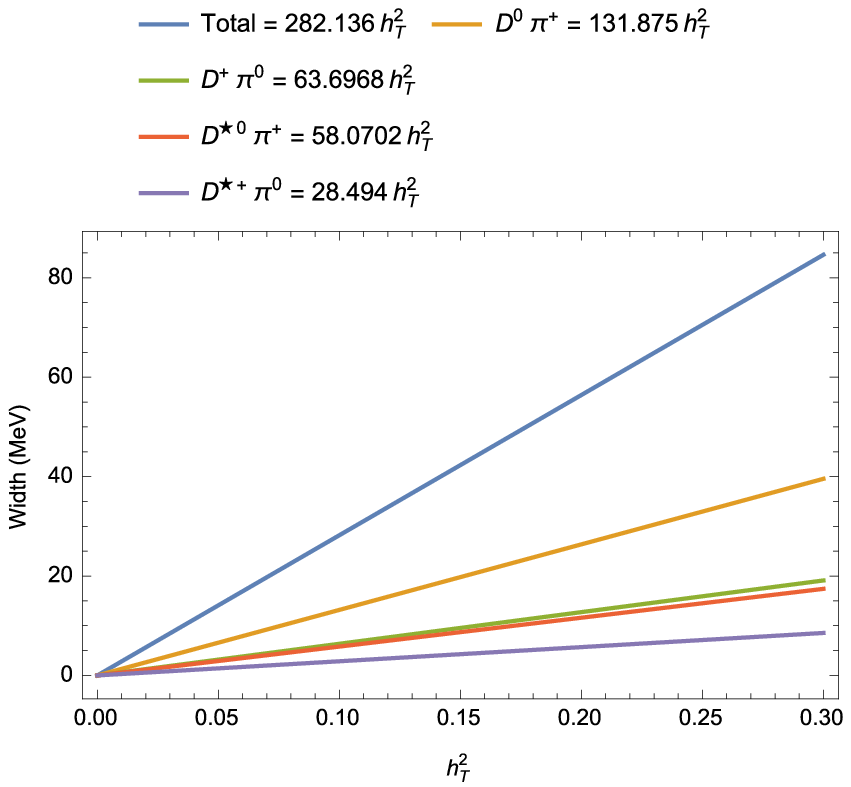}
      \includegraphics[width=0.49\textwidth]{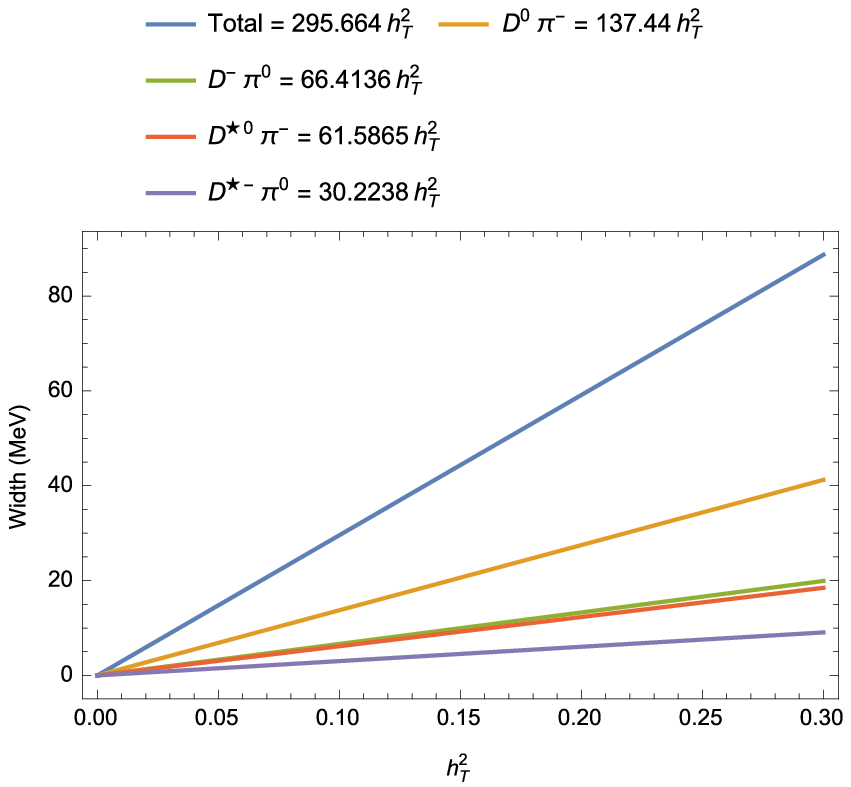}
\caption{Strong decay widths of $D_2^*(2460)^+$ (left) and $D_2^*(2460)^-$ (right) (in MeV) changing with the square of the coupling $h_T^2$ in HQET. The mass of $D_2^*(2460)^+$ observed (in the decay mode $D^0\pi^+$) by LHCb(2013) \cite{Aaij2013} (upper) and the mass of $D_2^*(2460)^-$ observed (in the decay mode $D^0\pi^-$) by LHCb(2015) \cite{Aaij2015} (lower) are used.}
\label{fig3}       
\end{figure}

\begin{figure}
  \includegraphics[width=0.49\textwidth]{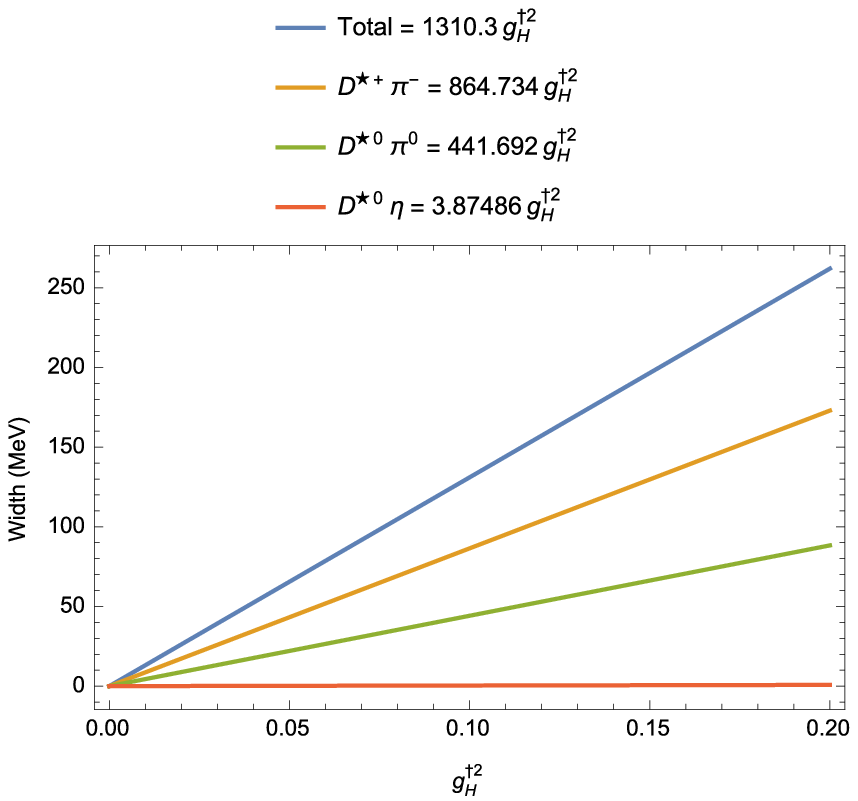}
    \includegraphics[width=0.49\textwidth]{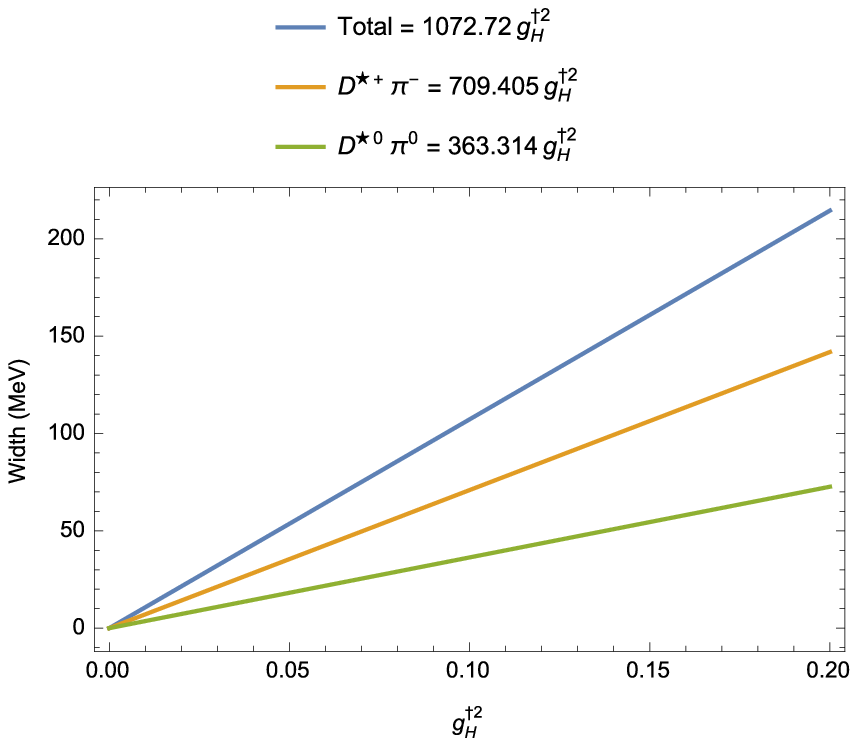}
\caption{Strong decay widths of $D(2550)^0$ (in MeV) changing with the square of the coupling $g_H^{\dag2}$ in HQET. The masses of $D(2550)^0$ observed (in the decay mode $D^{*+}\pi^-$) by LHCb(2013) \cite{Aaij2013} (upper) and $BABAR$(2010) \cite{del2010} (lower) are used.}
\label{fig4}       
\end{figure}

\begin{figure*}
  \includegraphics[width=0.49\textwidth]{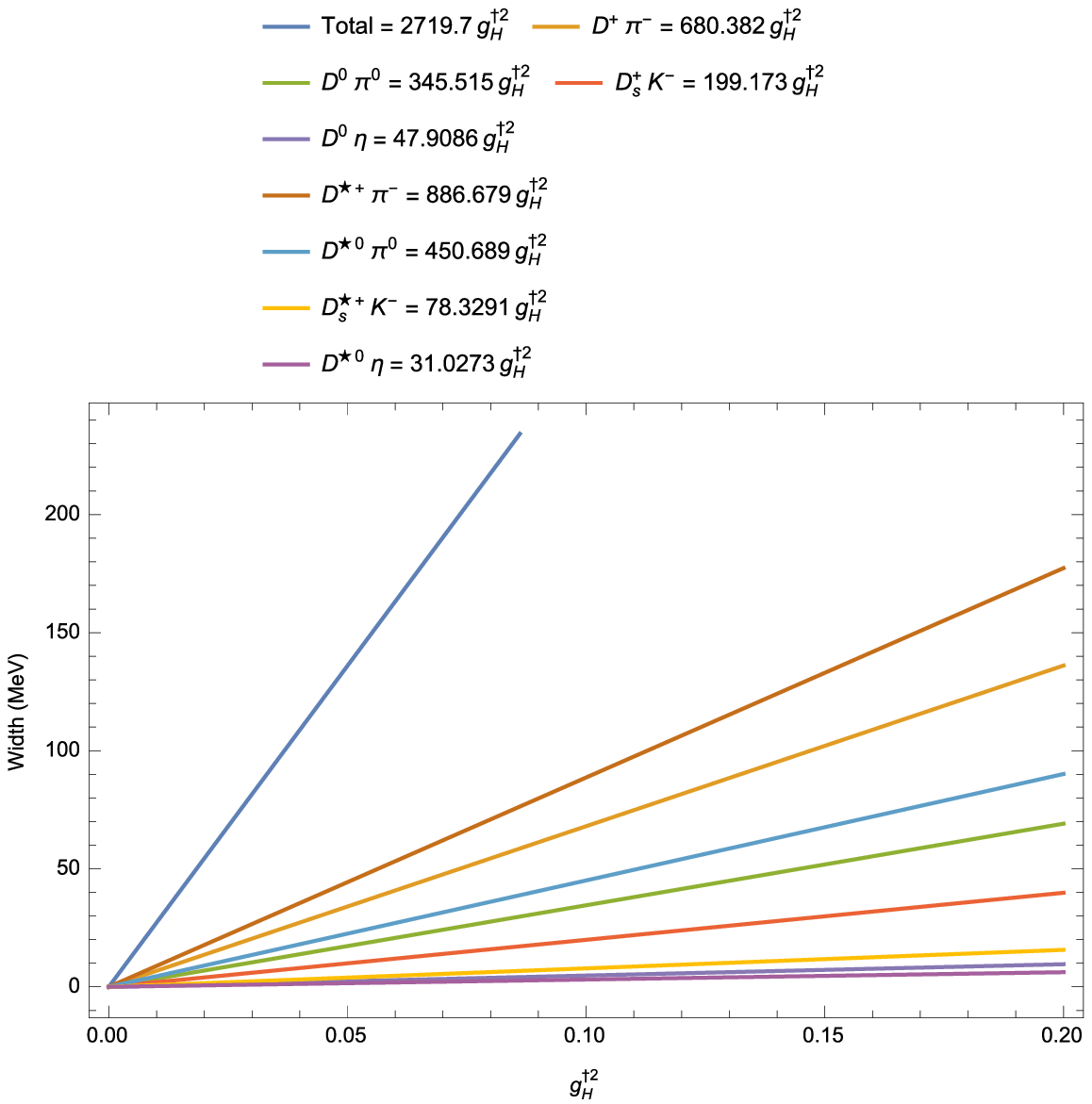}
    \includegraphics[width=0.49\textwidth]{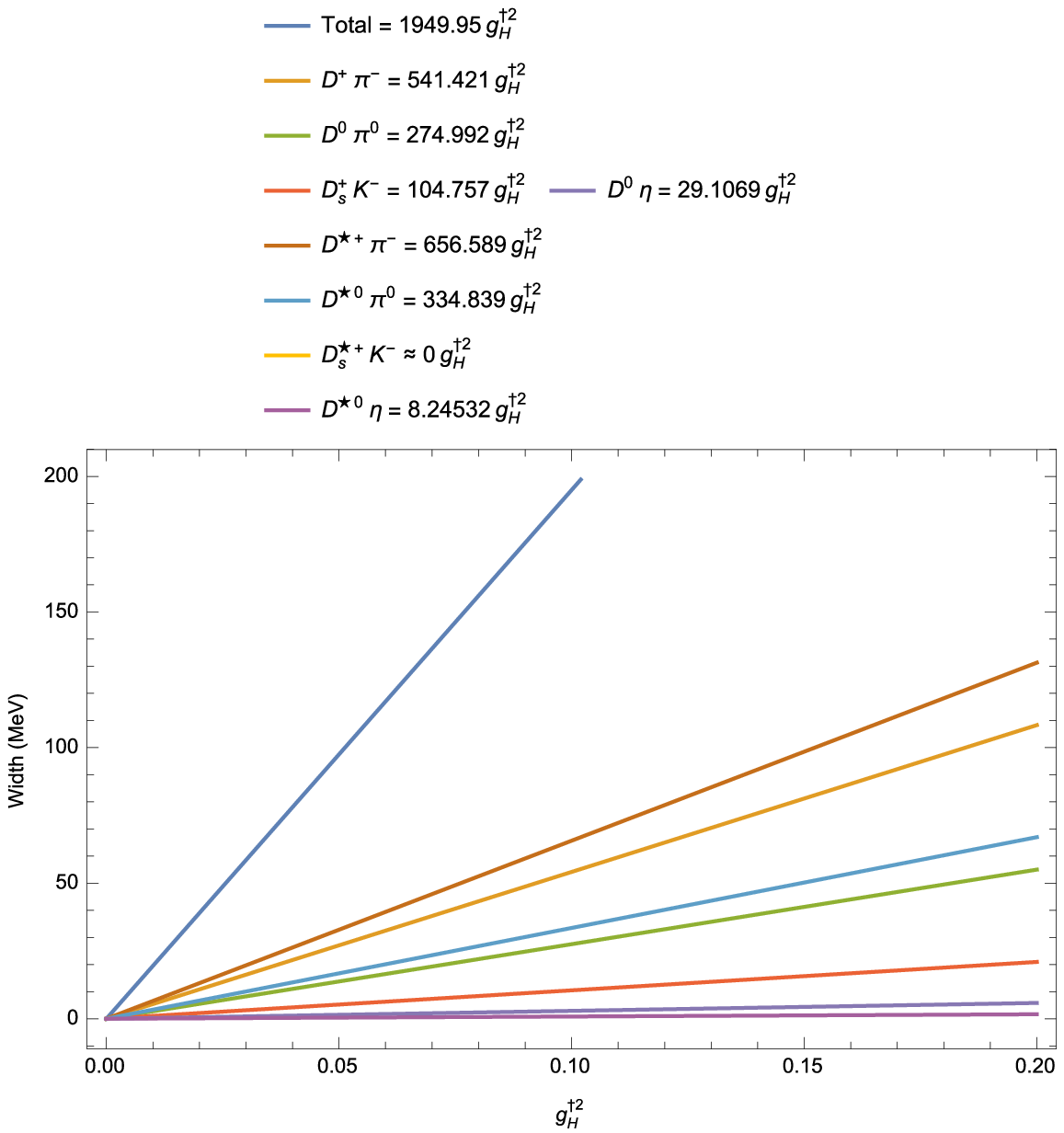}
      \includegraphics[width=0.49\textwidth]{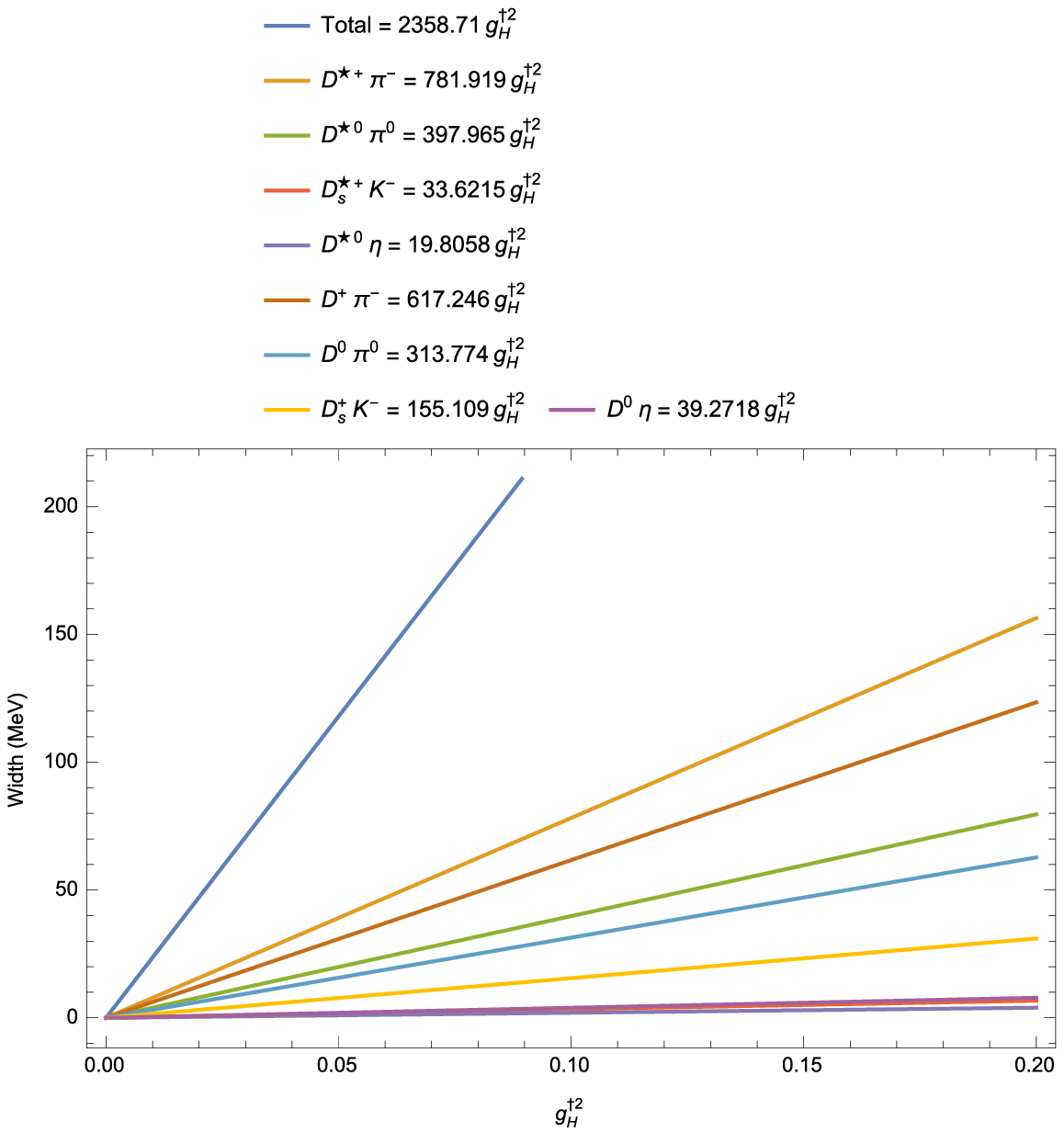}
\caption{Strong decay widths of $D_J^*(2600)^0$ (in MeV) changing with the square of the coupling $g_H^{\dag2}$ in HQET. The masses of $D_J^*(2600)^0$ observed (in the decay mode $D^+\pi^-$) by LHCb(2016) \cite{Aaij2016} (upper left) and $BABAR$(2010) \cite{del2010} (upper right), and (in the decay mode $D^{*+}\pi^-$) by LHCb(2013) \cite{Aaij2013} (lower) are used.}
\label{fig5}       
\end{figure*}

\begin{figure}
  \includegraphics[width=0.5\textwidth]{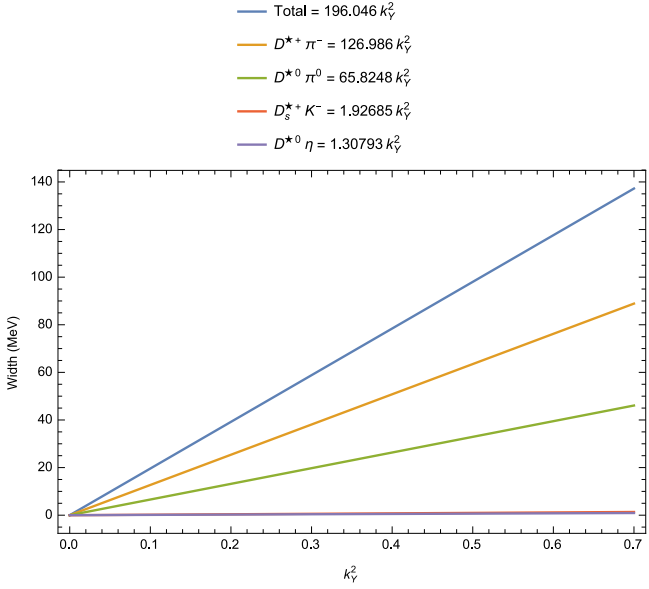}
\caption{Strong decay widths of $D(2740)^0$ (in MeV) changing with the square of the coupling $k_Y^2$ in HQET. The mass of $D(2740)^0$ observed (in the decay mode $D^{*+}\pi^-$) by LHCb(2013) \cite{Aaij2013} is used.}
\label{fig6}       
\end{figure}

\begin{figure*}
  \includegraphics[width=0.49\textwidth]{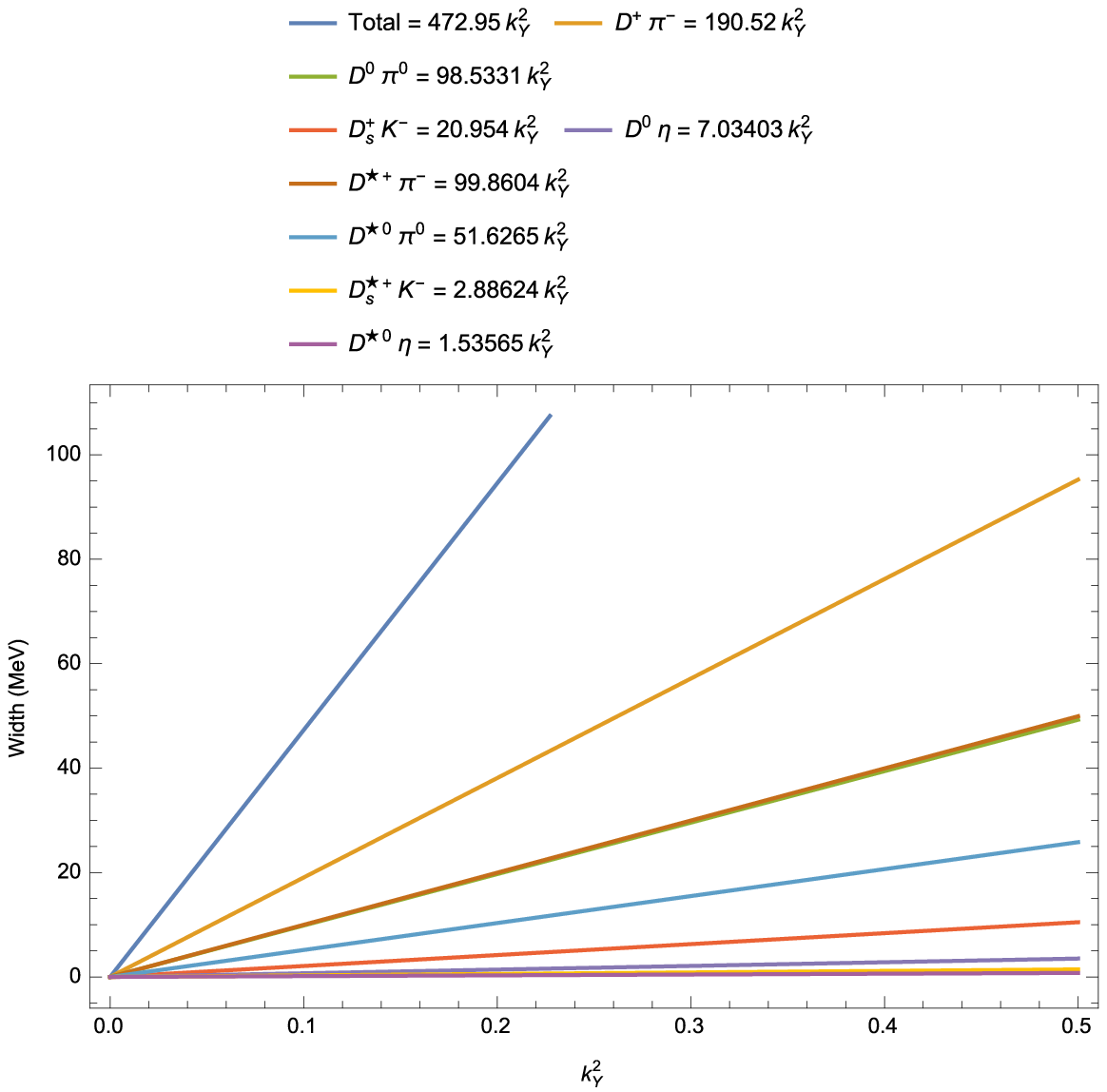}
    \includegraphics[width=0.49\textwidth]{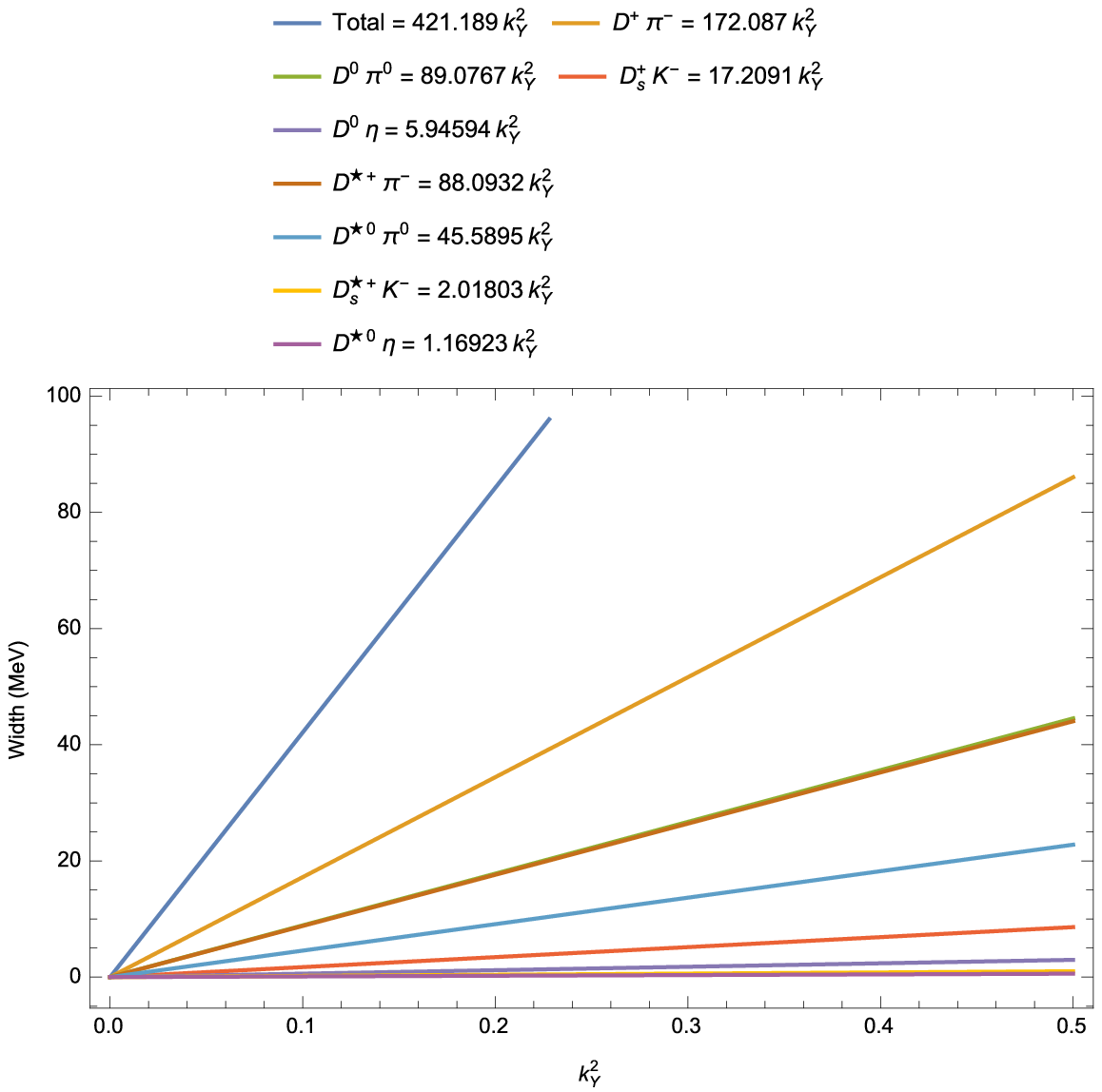}
  \includegraphics[width=0.49\textwidth]{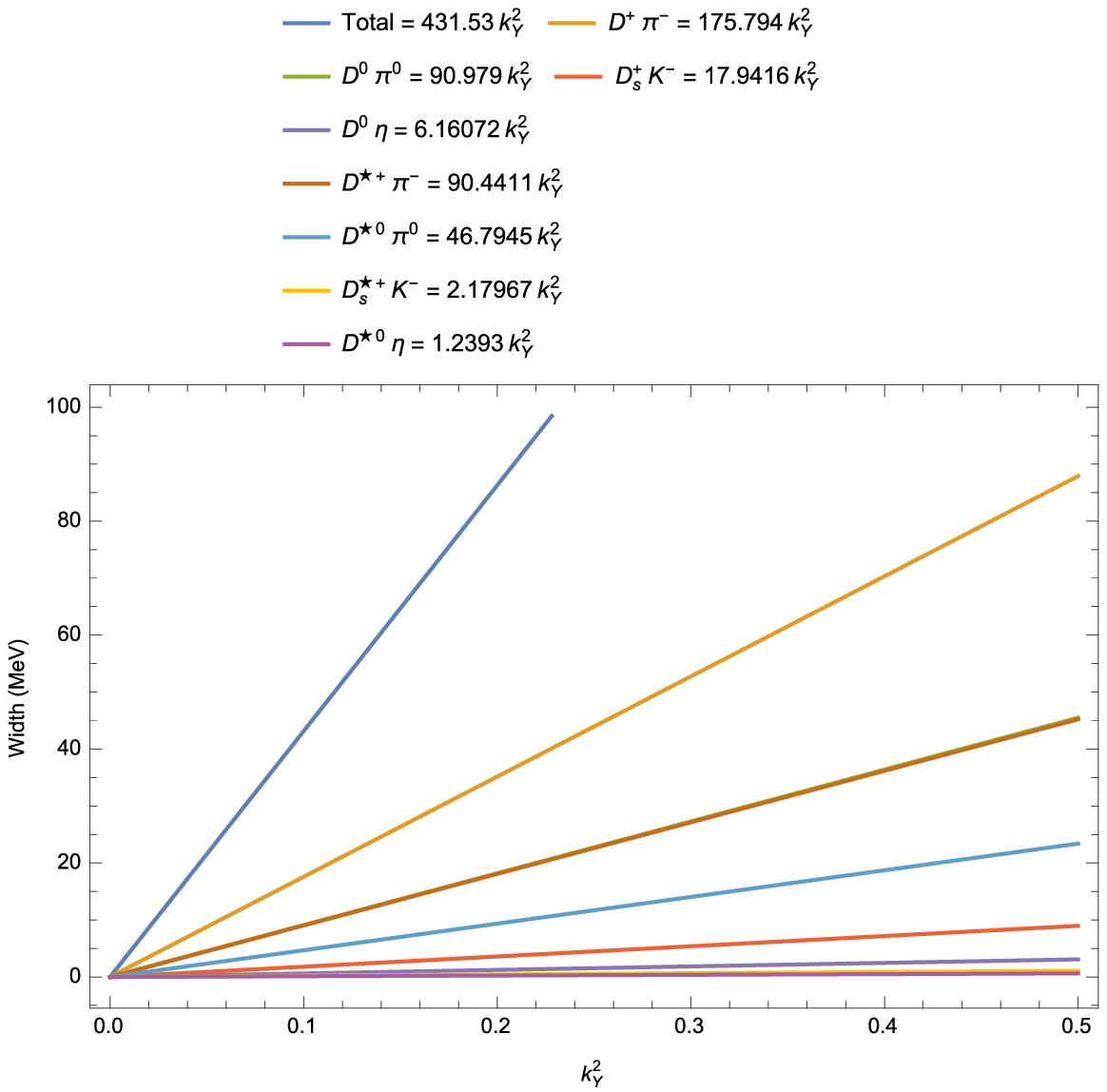}
\caption{Strong decay widths of $D_3^*(2750)^0$ (in MeV) changing with the square of the coupling $k_Y^2$ in HQET. The masses of $D_3^*(2750)^0$ observed (in the decay mode $D^+\pi^-$) by LHCb(2016) \cite{Aaij2016} (upper left), LHCb(2013) \cite{Aaij2013} (upper right) and $BABAR$(2010) \cite{del2010} (lower) are used.}
\label{fig7}       
\end{figure*}

\begin{figure}
  \includegraphics[width=0.49\textwidth]{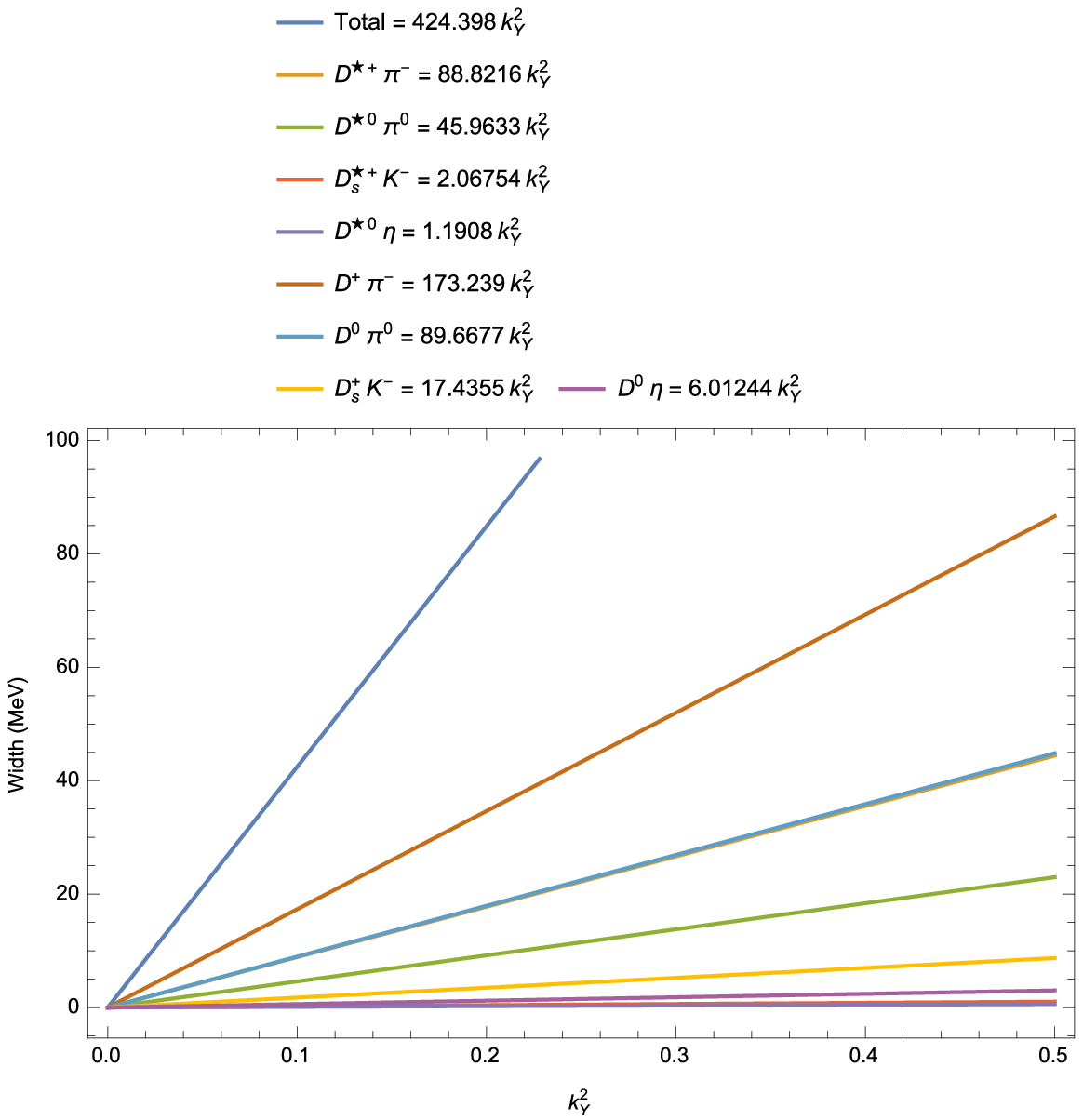}
    \includegraphics[width=0.49\textwidth]{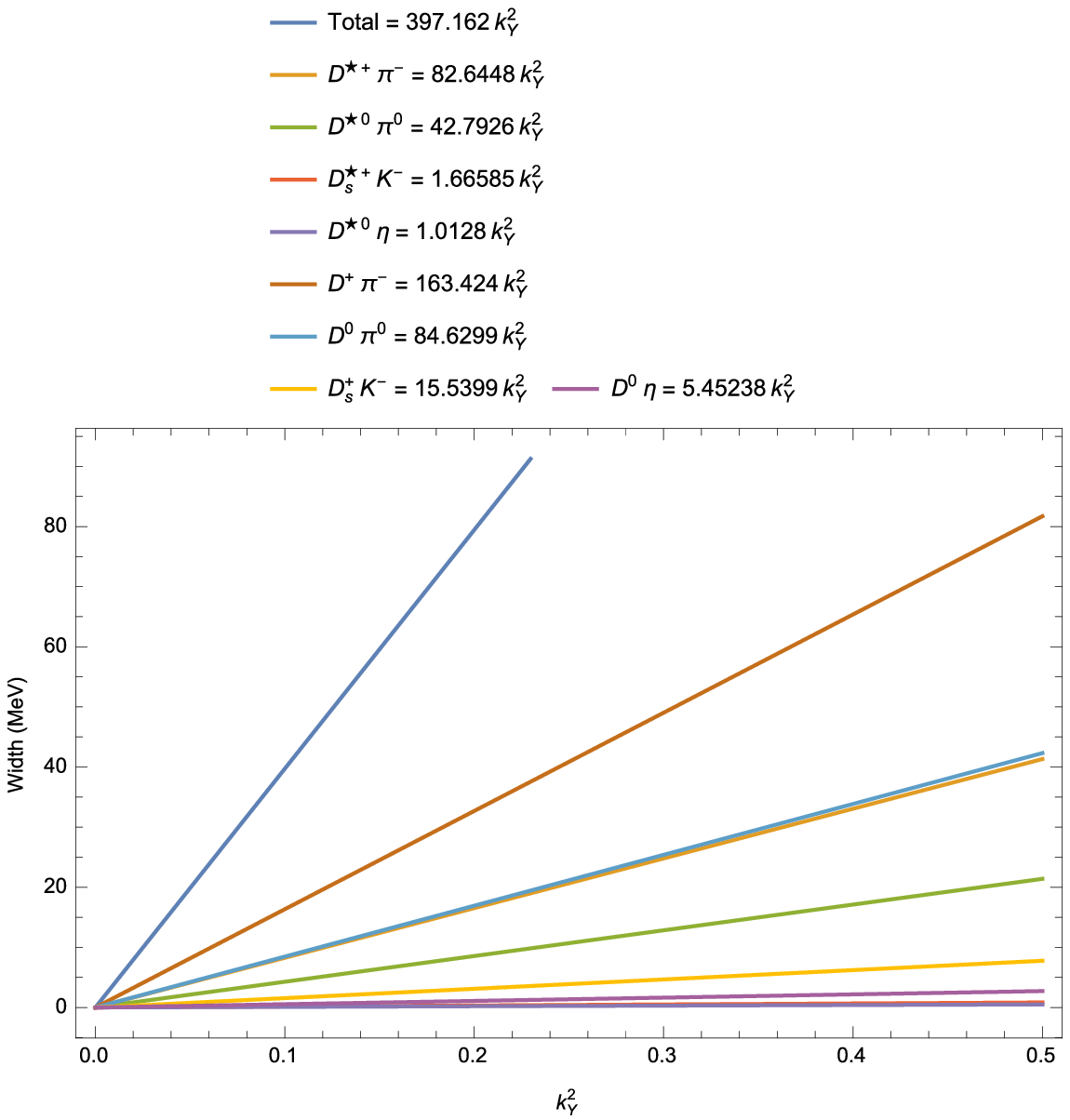}
\caption{Strong decay widths of $D_3^*(2750)^0$ (in MeV) changing with the square of the coupling $k_Y^2$ in HQET. The masses of $D_3^*(2750)^0$ observed (in the decay mode $D^{*+}\pi^-$) by LHCb(2013) \cite{Aaij2013} (upper) and $BABAR$(2010) \cite{del2010} (lower) are used.}
\label{fig8}       
\end{figure}

\begin{figure*}
  \includegraphics[width=0.49\textwidth]{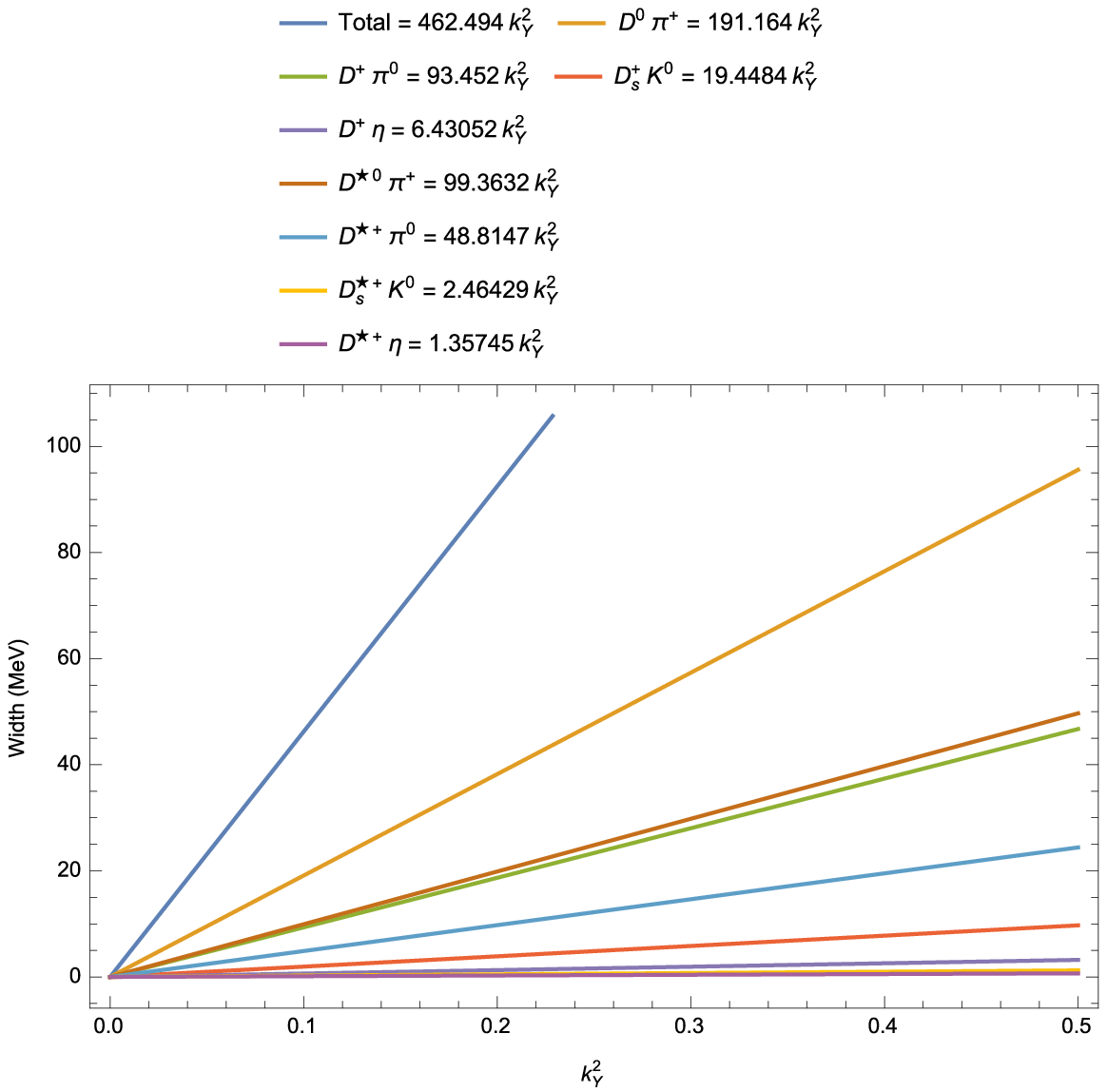}
    \includegraphics[width=0.49\textwidth]{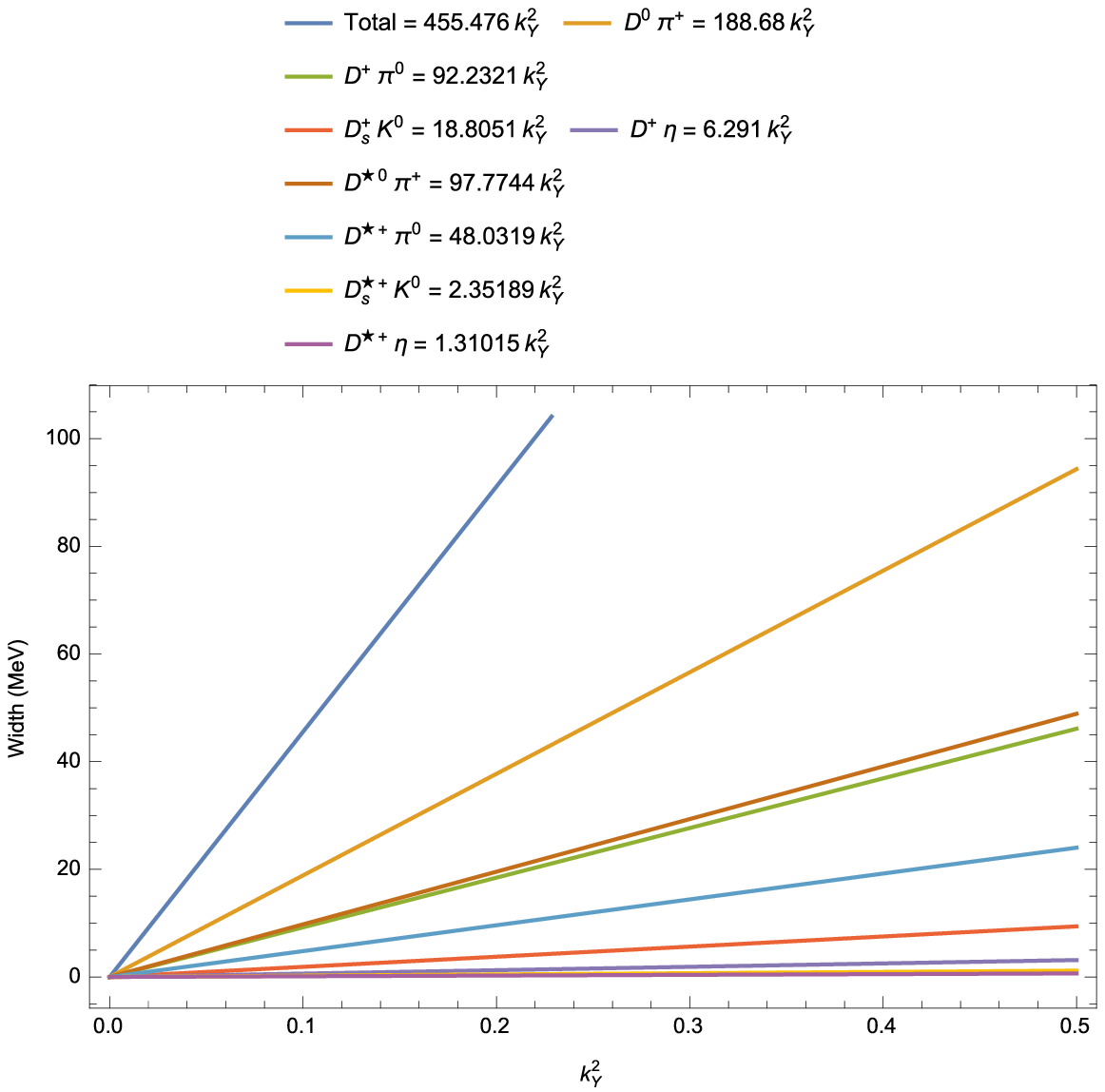}
  \includegraphics[width=0.49\textwidth]{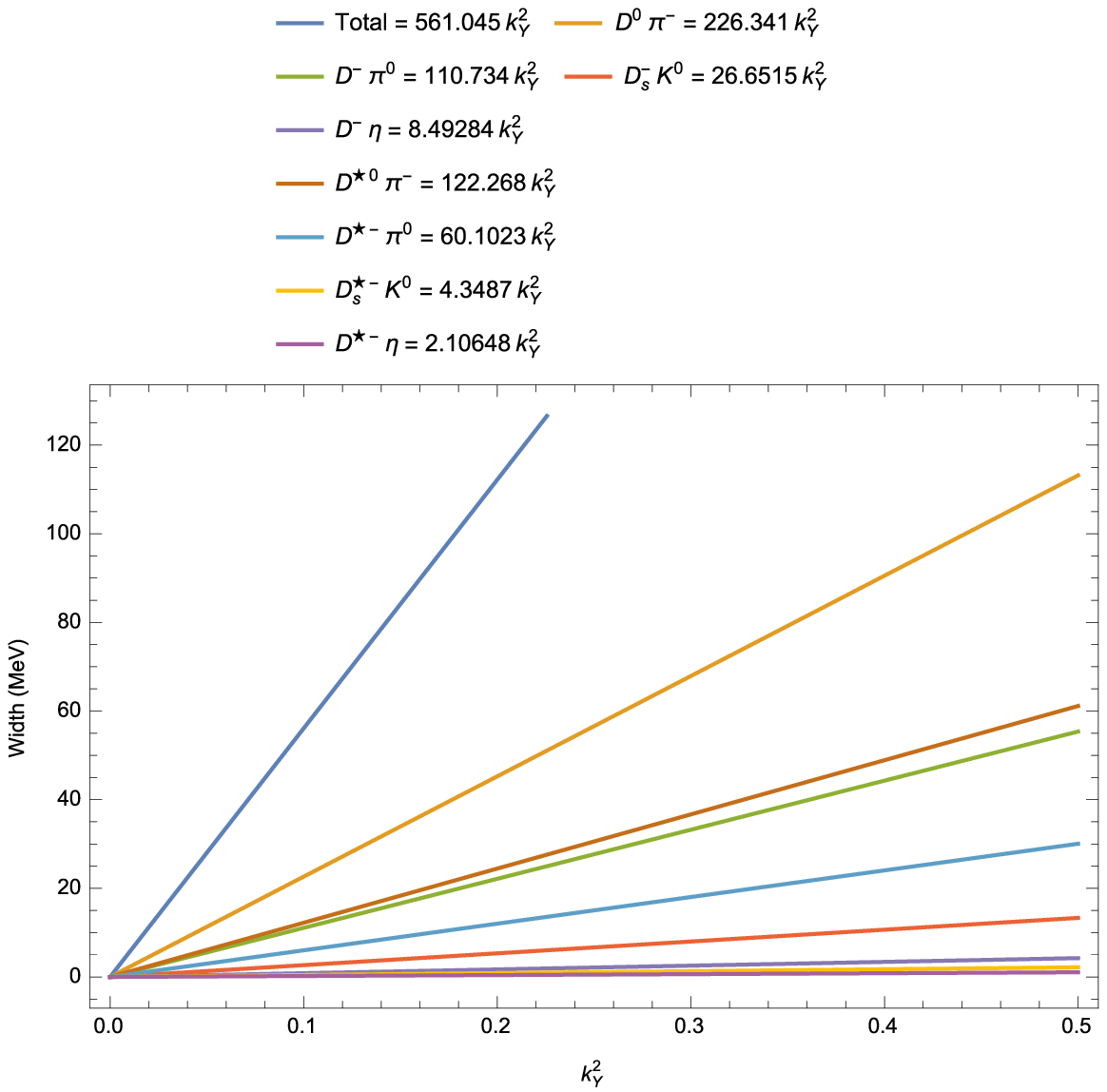}
\caption{Strong decay widths of $D_3^*(2750)^+$ (upper) and $D_3^*(2750)^-$ (lower) (in MeV) changing with the square of the coupling $k_Y^2$ in HQET. The masses of $D_3^*(2750)^+$ observed (in the decay mode $D^0\pi^+$) by LHCb(2013) \cite{Aaij2013} (upper left) and $BABAR$(2010) \cite{del2010} (upper right), and (in the decay mode $D^0\pi^-$) by LHCb(2015) \cite{Aaij2015} (lower) are used.}
\label{fig9}       
\end{figure*}

\begin{figure}
  \includegraphics[width=0.49\textwidth]{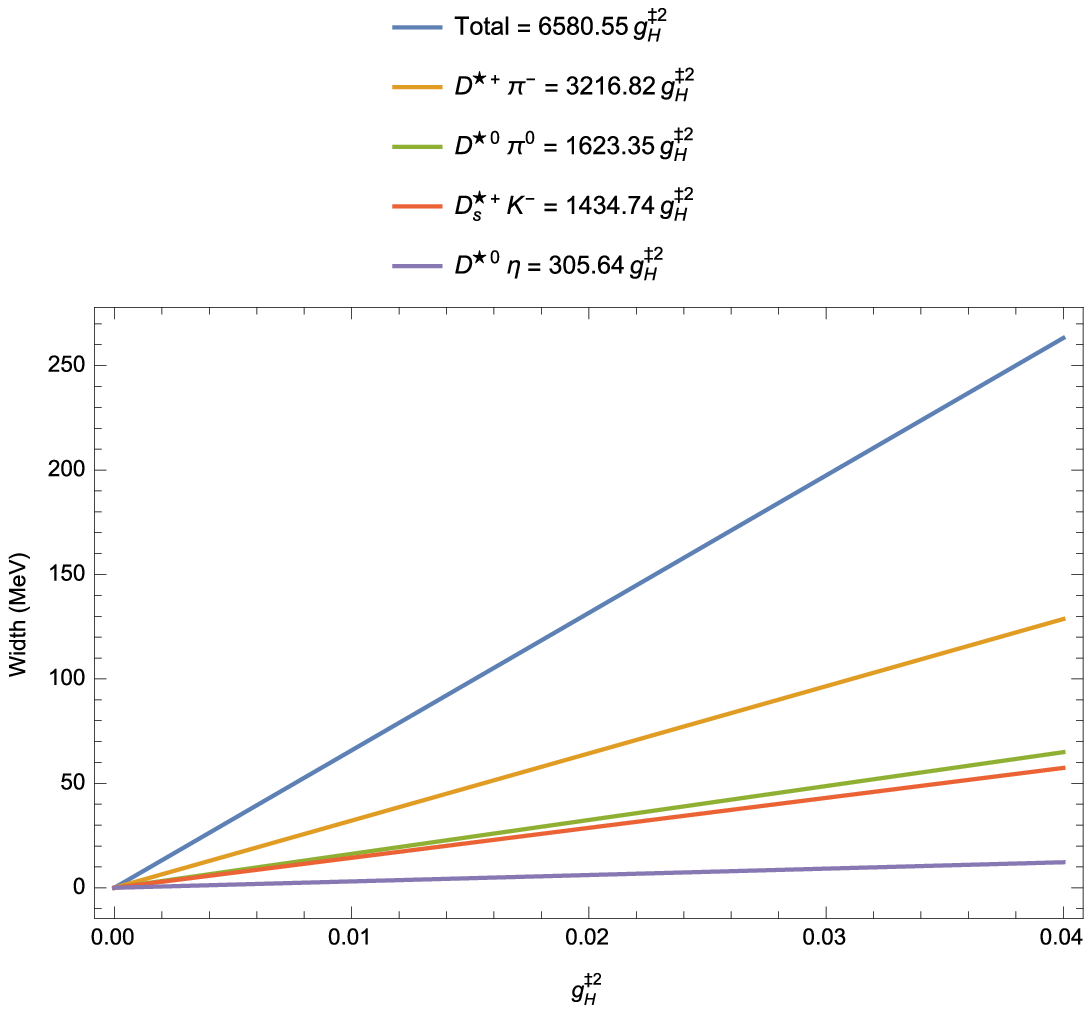}
      \includegraphics[width=0.49\textwidth]{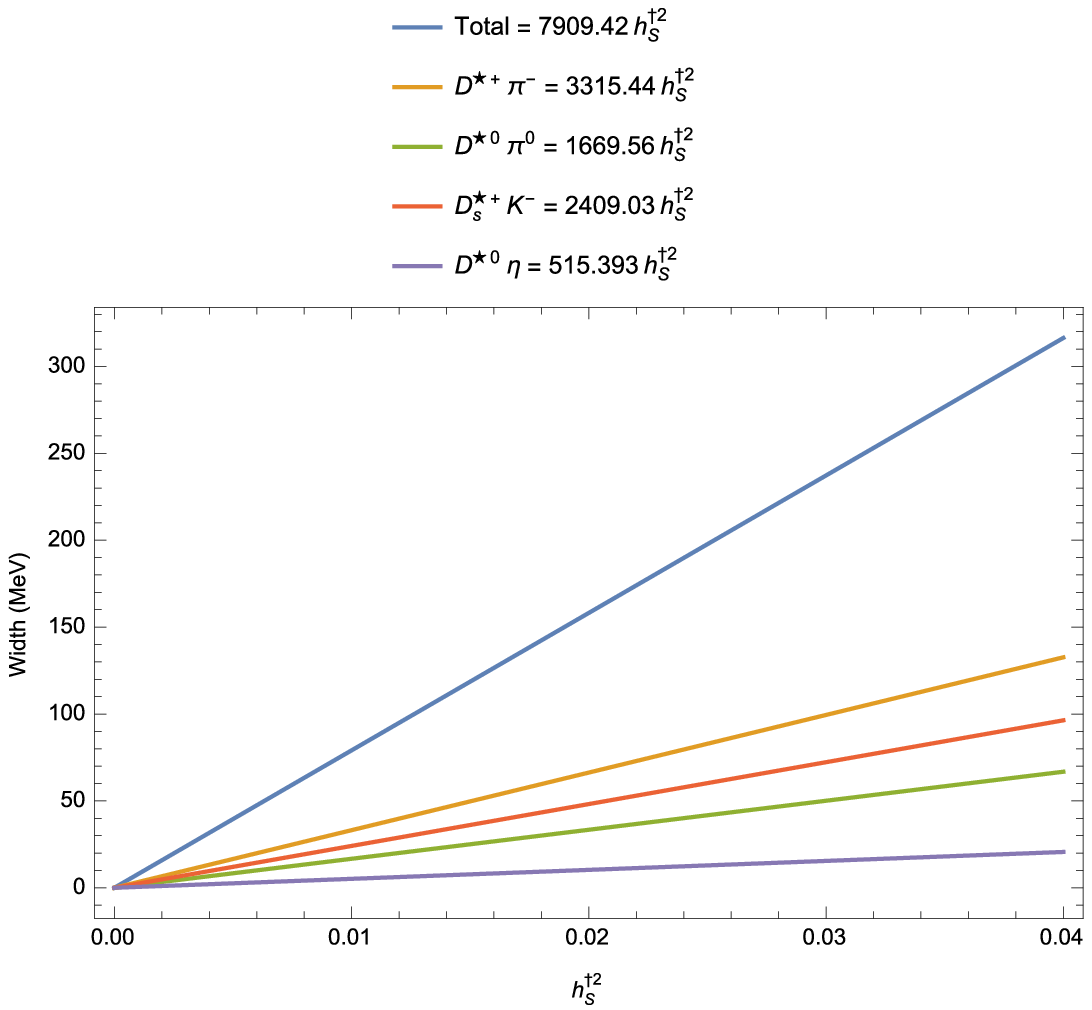}
\caption{Strong decay widths of $D_J(3000)^0$ as 3$^1S_0$ (upper) and 2$^3P_1$ (lower) changing with the square of the couplings $g_H^{\ddag2}$ and $h_S^{\dag2}$ respectively in HQET. The mass of $D_J(3000)^0$ observed (in the decay mode $D^{*+}\pi^-$) by LHCb(2013) \cite{Aaij2013} is used.}
\label{fig10}       
\end{figure}

\begin{figure*}
  \includegraphics[width=0.49\textwidth]{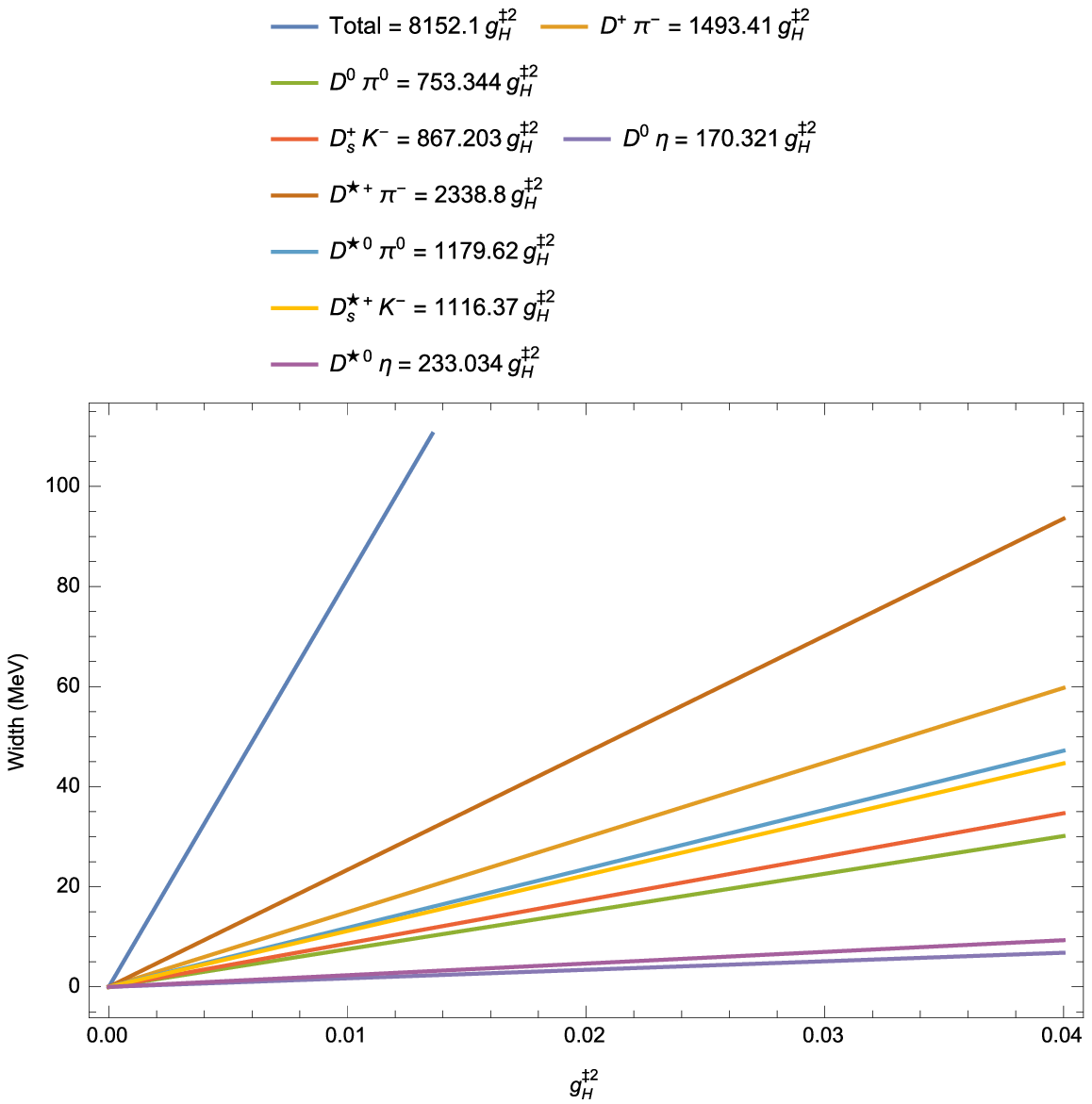}
    \includegraphics[width=0.49\textwidth]{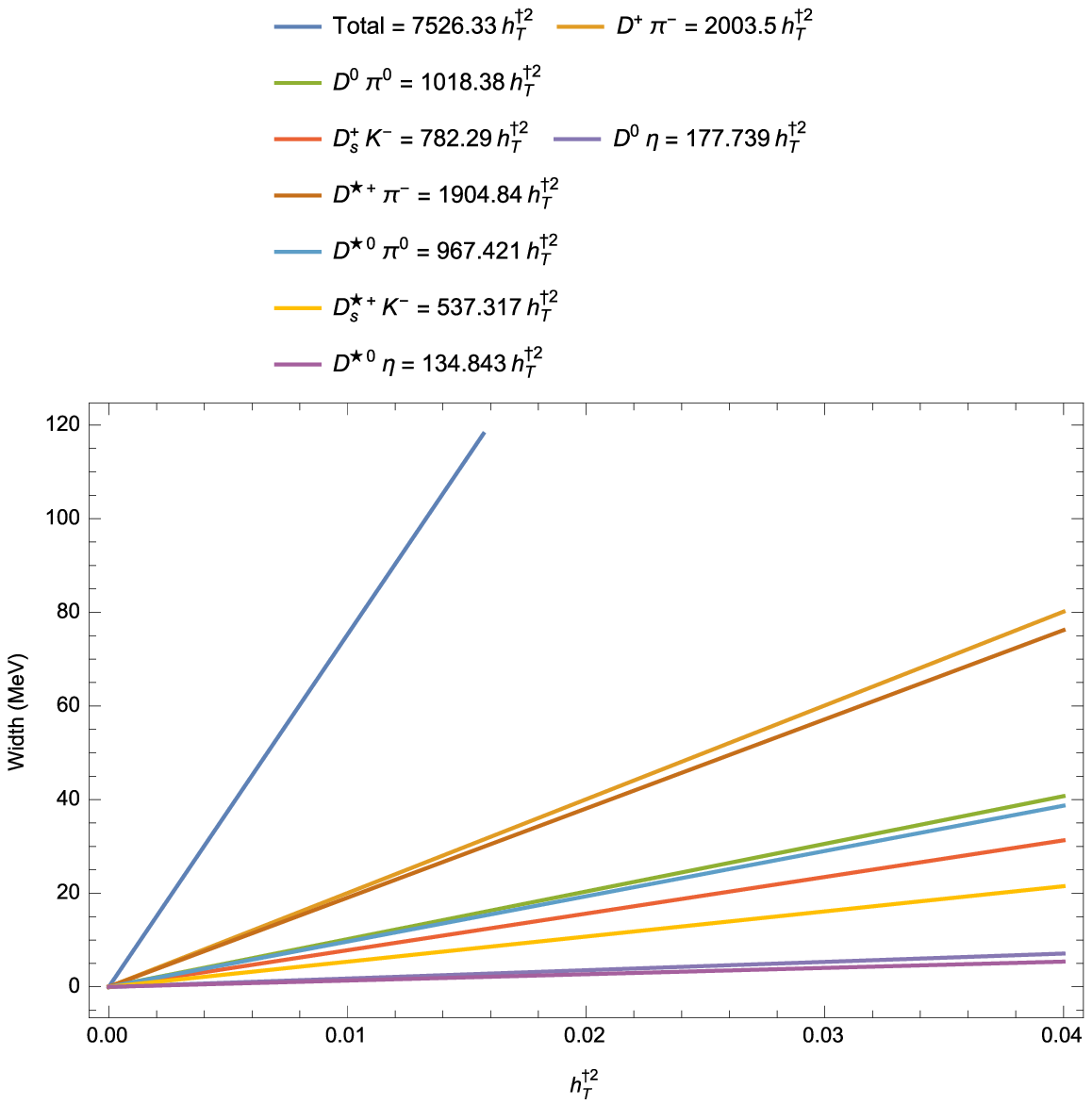}
    \includegraphics[width=0.49\textwidth]{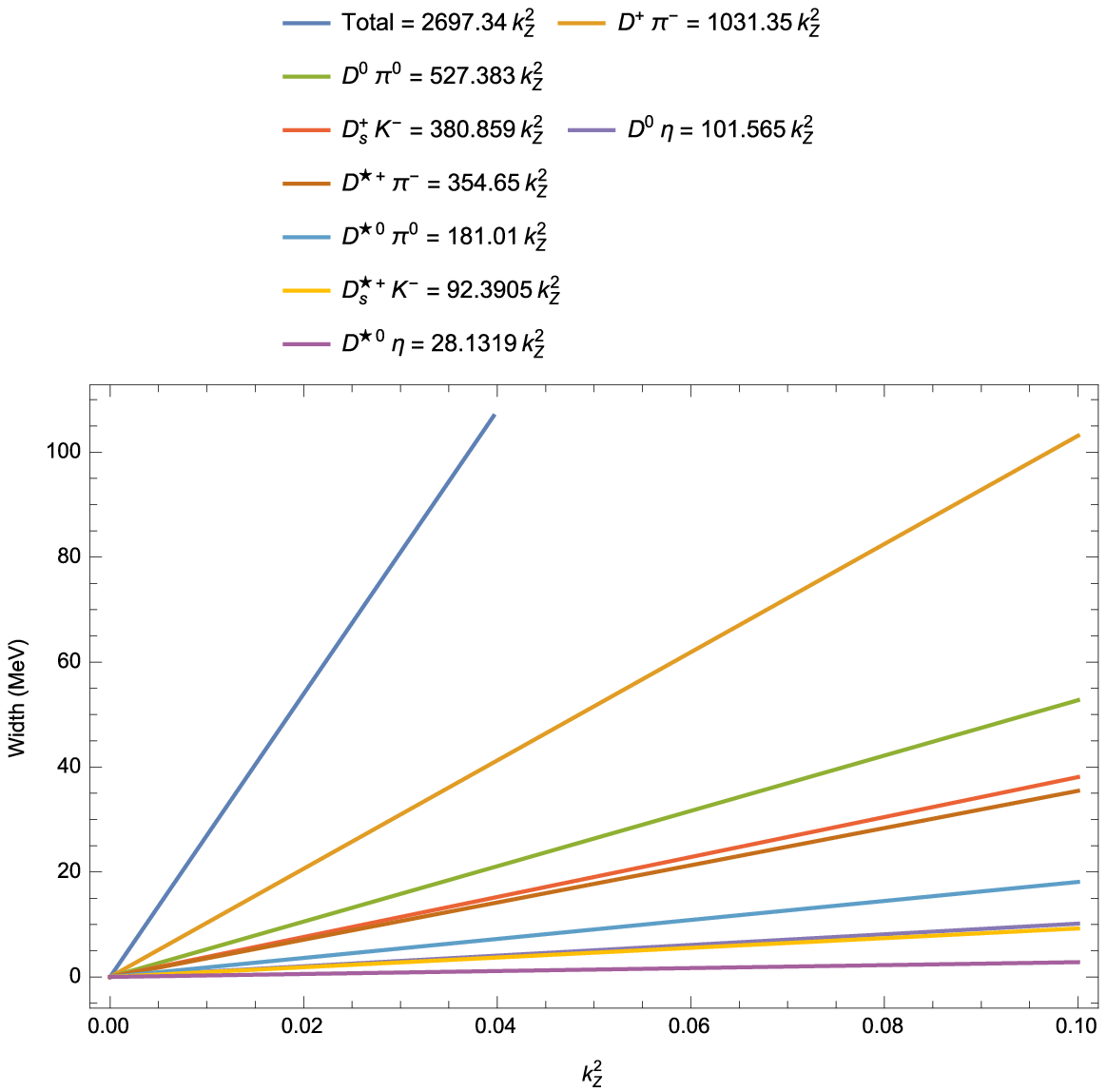}
    \includegraphics[width=0.49\textwidth]{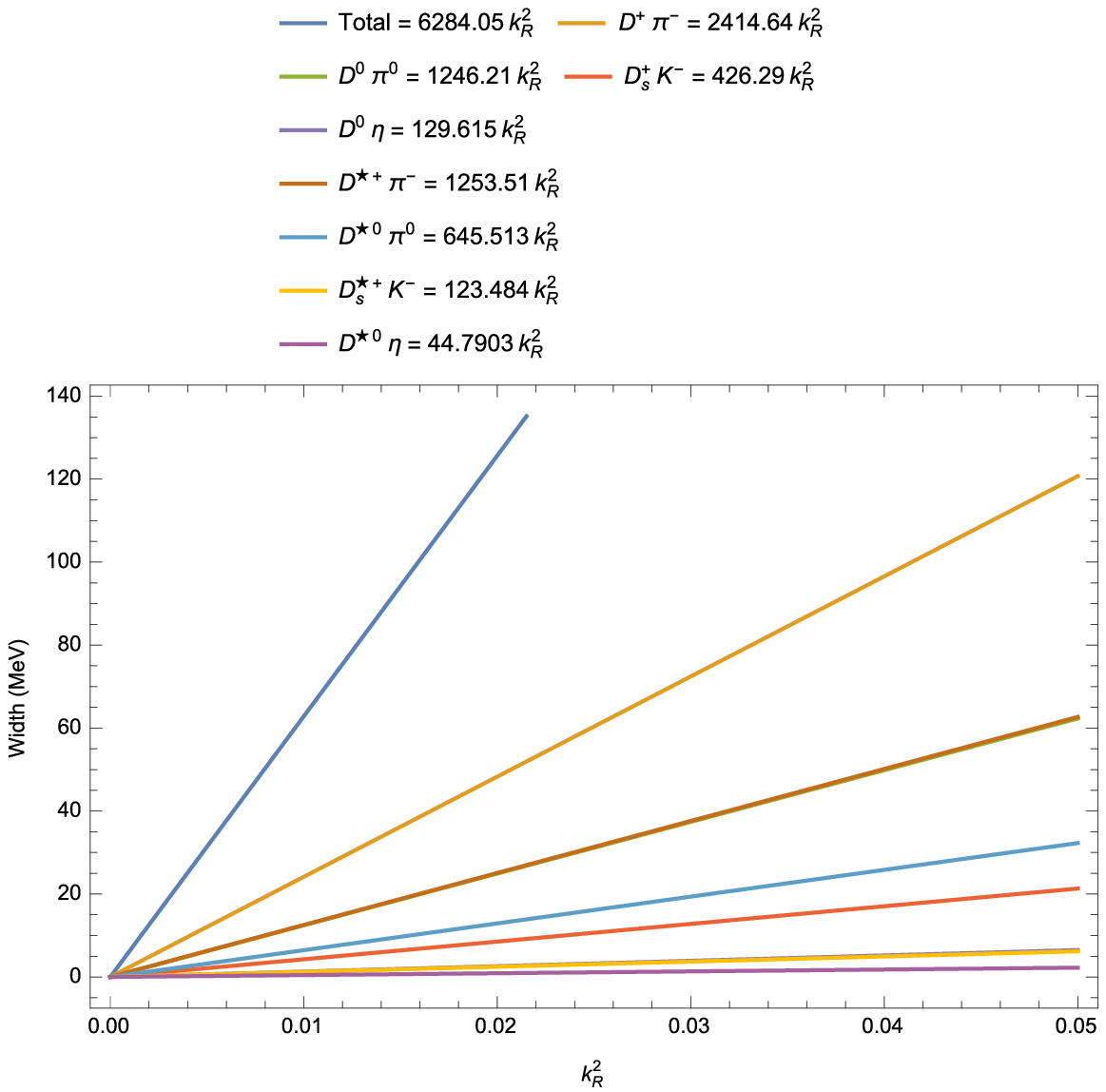}
\caption{Strong decay widths of $D_J^*(3000)^0$ as 3$^3S_1$ (upper left), 2$^3P_2$ (upper right), 1$^3F_2$ (lower left) and 1$^3F_4$ (lower right) changing with the square of the couplings $g_H^{\ddag2}$, $h_T^{\dag2}$, $k_Z^2$ and $k_R^2$ respectively. The mass of $D_J^*(3000)^0$ observed (in the decay mode $D^+\pi^-$) by LHCb(2013) \cite{Aaij2013} is used.}
\label{fig11}       
\end{figure*}

\begin{figure}
  \includegraphics[width=0.49\textwidth]{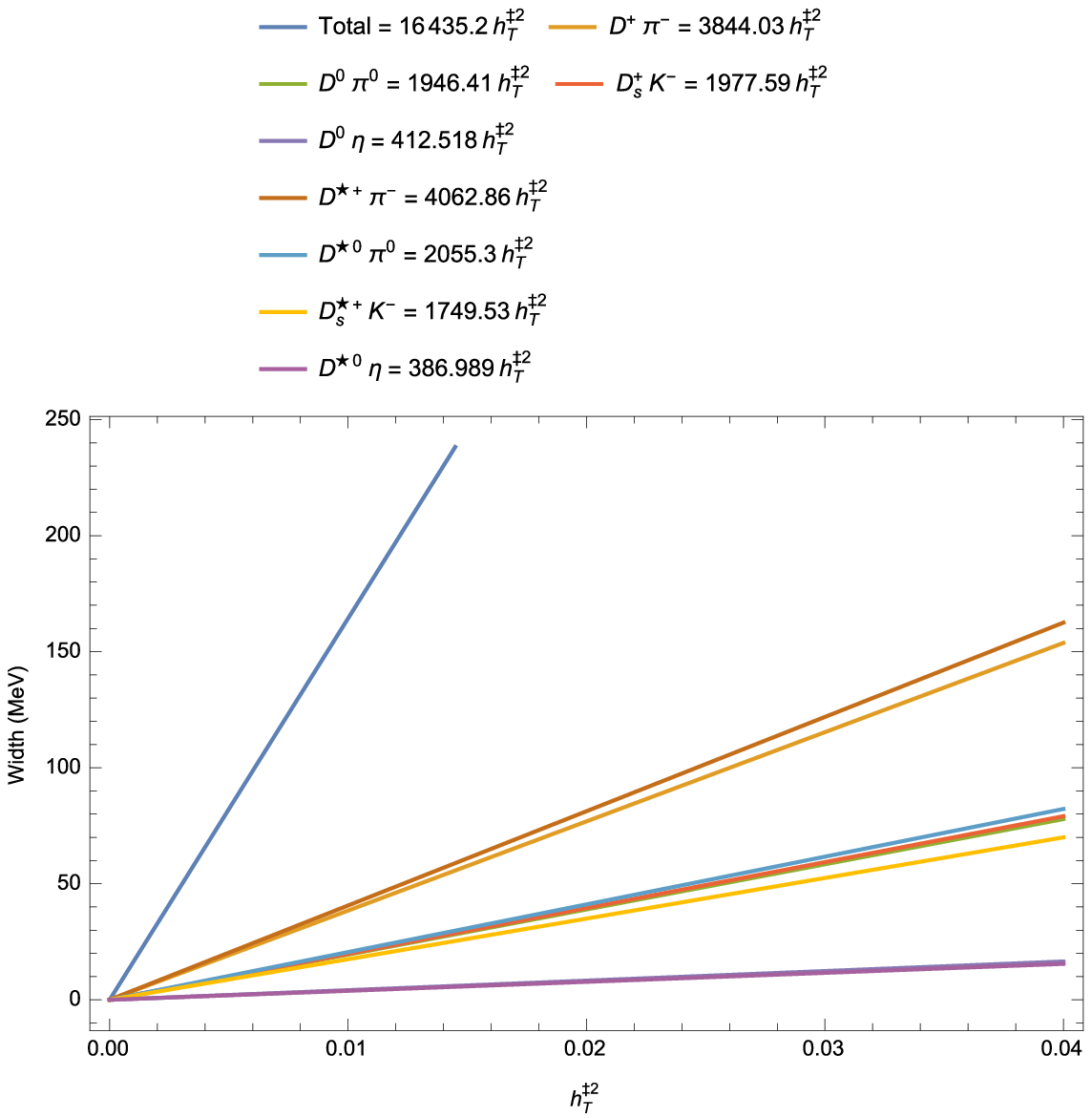}
    \includegraphics[width=0.49\textwidth]{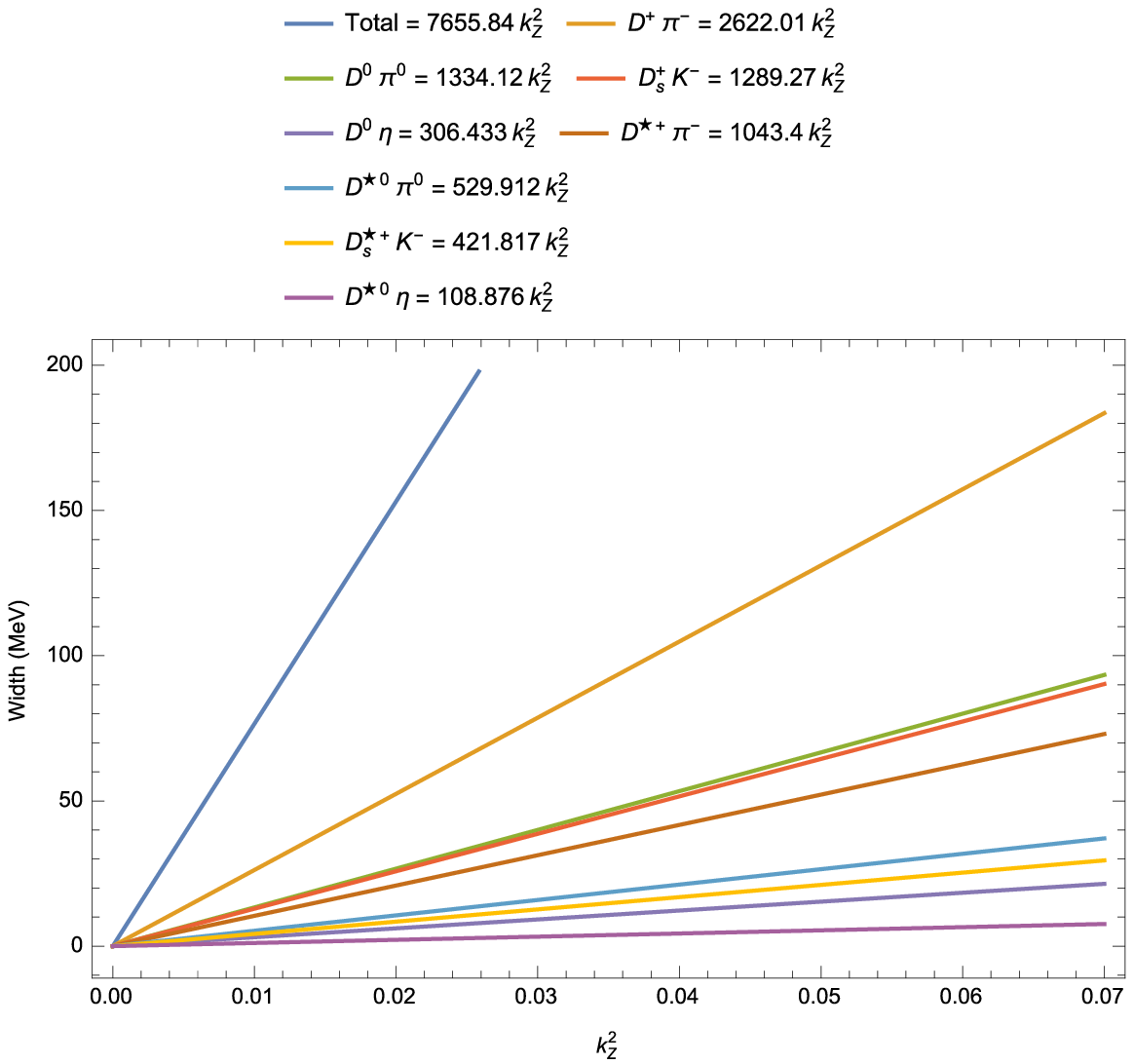}
\caption{Strong decay widths of $D_2^*(3000)^0$ as 3$^3P_2$ (upper) and 1$^3F_2$ (lower) changing with the square of the couplings $h_T^{\ddag2}$ and $k_Z^2$ respectively. The mass of $D_2^*(3000)^0$ observed (in the decay mode $D^+\pi^-$) by LHCb(2016) \cite{Aaij2016} is used.}
\label{fig12}       
\end{figure}

\begin{figure}
  \includegraphics[width=0.49\textwidth]{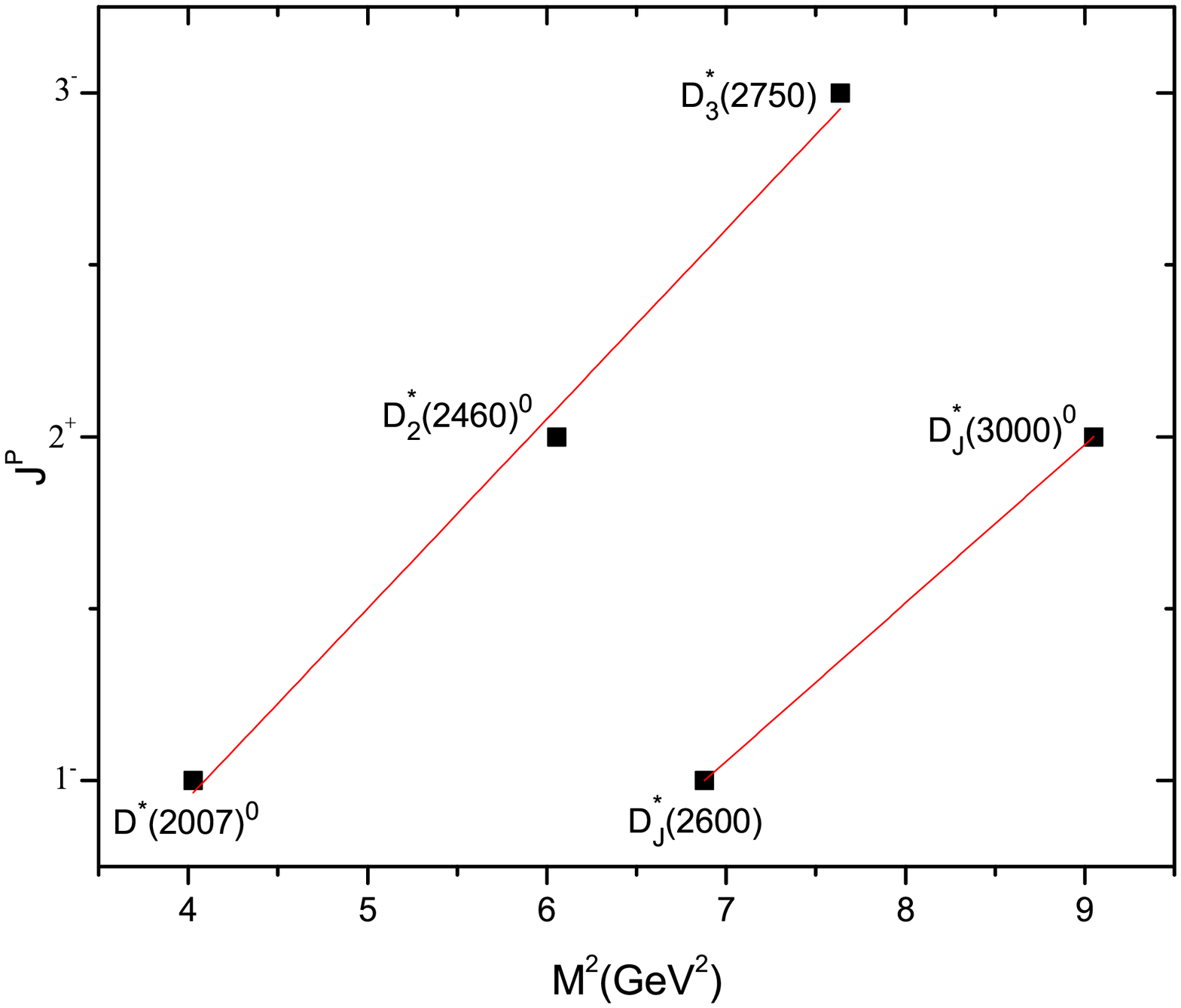}
  \includegraphics[width=0.49\textwidth]{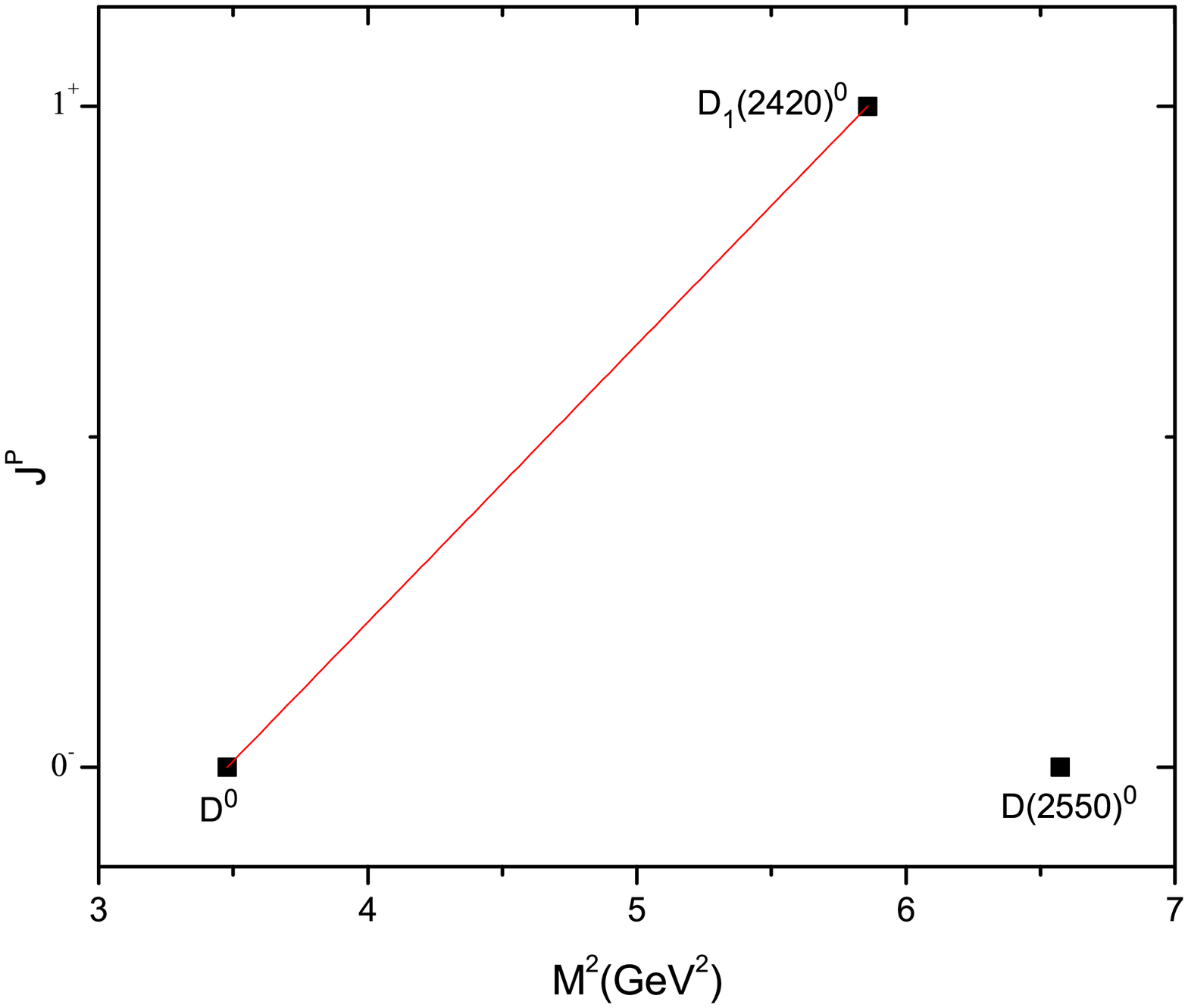}
\caption{Regge trajectory of nonstrange charmed mesons in $(J, M^2)$ plane with natural parity (upper) unnatural parity (lower).}
\label{fig13}       
\end{figure}

\begin{figure}
  \includegraphics[width=0.49\textwidth]{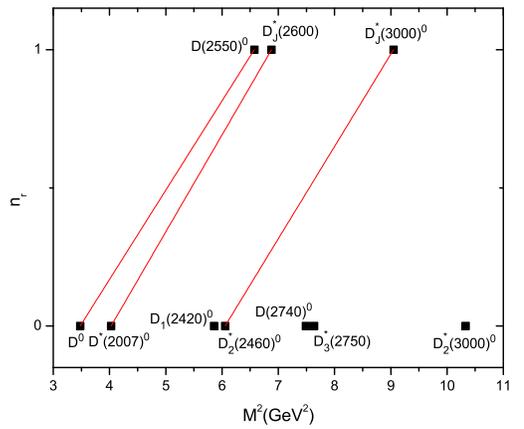}
\caption{Regge trajectory of nonstrange charmed mesons in $(n_r, M^2)$ plane.}
\label{fig14}       
\end{figure}

\begin{table*}
\addtocounter{table}{-1}
\caption{\label{tab3}
The strong decay widths of nonstrange charmed mesons with possible quantum state assignments (in MeV).}
\begin{ruledtabular}
\begin{tabular}{ccccccccccccc}
Meson & $\cal{N}$$^{2S+1}L_J$ & Decay mode & LHCb(2016) \cite{Aaij2016} & LHCb(2015) \cite{Aaij2015} & LHCb(2013) \cite{Aaij2013} & $BABAR$(2010) \cite{del2010}\\
\noalign{\smallskip}\hline\noalign{\smallskip} 
$D_J^*(3000)^0$& 1$^3F_2$ & $D^+ \pi^-$ &&& 1031.35$k_Z^2$ \\
&& $D^0 \pi^0$ &&& 527.383$k_Z^2$ \\
&& $D^+_s K^-$ &&& 380.859$k_Z^2$ \\
&& $D^0 \eta$ &&& 101.565$k_Z^2$ \\
&& $D^{*+} \pi^-$ &&& 354.650$k_Z^2$\\
&& $D^{*0} \pi^0$ &&& 181.010$k_Z^2$\\
&& $D{_{s}^{*+}} K^-$ &&& 92.3905$k_Z^2$ \\
&& $D^{*0} \eta$ &&& 28.1319$k_Z^2$ \\
&& Total &&& 2697.34$k_Z^2$ \\
&& $k_Z$ &&& 0.202 \\
\noalign{\smallskip}\hline\noalign{\smallskip} 
$D_J^*(3000)^0$& 1$^3F_4$ & $D^+ \pi^-$ &&& 2414.64$k_R^2$ \\
&& $D^0 \pi^0$ &&& 1246.21$k_R^2$\\
&& $D^+_s K^-$ &&& 426.29$k_R^2$ \\
&& $D^0 \eta$ &&& 129.615$k_R^2$ \\
&& $D^{*+} \pi^-$ &&& 1253.51$k_R^2$\\
&& $D^{*0} \pi^0$ &&& 645.513$k_R^2$ \\
&& $D{_{s}^{*+}} K^-$ &&& 123.484$k_R^2$ \\
&& $D^{*0} \eta$ &&& 44.7903$k_R^2$ \\
&& Total &&& 6284.05$k_R^2$ \\
&& $k_R$ &&& 0.132 \\
\noalign{\smallskip}\hline\noalign{\smallskip} 
$D_2^*(3000)^0$& 3$^3P_2$ & $D^+ \pi^-$ & 3844.03$h_T^{\ddag2}$\\
&& $D^0 \pi^0$ & 1946.41$h_T^{\ddag2}$ \\
&& $D^+_s K^-$ & 1977.59$h_T^{\ddag2}$ \\
&& $D^0 \eta$ & 412.518$h_T^{\ddag2}$ \\
&& $D^{*+} \pi^-$ & 4062.86$h_T^{\ddag2}$ \\
&& $D^{*0} \pi^0$ & 2055.30$h_T^{\ddag2}$ \\
&& $D{_{s}^{*+}} K^-$ & 1749.53$h_T^{\ddag2}$ \\
&& $D^{*0} \eta$ & 386.989$h_T^{\ddag2}$ \\
&& Total & 16435.2$h_T^{\ddag}$ \\
&& $h_T^{\ddag}$ & 0.106 \\
\noalign{\smallskip}\hline\noalign{\smallskip} 
$D_2^*(3000)^0$& 1$^3F_2$ & $D^+ \pi^-$ & 2622.01$k_Z^2$ \\
&& $D^0 \pi^0$ & 1334.12$k_Z^2$ \\
&& $D^+_s K^-$ & 1289.27$k_Z^2$ \\
&& $D^0 \eta$ & 306.433$k_Z^2$ \\
&& $D^{*+} \pi^-$ & 1043.40$k_Z^2$ \\
&& $D^{*0} \pi^0$ & 529.912$k_Z^2$\\
&& $D{_{s}^{*+}} K^-$ & 421.817$k_Z^2$ \\
&& $D^{*0} \eta$ & 108.876$k_Z^2$ \\
&& Total & 7655.84$k_Z^2$ \\
&& $k_Z$ & 0.156\\
\end{tabular}
\end{ruledtabular}
\end{table*}

\begin{table}
\caption{\label{tab4}
Quantum number assignment of excited $D$ mesons through strong decays analysis.}
\begin{ruledtabular}
\begin{tabular}{ccccccccccccc}
$\cal{N}$$^{2S+1}L_J$ & $J^P$ & Exp. \cite{Tanabashi2018-19} (in GeV)  \\
\noalign{\smallskip}\hline\noalign{\smallskip} 
$1^1P_1$ & $1^+$ & 2.420 $D_1(2420)^0$ \\
\noalign{\smallskip}
& & 2.423 $D_1(2420)^{\pm}$ \\ 
\noalign{\smallskip}
$1^3P_2$ & $2^+$ & 2.460 $D_2^*(2460)^0$  \\
\noalign{\smallskip}
&& 2.465 $D_2^*(2460)^{\pm}$\\
\noalign{\smallskip}
$2^1S_0$ & $0^-$ & 2.564 $D(2550)^0$ \\
\noalign{\smallskip}
$2^3S_1$ & $1^-$ & 2.623 $D_J^*(2600)$ \\
\noalign{\smallskip}
$1^3D_2$ & $2^-$ & 2.737 $D(2740)^0$ \cite{Aaij2013}\\
\noalign{\smallskip}
$1^3D_3$ & $3^-$ & 2.764 $D_3^*(2750)$\\
\noalign{\smallskip}
$2^3P_1$ & $1^+$ & 2.972 $D_J(3000)^0$ \cite{Aaij2013}\\
\noalign{\smallskip}
$2^3P_2$ & $2^+$ & 3.008 $D_J^*(3000)^0$ \cite{Aaij2013} \\
\noalign{\smallskip}
$1^3F_2$ & $2^+$ & 3.214 $D_2^*(3000)^0$ \cite{Aaij2016} \\
\end{tabular}
\end{ruledtabular}
\end{table}

\begin{table}
\caption{\label{tab5}
Fitted parameters of the $D$ mesons parent and daughter Regge trajectories in $(J, M^2)$ plane with natural and unnatural parity.}
\begin{ruledtabular}
\begin{tabular}{ccccccccccccc}
& $\alpha$ (GeV$^{-2}$) & $\alpha_0$ & $\alpha$ (GeV$^{-2}$) & $\alpha_0$ &\\ 
\noalign{\smallskip}
\cline{2-3}
\cline{4-5}
\noalign{\smallskip}
Parent & 0.49366 & -0.98699 & 0.41996 & -1.46046 \\
Daughter & 0.46114 & -2.17271 & $-$ & $-$ \\
\end{tabular}
\end{ruledtabular}
\end{table}

\begin{table}
\caption{\label{tab6}
Fitted parameters of the $D$ mesons Regge trajectories in $(n_r, M^2)$ plane.}
\begin{ruledtabular}
\begin{tabular}{ccccccccccccc}
Meson & $\beta$ (GeV$^{-2}$) & $\beta_0$\\
\noalign{\smallskip}\hline\noalign{\smallskip}
$D^0$ & 0.32295 & -1.12310 \\
$D^*(2007)^0$ & 0.35055 & -1.41181 \\
$D_2^*(2460)^0$ & 0.33397 & -2.02195 \\
\end{tabular}
\end{ruledtabular}
\end{table}

\begin{table}
\caption{\label{tab7}
The masses of nonstrange charmed meson states (in GeV) lying on the $1^3S_1$, $2^3S_1$ and $1^1S_0$ Regge lines in $(J, M^2)$ plane.}
\begin{ruledtabular}
\begin{tabular}{ccccccccccccc}
State & $1^3S_1$ & $1^3P_2$ & $1^3D_3$ & $1^3F_4$\\
\noalign{\smallskip}\hline\noalign{\smallskip}
Present & 2.007 \cite{Tanabashi2018-19} & 2.460 \cite{Tanabashi2018-19} & 2.843 & 3.179\\
Ref. \cite{Godfrey2016} & 2.041 & 2.502 & 2.833 & 3.132 \\ 
Ref. \cite{Sun2013} & 2.038 & 2.501 & 2.833 & 3.113 \\
Ref. \cite{Ebert2010} & 2.010 & 2.460 & 2.863 & 3.187 \\
Ref. \cite{DiPierro2001} & 2.005 & 2.460 & 2.799 & 3.101 \\
\noalign{\smallskip}\hline\noalign{\smallskip}
State & $2^3S_1$ & $2^3P_2$ & $2^3D_3$ & $2^3F_4$\\
\noalign{\smallskip}\hline\noalign{\smallskip}
Present & 2.623 \cite{Tanabashi2018-19} & 3.008 \cite{Tanabashi2018-19} & 3.349 & 3.659\\
Ref. \cite{Godfrey2016} & 2.643 & 2.957 & 3.226 & 3.466 \\ 
Ref. \cite{Sun2013} & 2.645 & 2.957 & 3.226 & $-$ \\
Ref. \cite{Ebert2010} & 2.632 & 3.012 & 3.335 & 3.610 \\
\noalign{\smallskip}\hline\noalign{\smallskip}
State & $1^1S_0$ & $1^1P_1$ & $1^1D_2$ & $1^1F_3$\\
\noalign{\smallskip}\hline\noalign{\smallskip}
Present & 1.865 \cite{Tanabashi2018-19} & 2.420 \cite{Tanabashi2018-19} & 2.870 & 3.259\\
Ref. \cite{Godfrey2016} & 1.877 & 2.456 & 2.816 & 3.108 \\ 
Ref. \cite{Sun2013} & 1.874 & 2.457 & 2.827 & 3.123 \\
Ref. \cite{Ebert2010} & 1.871 & 2.426 & 2.806 & 3.129 \\
Ref. \cite{DiPierro2001} & 1.868 & 2.417 & 2.775 & 3.074 \\
\end{tabular}
\end{ruledtabular}
\end{table}

\begin{table}
\caption{\label{tab8}
The masses of nonstrange charmed meson states (in GeV) lying on the $1^1S_0$, $1^3S_1$ and $1^3P_2$ Regge lines in $(n_r, M^2)$ plane.}
\begin{ruledtabular}
\begin{tabular}{ccccccccccccc}
State & $1^1S_0$ & $2^1S_0$ & $3^1S_0$\\
\noalign{\smallskip}\hline\noalign{\smallskip}
Present & 1.865 \cite{Tanabashi2018-19} & 2.564 \cite{Tanabashi2018-19} & 3.110\\
Ref. \cite{Godfrey2016} & 1.877 & 2.581 & 3.068 & \\ 
Ref. \cite{Sun2013} & 1.874 & 2.583 & 2.827 &  \\
Ref. \cite{Ebert2010} & 1.871 & 2.581 & 3.062 &  \\
Ref. \cite{DiPierro2001} & 1.868 & 2.589 & 2.775 &  \\
\noalign{\smallskip}\hline\noalign{\smallskip}
State & $1^3S_1$ & $2^3S_1$ & $3^3S_1$\\
\noalign{\smallskip}\hline\noalign{\smallskip}
Present & 2.007 \cite{Tanabashi2018-19} & 2.623 \cite{Tanabashi2018-19} & 3.120\\
Ref. \cite{Godfrey2016} & 2.041 & 2.643 & 3.110  \\ 
Ref. \cite{Sun2013} & 2.038 & 2.645 & 3.111 \\
Ref. \cite{Ebert2010} & 2.010 & 2.632 & 3.096   \\
Ref. \cite{DiPierro2001} & 2.005 & 2.692 & 3.226  \\
\noalign{\smallskip}\hline\noalign{\smallskip}
State & $1^3P_2$ & $2^3P_2$ & $3^3P_2$\\
\noalign{\smallskip}\hline\noalign{\smallskip}
Present & 2.460 \cite{Tanabashi2018-19} & 3.008 \cite{Tanabashi2018-19} & 3.470\\
Ref. \cite{Godfrey2016} & 2.502 & 2.957 & 3.353 \\ 
Ref. \cite{Sun2013} & 2.501 & 2.957 & $-$  \\
Ref. \cite{Ebert2010} & 2.460 & 3.012 & 3407 \\
Ref. \cite{DiPierro2001} & 2.460 & 3.035 & $-$ \\
\end{tabular}
\end{ruledtabular}
\end{table}

\begin{table*}
\caption{\label{tab9}
Strong decay widths (in MeV), ratio and branching fraction of nonstrange charmed mesons lying on the Regge lines with possible quantum number assignments.}
\begin{ruledtabular}
\begin{tabular}{ccccccccccccc}
$\cal{N}$$^{2S+1}L_J$ & Decay & Decay & Ratio & Branching\\
& mode & width & & fraction\\
\noalign{\smallskip}\hline\noalign{\smallskip}
1$^1D_2$ & $D^{*+} \pi^-$ & 1772.11$k_X^2$ & 1 & 52.3 \\
& $D^{*0} \pi^0$ & 901.224$k_X^2$ & 0.51 & 26.6 \\
& $D_s^{*+} k^-$ & 553.089$k_X^2$ & 0.31 & 16.32 \\
& $D^{*0} \eta$ & 160.782$k_X^2$ & 0.09 & 4.74 \\
& Total & 3388.20$k_X^2$\\
\noalign{\smallskip}\hline\noalign{\smallskip} 
1$^3D_3$ & $D^+ \pi^-$ & 290.925$k_Y^2$ & 1 & 37.89\\
& $D^0 \pi^0$ & 149.953$k_Y^2$ & 0.52 & 19.53 \\
& $D^+_s K^-$ & 45.1974$k_Y^2$ & 0.16 & 5.89 \\
& $D^0 \eta$ & 13.6889$k_Y^2$ & 0.05 & 4.7 \\
& $D^{*+} \pi^-$ & 167.347$k_Y^2$ & 0.58 & 21.79 \\
& $D^{*0} \pi^0$ & 86.1774$k_Y^2$ & 0.3 & 29.62  \\
& $D{_{s}^{*+}} K^-$ & 10.3364$k_Y^2$ &0.04 & 3.55 \\
& $D^{*0} \eta$ & 4.24088$k_Y^2$ & 0.01 & 1.46 \\
& Total & 767.866$k_Y^2$ \\
\noalign{\smallskip}\hline\noalign{\smallskip} 
3$^1S_0$ & $D^{*+} \pi^-$ & 4389.38$g_H^{\ddag2}$ & 1 & 46.14 \\
& $D^{*0} \pi^0$ & 2210.97$g_H^{\ddag2}$ & 0.5 & 23.24 \\
& $D{_{s}^{*+}} K^-$ & 2427.41$g_H^{\ddag2}$ & 0.55 & 25.52 \\
& $D^{*0} \eta$ & 484.987$g_H^{\ddag2}$ & 0.11 & 11.05 \\
& Total & 9512.75$g_H^{\ddag2}$ \\
\noalign{\smallskip}\hline\noalign{\smallskip} 
3$^3S_1$ & $D^+ \pi^-$ & 1837.20$g_H^{\ddag2}$ & 1 & 17.23 \\
& $D^0 \pi^0$ & 925.499$g_H^{\ddag2}$ & 0.5 & 8.68 \\
& $D^+_s K^-$ & 1180.56$g_H^{\ddag2}$ & 0.64 & 11.07 \\
& $D^0 \eta$ & 224.751$g_H^{\ddag2}$ & 0.12 & 2.11 \\
& $D^{*+} \pi^-$ & 2987.19$g_H^{\ddag2}$ & 1.62 & 28.01 \\
& $D^{*0} \pi^0$ & 1504.51$g_H^{\ddag2}$ & 0.82 & 14.11 \\
& $D{_{s}^{*+}} K^-$ & 1671.63$g_H^{\ddag2}$ & 0.91 & 15.67 \\
& $D^{*0} \eta$ & 332.810$g_H^{\ddag2}$ & 0.18 & 3.12 \\
& Total & 10664.15$g_H^{\ddag2}$ \\
\noalign{\smallskip}\hline\noalign{\smallskip} 
1$^1F_3$ & $D^{*+} \pi^-$ & 3211.24$k_Z^2$ & 1 & 48.85 \\
& $D^{*0} \pi^0$ & 1629.54$k_Z^2$ & 0.51 & 24.79 \\
& $D{_{s}^{*+}} K^-$ & 1384.61$k_Z^2$ & 0.43 & 21.06 \\
& $D^{*0} \eta$ & 348.403$k_Z^2$ & 0.11 & 5.3 \\
& Total & 6573.79$k_Z^2$ \\
\noalign{\smallskip}\hline\noalign{\smallskip} 
1$^3F_4$ & $D^+ \pi^-$ & 6784.48$k_R^2$ & 1 & 34.1 \\
& $D^0 \pi^0$ & 3480.78$k_R^2$ & 0.51 & 17.49 \\
& $D^+_s K^-$ & 1841.32$k_R^2$ & 0.27 & 9.25 \\
& $D^0 \eta$ & 490.200$k_R^2$ & 0.07 & 2.46  \\
& $D^{*+} \pi^-$ & 4145.64$k_R^2$ & 0.61 & 20.84 \\
& $D^{*0} \pi^0$ & 2121.85$k_R^2$ & 0.31 & 10.66  \\
& $D{_{s}^{*+}} K^-$ & 794.006$k_R^2$ & 0.12 & 3.99 \\
& $D^{*0} \eta$ & 236.237$k_R^2$ & 0.03 & 1.19 \\
& Total & 19896.51$k_R^2$ \\
\end{tabular}
\end{ruledtabular}
{continued...}
\end{table*}

\begin{table*}
\addtocounter{table}{-1}
\caption{\label{tab9}
Strong decay widths (in MeV), ratio and branching fraction of nonstrange charmed mesons lying on the Regge lines with possible quantum number assignments.}
\begin{ruledtabular}
\begin{tabular}{ccccccccccccc}
$\cal{N}$$^{2S+1}L_J$ & Decay & Decay & Ratio & Branching\\
& mode & width & & fraction\\
\noalign{\smallskip}\hline\noalign{\smallskip}
2$^3D_3$ & $D^+ \pi^-$ & 3130.22$k_Y^{\dag2}$ & 1 & 26.65 \\
& $D^0 \pi^0$ & 1590.47$k_Y^{\dag2}$ & 0.51 & 13.54 \\
& $D^+_s K^-$ & 1436.88$k_Y^{\dag2}$ & 0.46 & 12.23 \\
& $D^0 \eta$ & 315.141$k_Y^{\dag2}$ & 0.1 & 2.68 \\
& $D^{*+} \pi^-$ & 2668.36$k_Y^{\dag2}$ & 0.85 & 22.72  \\
& $D^{*0} \pi^0$ & 1353.63$k_Y^{\dag2}$ & 0.43 & 11.52 \\
& $D{_{s}^{*+}} K^-$ & 1012.98$k_Y^{\dag2}$ & 0.32 & 8.62 \\
& $D^{*0} \eta$ & 238.644$k_Y^{\dag2}$ & 0.08 & 2.03 \\
& Total & 11746.32$k_Y^{\dag2}$ \\
\noalign{\smallskip}\hline\noalign{\smallskip}
3$^3P_2$ & $D^+ \pi^-$ & 7478.9$h_T^{\ddag2}$ & 1 & 20.93 \\
& $D^0 \pi^0$ & 3774.68$h_T^{\ddag2}$ & 0.5 & 10.56 \\
& $D^+_s K^-$ & 4684.67$h_T^{\ddag2}$ & 0.63 & 13.11 \\
& $D^0 \eta$ & 916.581$h_T^{\ddag2}$ & 0.12 & 2.69 \\
& $D^{*+} \pi^-$ & 8640.15$h_T^{\ddag2}$ & 1.16 & 24.18 \\
& $D^{*0} \pi^0$ & 4357.18$h_T^{\ddag2}$ & 0.58 & 12.19  \\
& $D{_{s}^{*+}} K^-$ & 4886.89$h_T^{\ddag2}$ & 0.65 & 13.68  \\
& $D^{*0} \eta$ & 993.325$h_T^{\ddag2}$ & 0.13 & 2.78 \\
& Total & 35732.4$h_T^{\ddag}$ \\
\noalign{\smallskip}\hline\noalign{\smallskip} 
2$^3F_4$ & $D^+ \pi^-$ & 64151$k_R^{\dag2}$ & 1 & 26.75 \\
& $D^0 \pi^0$ & 32588$k_R^{\dag2}$ & 0.51 & 13.59 \\
& $D^+_s K^-$ & 31378.1$k_R^{\dag2}$ & 0.49 & 13.08 \\
& $D^0 \eta$ & 6879.94$k_R^{\dag2}$ & 0.11 & 2.87 \\
& $D^{*+} \pi^-$ & 51500.9$k_R^{\dag2}$ & 0.8 & 21.48 \\
& $D^{*0} \pi^0$ & 26110.5$k_R^{\dag2}$ & 0.41 & 10.89  \\
& $D{_{s}^{*+}} K^-$ & 22101.5$k_R^{\dag2}$ & 0.34 & 9.21  \\
& $D^{*0} \eta$ & 5103.52$k_R^{\dag2}$ & 0.08 & 2.13 \\
& Total & 239813$k_R^{\dag2}$ \\
\end{tabular}
\end{ruledtabular}
\end{table*}

Here we discuss the $D$ mesons doublets corresponding to $s$, $p$, $d$ and $f$ waves for $l = 0, 1, 2$ and $3$ respectively. For the $s$ wave, $l = 0$ gives $\vec{s}_l^P = {\frac{1}{2}}^-$, the ground state doublet, which consists of two states represented by $(P, P^*)$, having $J^P_{s_l} = (0^-, 1^-)_{{\frac{1}{2}}^-}$. For the $p$ wave, $l = 1$, the first orbital excited states have two doublets $\vec{s}_l^P = {\frac{1}{2}}^+$ and $\vec{s}_l^P = {\frac{3}{2}}^+$, having $J^P_{s_l} = (0^+, 1^+)_{{\frac{1}{2}}^+}$ and $J^P_{s_l} = (1^+, 2^+)_{{\frac{3}{2}}^+}$ represented by $(P^*_0, P^{\prime}_1)$ and $(P_1, P_2^*)$ respectively. Similarly, for the $d$ wave, $l = 2$, two doublets $\vec{s}_l^P = {\frac{3}{2}}^-$ and $\vec{s}_l^P = {\frac{5}{2}}^-$, having $J^P_{s_l} = (1^-, 2^-)_{{\frac{3}{2}}^-}$ and $J^P_{s_l} = (2^-, 3^-)_{{\frac{5}{2}}^-}$ are represented by $(P^*_1, P_2)$ and $(P^{\prime}_2, P^*_3)$ respectively. And, for the $f$ wave, $l = 3$, two doublets $\vec{s}_l^P = {\frac{5}{2}}^+$ and $\vec{s}_l^P = {\frac{7}{2}}^+$, having $J^P_{s_l} = (2^+, 3^+)_{{\frac{5}{2}}^+}$ and $J^P_{s_l} = (3^+, 4^+)_{{\frac{7}{2}}^+}$ are represented by $(P^{\prime*}_2, P_3)$ and $(P^{\prime}_3, P^*_4)$ respectively. The above symbols $(P, P^*,...)$ are used for radial quantum number $n = 1$ and the same classifications follows for higher radial excitations ($n = 2, 3, ...$). For $n = 2$, these symbols are denoted with dagger $(P^{\dag}, P^{\dag*},...)$ and for $n = 3$ they are $(P^{\ddag}, P^{\ddag*},...)$. Hence, each doublet contains two states (or two spin partners) with total spin $J = s_l \pm \frac{1}{2}$ and parity $P = (-1)^{l+1}$ and can be described by the superfields $H_a$, $S_a$, $T_a$, $X_a$, $Y_a$, $Z_a$ and $R_a$, written as \cite{Falk1992,FalkandLuke1992},

{\small{\begin{equation}
\label{eq:1} 
H_a = \frac{1 + {\rlap{v}/}}{2} [P^*_{a\mu}\gamma^{\mu} - P_a\gamma_5],
\end{equation}}}
{\small{\begin{equation}
\label{eq:2} 
S_a = \frac{1 + {\rlap{v}/}}{2} [P^{\mu}_{1a}\gamma_{\mu}\gamma_5 - P^*_{0a}],
\end{equation}}}
{\small{\begin{equation}
\label{eq:3} 
T_a^{\mu} = \frac{1 + {\rlap{v}/}}{2} \Bigg\{P_{2a}^{*\mu\nu} \gamma_{\nu} - P_{1a\nu} \sqrt{\frac{3}{2}} \gamma_5 \bigg[g^{\mu\nu} - \frac{\gamma^{\nu}(\gamma^{\mu} - v^{\mu})}{3} \bigg]\Bigg\},
\end{equation}}}
{\small{\begin{equation}
\label{eq:4} 
X_a^{\mu} = \frac{1 + {\rlap{v}/}}{2} \Bigg\{P_{2a}^{\mu\nu} \gamma_{5} \gamma_{\nu} - P_{1a\nu}^{*} \sqrt{\frac{3}{2}} \bigg[g^{\mu\nu} - \frac{\gamma^{\nu}(\gamma^{\mu} + v^{\mu})}{3} \bigg]\Bigg\},
\end{equation}}}
{\small{\begin{equation}
\begin{aligned}
\label{eq:5} 
Y_a^{\mu\nu} = & \frac{1 + {\rlap{v}/}}{2} \Bigg\{P_{3a}^{*\mu\nu\sigma} \gamma_{\sigma} - P_{2a}^{\alpha\beta} \sqrt{\frac{5}{3}} \gamma_5\\
& \bigg[g^{\mu}_{\alpha} g^{\nu}_{\beta} - \frac{g^{\nu}_{\beta}\gamma_{\alpha}(\gamma^{\mu} - v^{\mu})}{5} - \frac{g_{\alpha}^{\mu}\gamma_{\beta}(\gamma^{\mu} - v^{\nu})}{5} \bigg]\Bigg\},
 \end{aligned}
\end{equation}}}
{\small{\begin{equation}
\begin{aligned}
\label{eq:6} 
Z_a^{\mu\nu} = & \frac{1 + {\rlap{v}/}}{2} \Bigg\{P_{3a}^{\mu\nu\sigma} \gamma_5 \gamma_{\sigma} - P_{2a}^{*\alpha\beta} \sqrt{\frac{5}{3}}\\
& \bigg[g^{\mu}_{\alpha} g^{\nu}_{\beta} - \frac{g^{\nu}_{\beta}\gamma_{\alpha}(\gamma^{\mu} + v^{\mu})}{5} - \frac{g_{\alpha}^{\mu}\gamma_{\beta}(\gamma^{\mu} + v^{\nu})}{5} \bigg]\Bigg\},
\end{aligned}
\end{equation}}}
{\small{\begin{equation}
\begin{aligned}
\label{eq:7} 
R_a^{\mu\nu\rho} = & \frac{1 + {\rlap{v}/}}{2} \Bigg\{P_{4a}^{*\mu\nu\rho\sigma} \gamma_5 \gamma_{\sigma} - P_{3a}^{\alpha\beta\tau} \sqrt{\frac{7}{4}} \\ 
& \bigg[g^{\mu}_{\alpha} g^{\nu}_{\beta} g^{\rho}_{\tau} - \frac{g^{\nu}_{\beta}g^{\rho}_{\tau}\gamma_{\alpha}(\gamma^{\mu} - v^{\mu})}{7} \\
& - \frac{g_{\alpha}^{\mu}g^{\rho}_{\tau}\gamma_{\beta}(\gamma^{\nu} - v^{\nu})}{7} - \frac{g^{\mu}_{\alpha}g^{\nu}_{\beta}\gamma_{\tau}(\gamma^{\rho} -v^{\rho})}{7} \bigg]\Bigg\}.
\end{aligned}
\end{equation}}}

\noindent where $a$ ($= u, d$ or $s$) is the $SU(3)$ light quark flavor representation and $\nu$ gives the meson four velocity and is conserved in strong interactions. The heavy meson field operators $P$ and $P^*$ (see Eqs. (\ref{eq:1}) to (\ref{eq:7})) contain a factor $\sqrt{m_Q}$ having a mass dimension $\frac{3}{2}$, which annihilate the mesons with four-velocity $\nu$. Eq. (\ref{eq:1}) is for $s$ wave mesons; Eq. (\ref{eq:2}) and (\ref{eq:3}) for $p$ wave mesons; Eq. (\ref{eq:4}) and (\ref{eq:5}) for $d$ wave mesons, and Eq. (\ref{eq:6}) and (\ref{eq:7}) for $f$ wave mesons. The strong decays take place with the emission of light pseudoscalar octet mesons. We write the matrix $\cal{M}$ of light pseudoscalar mesons described by the fields $\xi = e^{\frac{i\cal{M}}{f_{\pi}}}$ as, 

{\small{\begin{equation}
\label{eq:8} 
 \cal{M} = 
\begin{pmatrix}
\frac{1}{\sqrt{2}}\pi^0 + \frac{1}{\sqrt{6}}\eta & \pi^+ & K^+ \\
\pi^- & -\frac{1}{\sqrt{2}}\pi^0 + \frac{1}{\sqrt{6}}\eta & K^0 \\
K^- & \bar{K}^0 & -\sqrt{\frac{2}{3}}\eta \\
\end{pmatrix}
\end{equation}}}

\noindent where $f_{\pi} = 130.2$ MeV.  Refs. \cite{Casalbuoni1997,Campanella2018} also study the strong decays of heavy mesons along with the light vector mesons ($\rho$, $\omega$, $K$ and $\phi$). The effective heavy meson chiral Lagrangians $\cal{L}$$_H$, $\cal{L}$$_S$, $\cal{L}$$_T$, $\cal{L}$$_X$ $\cal{L}$$_Y$, $\cal{L}$$_Z$ and $\cal{L}$$_R$ describe the two body strong interactions by an exchange of light pseudoscalar mesons, are taken from \cite{Lagrangianall},

{\small{\begin{equation}
\label{eq:9} 
{\cal{L}}_H = g_HTr[\bar{H}_aH_b\gamma_{\mu}\gamma_5{\cal{A}}^{\mu}_{ba}],
\end{equation}}}
{\small{\begin{equation}
\label{eq:10} 
{\cal{L}}_S = h_STr[\bar{H}_aS_b\gamma_{\mu}\gamma_5{\cal{A}}^{\mu}_{ba}] + H.C.,
\end{equation}}}
{\small{\begin{equation}
\label{eq:11} 
{\cal{L}}_T = \frac{h_T}{{\Lambda}_{\chi}}Tr[\bar{H}_aT^{\mu}_b(iD_{\mu} {\not\! {\cal A}} + i {\not\! {\cal D}} {\cal{A}}^{\mu})_{ba}\gamma_5] + H.C., 
\end{equation}}
\small{\begin{equation}
\label{eq:12} 
{\cal{L}}_X = \frac{k_X}{{\Lambda}_{\chi}}Tr[\bar{H}_aX^{\mu}_b(iD_{\mu} {\not\! {\cal A}} + i {\not\! {\cal D}} {\cal{A}}^{\mu})_{ba}\gamma_5] + H.C., 
\end{equation}}}
{\small{\begin{equation}
\begin{aligned}
\label{eq:13} 
{\cal{L}}_Y = & \frac{1}{{\Lambda}_{\chi}^2}Tr[\bar{H}_aY^{\mu\nu}_b [k_1^Y\{D_{\mu}, D_{\nu}\} {\cal{A}}_{\lambda}\\  
& + k_2^Y(D_{\mu}D_{\lambda}{\cal{A}}_{\nu} + D_{\nu}D_{\lambda}{\cal{A}}_{\mu})]_{ba}\gamma^{\lambda}\gamma_5] + H.C., 
\end{aligned}
\end{equation}}}
{\small{\begin{equation}
\begin{aligned}
\label{eq:14} 
{\cal{L}}_Z = & \frac{1}{{\Lambda}_{\chi}^2}Tr[\bar{H}_aZ^{\mu\nu}_b [k_1^Z\{D_{\mu}, D_{\nu}\} {\cal{A}}_{\lambda}\\
& + k_2^Z(D_{\mu}D_{\lambda}{\cal{A}}_{\nu} + D_{\nu}D_{\lambda}{\cal{A}}_{\mu})]_{ba}\gamma^{\lambda}\gamma_5] + H.C., 
\end{aligned}
\end{equation}}}
{\small{\begin{equation}
\begin{aligned}
\label{eq:15} 
{\cal{L}}_R = & \frac{1}{{\Lambda}_{\chi}^3}Tr[\bar{H}_aR^{\mu\nu\rho}_b [k_1^R\{D_{\mu}, D_{\nu}, D_{\rho}\} {\cal{A}}_{\lambda}\\
& + k_2^R (\{D_{\mu}, D_{\rho}\}D_{\lambda}{\cal{A}}_{\nu} \\
& + \{D_{\nu}, D_{\rho}\}D_{\lambda} {\cal{A}}_{\mu} \{D_{\mu}, D_{\nu}\} D_{\lambda} {\cal{A}}_{\rho})]_{ba} \gamma^{\lambda} \gamma_5 \\
& + H.C.,
\end{aligned} 
\end{equation}}}

\noindent where vector and axial-vector operators are,

{\small{\begin{equation}
\begin{aligned}
\label{eq:16} 
{\cal{V}}_{{\mu}ba} = \frac{1}{2} (\xi^{\dag} \partial_{\mu} \xi + \xi \partial_{\mu} \xi^{\dag})_{ba},
\end{aligned} 
\end{equation}}}
{\small{\begin{equation}
\begin{aligned}
\label{eq:17} 
{\cal{A}}_{{\mu}ba} = \frac{i}{2} (\xi^{\dag} \partial_{\mu} \xi - \xi \partial_{\mu} \xi^{\dag})_{ba};
\end{aligned} 
\end{equation}}}

\noindent and the operator, {\small{$D_{{\mu}ba} = -\delta_{ba} \partial_{\mu} + {\cal{V}}_{{\mu}ba}$}}. Also here, {\small{$\{D_{\mu}, D_{\nu}\} = D_{\mu}D_{\nu}+D_{\nu}D_{\mu}$}} and {\small{$\{D_{\mu}, D_{\nu}, D_{\rho}\} = D_{\mu}D_{\nu}D_{\rho} + D_{\mu}D_{\rho}D_{\nu} + D_{\nu}D_{\mu}D_{\rho} + D_{\nu}D_{\rho}D_{\mu} + D_{\rho}D_{\mu}D_{\nu} + D_{\rho}D_{\nu}D_{\mu}$}}. ${\Lambda}_{\chi}$ is the chiral symmetry breaking scale and is fixed to 1 GeV. The mass parameters $\delta{m}_S = m_S - m_H$, $\delta{m}_T = m_T - m_H$, $\delta{m}_X = m_X - m_H$, $\delta{m}_Y = m_Y - m_H$, $\delta{m}_Z = m_Z - m_H$, and $\delta{m}_R = m_R - m_H$ represent the mass splittings between the higher and the lower mass doublets described by the field $H_a$ (see Eq. (\ref{eq:1})). The strong running coupling constants $g_H$, $h_S$, $h_T$, $k_X$, $k_Y = k_1^Y + k_2^Y$, $k_Z = k_1^Z + k_2^Z$, and $k_R = k_1^R + k_2^R$ can be fitted to the experimental data. For $n = 2$ the coupling constants are denoted by $g_H^{\dag}$, $h_S^{\dag}$, $h_T^{\dag}$, $k_X^{\dag}$, $k_Y^{\dag}$, $k_Z ^{\dag}$, and $k_R^{\dag}$ and for $n = 3$ they are $g_H^{\ddag}$, $h_S^{\ddag}$, $h_T^{\ddag}$, $k_X^{\ddag}$, $k_Y^{\ddag}$, $k_Z ^{\ddag}$, and $k_R^{\ddag}$. $g_H$ (in Eq. (\ref{eq:9})) controls the $s$ wave decays, $h_S$ and $h_T$ (in Eqs. (\ref{eq:10}) and (\ref{eq:11})) are governs the $p$ wave decays, $k_X$ and $k_Y$ (in Eqs. (\ref{eq:12}) and (\ref{eq:13})) describe the $d$ wave decays, and $k_Z$ and $k_R$ (in Eqs. (\ref{eq:14}) and (\ref{eq:15})) are responsible for the $f$ wave decays. Such chiral Lagrangians can determine the expressions of strong decays of heavy-light mesons into the lower mass charged and neutral $D^{(*)}$ and $D_S^{(*)}$ mesons along with the light pseudoscalar mesons ($\pi$, $\eta$ and $K$),      

{\begin{enumerate}[label=\Roman*.]
\item Decaying $s$ wave doublet $(P, P^*)$ or $(P^{\dag}, P^{\dag*})$ or $(P^{\ddag}, P^{\ddag*})$:
\begin{equation}
\label{eq:18}
{\small{\Gamma(P^{\dag} \rightarrow P^*{\cal{P}}) = C_{\cal{P}} \frac{g_H^{\dag2}}{2\pi f_{\pi}^2} \frac{P^*}{P^{\dag}} {|\vec{P}_{\cal{P}}|}^3}}
\end{equation}
\begin{equation}
\label{eq:19}
{\small{\Gamma(P^{\dag*} \rightarrow P{\cal{P}}) = C_{\cal{P}} \frac{g_H^{\dag2}}{6\pi f_{\pi}^2} \frac{P}{P^{\dag*}} {|\vec{P}_{\cal{P}}|}^3}}
\end{equation}
\begin{equation}
\label{eq:20}
{\small{\Gamma(P^{\dag*} \rightarrow P^*{\cal{P}}) = C_{\cal{P}} \frac{g_H^{\dag2}}{3\pi f_{\pi}^2} \frac{P^*}{P^{\dag*}} {|\vec{P}_{\cal{P}}|}^3}}
\end{equation}
\begin{equation}
\label{eq:21}
{\small{\Gamma(P^{\ddag} \rightarrow P^*{\cal{P}}) = C_{\cal{P}} \frac{g_H^{\ddag2}}{2\pi f_{\pi}^2} \frac{P^*}{P^{\ddag}} {|\vec{P}_{\cal{P}}|}^3}}
\end{equation}
\begin{equation}
\label{eq:22}
{\small{\Gamma(P^{\ddag*} \rightarrow P{\cal{P}}) = C_{\cal{P}} \frac{g_H^{\ddag2}}{6\pi f_{\pi}^2} \frac{P}{P^{\ddag*}} {|\vec{P}_{\cal{P}}|}^3}}
\end{equation}
\begin{equation}
\label{eq:23}
{\small{\Gamma(P^{\ddag*} \rightarrow P^*{\cal{P}}) = C_{\cal{P}} \frac{g_H^{\ddag2}}{3\pi f_{\pi}^2} \frac{P^*}{P^{\ddag*}} {|\vec{P}_{\cal{P}}|}^3}}
\end{equation}

\item Decaying $p$ wave doublets $(P^*_0, P^{\prime}_1)$ and $(P_1, P_2^*)$ or $(P^{\dag*}_0, P^{\dag\prime}_1)$ and $(P^{\dag}_1, P_2^{\dag*})$ or $(P^{\ddag*}_0, P^{\ddag\prime}_1)$ and $(P^{\ddag}_1, P_2^{\ddag*})$:

\begin{equation}
\label{eq:24}
{\small{\Gamma(P^{\dag{\prime}}_1 \rightarrow P^*{\cal{P}}) = C_{\cal{P}} \frac{h_S^{\dag2}}{2\pi f_{\pi}^2} \frac{P^*}{P^{\dag{\prime}}_1} [m_{\cal{P}}^2 + {|\vec{P}_{\cal{P}}|}^2]|\vec{P}_{\cal{P}}|}}
\end{equation}
\begin{equation}
\label{eq:25}
{\small{\Gamma(P_1 \rightarrow P^*{\cal{P}}) = C_{\cal{P}} \frac{2h_T^2}{3\pi f_{\pi}^2} \frac{P^*}{P_1} {|\vec{P}_{\cal{P}}|}^5}}
\end{equation}
\begin{equation}
\label{eq:25}
{\small{\Gamma(P^*_2 \rightarrow P{\cal{P}}) = C_{\cal{P}} \frac{4h_T^2}{15\pi f_{\pi}^2} \frac{P}{P^*_2} {|\vec{P}_{\cal{P}}|}^5}}
\end{equation}
\begin{equation}
\label{eq:26}
{\small{\Gamma(P^*_2 \rightarrow P^*{\cal{P}}) = C_{\cal{P}} \frac{2h_T^2}{5\pi f_{\pi}^2} \frac{P^*}{P^*_2} {|\vec{P}_{\cal{P}}|}^5}}
\end{equation}
\begin{equation}
\label{eq:27}
{\small{\Gamma(P^{\dag*}_2 \rightarrow P{\cal{P}}) = C_{\cal{P}} \frac{4h_T^{\dag2}}{15\pi f_{\pi}^2} \frac{P}{P^{\dag*}_2} {|\vec{P}_{\cal{P}}|}^5}}
\end{equation}
\begin{equation}
\label{eq:28}
{\small{\Gamma(P^{\dag*}_2 \rightarrow P^*{\cal{P}}) = C_{\cal{P}} \frac{2h_T^{\dag2}}{5\pi f_{\pi}^2} \frac{P^*}{P^{\dag*}_2} {|\vec{P}_{\cal{P}}|}^5}}
\end{equation}
\begin{equation}
\label{eq:29}
{\small{\Gamma(P^{\ddag*}_2 \rightarrow P{\cal{P}}) = C_{\cal{P}} \frac{4h_T^{\ddag2}}{15\pi f_{\pi}^2} \frac{P}{P^{\ddag*}_2} {|\vec{P}_{\cal{P}}|}^5}}
\end{equation}
\begin{equation}
\label{eq:30}
{\small{\Gamma(P^{\ddag*}_2 \rightarrow P^*{\cal{P}}) = C_{\cal{P}} \frac{2h_T^{\ddag2}}{5\pi f_{\pi}^2} \frac{P^*}{P^{\ddag*}_2} {|\vec{P}_{\cal{P}}|}^5}}
\end{equation}

\item Decaying $d$ wave doublets $(P^*_1, P_2)$ and $(P^{\prime}_2, P^*_3)$:

\begin{equation}
\label{eq:31}
{\small{\Gamma(P_2 \rightarrow P^*{\cal{P}}) = C_{\cal{P}} \frac{2k_X^2}{3\pi f_{\pi}^2} \frac{P^*}{P_2} [m_{\cal{P}}^2 + {|\vec{P}_{\cal{P}}|}^2]|\vec{P}_{\cal{P}}|^3}}
\end{equation}
\begin{equation}
\label{eq:32}
{\small{\Gamma(P^{\prime}_2 \rightarrow P^*{\cal{P}}) = C_{\cal{P}} \frac{4k_Y^2}{15\pi f_{\pi}^2} \frac{P^*}{P^{\prime}_2} {|\vec{P}_{\cal{P}}|}^7}}
\end{equation}
\begin{equation}
\label{eq:33}
{\small{\Gamma(P^{*}_3 \rightarrow P{\cal{P}}) = C_{\cal{P}} \frac{4k_Y^2}{35\pi f_{\pi}^2} \frac{P}{P^*_3} {|\vec{P}_{\cal{P}}|}^7}}
\end{equation}
\begin{equation}
\label{eq:34}
{\small{\Gamma(P^{*}_3 \rightarrow P^*{\cal{P}}) = C_{\cal{P}} \frac{16k_Y^2}{105\pi f_{\pi}^2} \frac{P^*}{P^*_3} {|\vec{P}_{\cal{P}}|}^7}}
\end{equation}
\begin{equation}
\label{eq:33}
{\small{\Gamma(P^{\dag*}_3 \rightarrow P{\cal{P}}) = C_{\cal{P}} \frac{4k_Y^{\dag2}}{35\pi f_{\pi}^2} \frac{P}{P^{\dag*}_3} {|\vec{P}_{\cal{P}}|}^7}}
\end{equation}
\begin{equation}
\label{eq:34}
{\small{\Gamma(P^{\dag*}_3 \rightarrow P^*{\cal{P}}) = C_{\cal{P}} \frac{16k_Y^{\dag^2}}{105\pi f_{\pi}^2} \frac{P^*}{P^{\dag*}_3} {|\vec{P}_{\cal{P}}|}^7}}
\end{equation}

\item Decaying $f$ wave doublets $(P^{\prime*}_2, P_3)$ and $(P^{\prime}_3, P^*_4)$:

\begin{equation}
\label{eq:35}
{\small{\Gamma(P^{\prime*}_2 \rightarrow P{\cal{P}}) = C_{\cal{P}} \frac{4k_Z^2}{25\pi f_{\pi}^2} \frac{P}{P^{\prime*}_2} [m_{\cal{P}}^2 + {|\vec{P}_{\cal{P}}|}^2]|\vec{P}_{\cal{P}}|^5}}
\end{equation}
\begin{equation}
\label{eq:36}
{\small{\Gamma(P^{\prime*}_2 \rightarrow P^*{\cal{P}}) = C_{\cal{P}} \frac{8k_Z^2}{75\pi f_{\pi}^2} \frac{P^*}{P^{\prime*}_2} [m_{\cal{P}}^2 + {|\vec{P}_{\cal{P}}|}^2]|\vec{P}_{\cal{P}}|^5}}
\end{equation}
\begin{equation}
\label{eq:35}
{\small{\Gamma(P_3 \rightarrow P^*{\cal{P}}) = C_{\cal{P}} \frac{4k_Z^2}{15\pi f_{\pi}^2} \frac{P^*}{P_3} [m_{\cal{P}}^2 + {|\vec{P}_{\cal{P}}|}^2]|\vec{P}_{\cal{P}}|^5}}
\end{equation}
\begin{equation}
\label{eq:37}
{\small{\Gamma(P^*_4 \rightarrow P{\cal{P}}) = C_{\cal{P}} \frac{16k_R^2}{35\pi f_{\pi}^2} \frac{P}{P^*_4} |\vec{P}_{\cal{P}}|^9}}
\end{equation}
\begin{equation}
\label{eq:38}
{\small{\Gamma(P^*_4 \rightarrow P^*{\cal{P}}) = C_{\cal{P}} \frac{4k_R^2}{7\pi f_{\pi}^2} \frac{P^*}{P^*_4} |\vec{P}_{\cal{P}}|^9}}.
\end{equation}
\begin{equation}
\label{eq:37}
{\small{\Gamma(P^{\dag*}_4 \rightarrow P{\cal{P}}) = C_{\cal{P}} \frac{16k_R^{\dag2}}{35\pi f_{\pi}^2} \frac{P}{P^{\dag*}_4} |\vec{P}_{\cal{P}}|^9}}
\end{equation}
\begin{equation}
\label{eq:38}
{\small{\Gamma(P^{\dag*}_4 \rightarrow P^*{\cal{P}}) = C_{\cal{P}} \frac{4k_R^{\dag2}}{7\pi f_{\pi}^2} \frac{P^*}{P^{\dag*}_4} |\vec{P}_{\cal{P}}|^9}}.
\end{equation}
\end{enumerate}}

\noindent For the decay mode $P_a \rightarrow P_b + {\cal{P}}$ we have  $|\vec{P}_{\cal{P}}| = \frac{\sqrt{m_{P_a}^2 + m_{P_b}^2 + m_{\cal{P}}^2 - 2m_{P_a} m_{P_b} - 2m_{P_a} m_{\cal{P}} - 2m_{P_b} m_{\cal{P}}}}{2m_{p_a}} $; where $m_{P_a}$, $m_{P_b}$ and $m_{\cal{P}}$ are their respective masses. The coefficients ${\cal{P}}$ of the light pseudoscalar  mesons are: $C_{{\pi}^{\pm}}, C_{{K}^{\pm}} = 1$, $C_{{\pi}^0} = \frac{1}{2}$ and $C_{\eta} = \frac{1}{6}$. The masses of the light pseudoscalar mesons and the ground state charmed mesons are taken from PDG-2018 \cite{Tanabashi2018-19}: $M_{\pi^{\pm}} = 139.57061$ MeV, $M_{\pi^0} = 134.9770$ MeV, $M_{K^{\pm}} = 493.677$ MeV, $M_{K^{0}} = 497.611$ MeV, $M_{\eta} = 547.862$ MeV, $M_{D^{\pm}} = 1869.65$, $M_{D^{0}} = 1864.84$ MeV, $M_{D^{*\pm}} = 2010.26$ MeV, $M_{D^{*0}} = 2006.85$ MeV, $M_{D_s^{\pm}} = 1969.0$ MeV, $M_{D_s^{*\pm}} = 2112.2$ MeV. In the heavy quark mass limit, the spin and flavor violations of order $\frac{1}{m_Q}$ are not taken into the consideration in this present study to avoid introducing new unknown coupling constants. The strong decay widths can provide some useful informations and are used for the classification of various mesonic states according to their total spin and parity. Also the ratio and the branching fractions among the decay widths, independent of the coupling constants, can help to identify the heavy mesons.

\section{Results and Discussion}
\label{sec3}

Using the Eqs. (\ref{eq:18}) to (\ref{eq:38}), the strong decay rates of nonstrange singly charmed mesons ($D{_{2}^*}(2460)$, $D(2550)^0$, $D{_{J}^*}(2600)^0$, $D(2740)^0$, $D{_{3}^*}(2750)^0$, $D_J(3000)^0$, $D{_{J}^*}(3000)^0$ and $D_2^*(3000)^0$ observed by the experimental Collaborations LHCb \cite{Aaij2016,Aaij2015,Aaij2013} and $BABAR$ \cite{del2010}) are computed. That are presented in Table \ref{tab3} in terms of the square of the coupling constants $h_T$, $g_H^{\dag}$, $k_Y$, $g_H^{\ddag}$, $h_S^{\dag}$, $h_T^{\dag}$, $k_R$ and $k_Z$. Such a wide range of couplings are not yet observed experimentally. The present experimental facility LHCb and an upcoming project $\overline{\mbox{\sffamily P}}${\sffamily ANDA} \cite{Singh(all),Barucca2019} will fit these strong couplings in near future. Theoretically, the Refs. \cite{Colangelo1995,Casalbuoni1997,Wang2006,Wang2007,Huang2010} have studied the strong coupling constants of $s$ and $p$ wave ground state heavy mesons. Comparing the calculated total decay widths shown in Table \ref{tab3} (also Figures (\ref{fig1}) to (\ref{fig12}) represents the strong decay rates that are changing with respect to the square of the couplings) with their respective experimentally observed decay widths listed in Table \ref{tab1}, we determine the strong coupling constants which are presented in Table \ref{tab3}.

The branching ratios avoid the unknown hadronic couplings and are compared with experimental observations where available. The branching ratio,

{\small{\begin{center}
$BR_{D_2^*(2460)^0}=\frac{\Gamma(D_2^*(2460)^0 \rightarrow D^{+} \pi^-)}{\Gamma(D_2^*(2460)^0 \rightarrow D^{*+} \pi^-)}\approx$ 2.3, 
\end{center}}}

\noindent calculated from Ref. \cite{Aaij2016}, \cite{Aaij2013} and \cite{del2010}. It is in good agreement with the measurements of CLEO Collaboration 2.3 $\pm$ 0.8 \cite{Avery1990}, underestimated to ZEUS 2.8 $\pm$ 0.8 \cite{Chekanov2009} and overestimated to $BABAR$ 1.47 $\pm$ 0.03 \cite{del2010} and ZEUS 1.4 $\pm$ 0.3 \cite{Abramowicz20013}. The ratio,      

{\small{\begin{center}
$R_{D_2^*(2460)^0}=\frac{\Gamma(D_2^*(2460)^0 \rightarrow D^{+} \pi^-)}{\Gamma(D_2^*(2460)^0 \rightarrow D^{+} \pi^-) + \Gamma(D_2^*(2460)^0 \rightarrow D^{*+} \pi^-)}$ 
\end{center}}}

\noindent $\approx$ 0.70 from Refs. \cite{Aaij2016,Aaij2013,del2010} and is close to 0.62 $\pm$ 0.03 $\pm$ 0.02 of $BABAR$ measurement \cite{Aubert2009}. The branching ratio,

{\small{\begin{center}
$BR_{D_2^*(2460)^+}=\frac{\Gamma(D_2^*(2460)^+ \rightarrow D^{0} \pi^+)}{\Gamma(D_2^*(2460)^+ \rightarrow D^{*0} \pi^+)}\approx$ 2.3 
\end{center}}}

\noindent from \cite{Aaij2013}, which is nearer to 1.9 $\pm$ 1.1 $\pm$ 0.3 of CLEO measurement \cite{Bergfeld1994} and overestimated to ZEUS 1.1 $\pm$ 0.4 \cite{Abramowicz20013}. And, also the ratio     

{\small{\begin{center}
$R_{D_2^*(2460)^+}=\frac{\Gamma(D_2^*(2460)^+ \rightarrow D^{0} \pi^+)}{\Gamma(D_2^*(2460)^+ \rightarrow D^{0} \pi^+) + \Gamma(D_2^*(2460)^+ \rightarrow D^{*0} \pi^+)}$
\end{center}}}

\noindent $\approx$ 0.7 from \cite{Aaij2013} close to Ref. \cite{Aubert2009}. The branching ratios,

{\small{\begin{center}
$BR_{D_J^*(2600)^0}=\frac{\Gamma(D_J^*(2600)^0 \rightarrow D^{+} \pi^-)}{\Gamma(D_J^*(2600)^0 \rightarrow D^{*+} \pi^-)}\approx$ 0.8 
\end{center}}}
{\small{\begin{center}
$BR_{D_J^*(2750)^0}=\frac{\Gamma(D_J^*(2750)^0 \rightarrow D^{+} \pi^-)}{\Gamma(D_J^*(2750)^0 \rightarrow D^{*+} \pi^-)}\approx$ 1.9
\end{center}}}

\noindent calculated from Ref. \cite{Aaij2016}, \cite{Aaij2013} and \cite{del2010}, which are overestimated to the $BABAR$ measurements $BR_{D_J^*(2600)^0} =$ 0.32 $\pm$ 0.02 $\pm$ 0.09 and $BR_{D_J^*(2750)^0}=$ 0.42 $\pm$ 0.05 $\pm$ 0.11 \cite{del2010}.

Therefore, the charmed mesons $D_2^*(2460)$ and $D_J^*(2750)$ belonging to 1$^3P_2$  and 1$^3D_3$ are dominant in $D\pi$ decay mode and, $D_J^*(2600)$ with 2$^3S_1$ dominant in $D^*\pi$ decay. That are in accessible with the experimental observations. Moreover, the $D_1(2420)$, $D(2550)$ and $D(2740)$ are found to be spin partners of $D_2^*(2460)$, $D_J^*(2600)$ and $D_J^*(2750)$ respectively. So we write,

{\small{\begin{equation}
\big(D_1(2420), D_2^*(2460)\big) = (1^+, 2^+)_{{\frac{3}{2}}^+} = \big(1^1P_1, 1^3P_2\big),
\end{equation}}}   
{\small{\begin{equation}
\big(D(2550), D_J^*(2600)\big) = (0^-, 1^-)_{{\frac{1}{2}}^-} = \big(2^1S_0, 2^3S_1\big),
\end{equation}}}
{\small{\begin{equation}
\big(D(2740), D_J^*(2750)\big) = (2^-, 3^-)_{{\frac{5}{2}}^-}  = \big(1^3D_2, 1^3D_3\big).
\end{equation}}}

The mass difference $M_{D_J^*(3000)^0} - M_{D_J(3000)^0} \approx 36$ MeV. They might be from the same wave family. Experimentally, $D_J(3000)$ is measured with unnatural parity and $D_J^*(3000)$ with natural parity. So they can have an isodoublet state either $(0^-, 1^-)_{{\frac{1}{2}}^-}$ or $(1^+, 2^+)_{{\frac{3}{2}}^+}$. The $\big(D_J(3000), D_J^*(3000)\big)$ is not of an isodoublet $(1^+, 0^+)_{{\frac{1}{2}}^+}$ because the $J^P = 0^+$ of $D_J^*(3000)$ is not possible to be heavier than $J^P = 1^+$ of $D_J(3000)$. The $D_J^*(3000)^0$ as $3^3S_1$ has

{\small{\begin{center}
$BR_{D_J^*(3000)^0}=\frac{\Gamma(D_J^*(3000)^0 \rightarrow D^{+} \pi^-)}{\Gamma(D_J^*(3000)^0 \rightarrow D^{*+} \pi^-)} \approx 0.64$, 
\end{center}}}

\noindent that means, the decay mode $D^{*+} \pi^-$ is dominant over $D^{+} \pi^-$. The $D_J^*(3000)^0$ as $2^3P_2$ has $BR_{D_J^*(3000)^0}\geq1$, which is in agreement with the experimental measurement. Hence,

{\small{\begin{equation}
\big(D_J(3000), D_J^*(3000)\big) = (1^+, 2^+)_{{\frac{3}{2}}^+} = \big(2^3P_1, 2^3P_2\big).
\end{equation}}}

The mass difference between $D_2^*(3000)^0$ and $D_J^*(3000)^0$ is approximately 206 MeV. Such a large mass difference indicate $D_2^*(3000)^0$ state is not of $2P$ state. Experimentally, its observed spin-parity is $2^+$. So it can be a candidate of $3^3P_2$ or $1^3F_2$. For $D_2^*(3000)^0$ as $3^3P_2$,
    
{\small{\begin{center}
$BR_{D_2^*(3000)^0}=\frac{\Gamma(D_2^*(3000)^0 \rightarrow D^{+} \pi^-)}{\Gamma(D_2^*(3000)^0 \rightarrow D^{*+} \pi^-)} \approx 0.95$, 
\end{center}}}

\noindent i.e. the decay $D^{*+} \pi^-$ is more dominant than $D^+ \pi^-$. For $1^3F_2$ state, the $BR_{D_2^*(3000)^0}$ is 2.51, which is most favorable to decay in $D^+ \pi^-$ and, it is in accordance with the experimental measurement. So,

{\small{\begin{equation}
D_2^*(3000)^0 = (2^+)_{{\frac{5}{2}}^+} = \big(1^3F_2\big).
\end{equation}}}

\section{Regge trajectory}
\label{sec4}

Spin and parity assignments of excited $D$ mesons from the strong decays analysis are presented in Table \ref{tab4} with their respective PDG-2018 \cite{Tanabashi2018-19} world average masses. Using these we construct the Regge trajectory in which the total spin (or principal quantum number $(n)$) and the mass of hadrons are related. This can help in predicting the possible quantum states of hadrons. An investigation of meson spectrum in the non-perturbative regime of quark-gluon interactions has a great importance for understanding the dynamics of strong interactions (for details see Refs. \cite{Collins1977,Godfrey1999}). We are using the following definitions: 
\begin{enumerate}[label=\Roman*.]
\item the Regge trajectory in $(J, M^2)$ plane,
{\small{\begin{equation}
J = \alpha M^2 + \alpha_0;
\end{equation}}}
\item and the Regge trajectory in $(n_r, M^2)$ plane,
{\small{\begin{equation}
n_r = \beta M^2 + \beta_0;
\end{equation}}}
\end{enumerate}
\noindent where $\alpha$, $\beta$ are slopes, $\alpha_0$, $\beta_0$ are intercepts and $n_r (= n -1) = 0, 1, 2, ...$ is the radial principal quantum number. The Regge trajectory in $(J, M^2)$ plane are available with the evenness and oddness of the total spin $J$ are respectively distinguished according to their parity $P = (-1)^J$ called natural parity and $P = (-1)^{J-1}$ called unnatural parity. Figures \ref{fig13} and \ref{fig14} shows the plots of Regge trajectories in $(J, M^2)$ and $(n_r, M^2)$ planes which are usually called Chew-Frautschi plots. The $D$ meson states are fitted on the Regge line with sufficiently good accuracy. The parameters like Regge slopes and the intercepts are extracted from the Regge trajectories (see in Table \ref{tab5} and \ref{tab6}), that estimate the masses of the states lying on these Regge trajectories. The Regge slope is assumed to be same for all $D$ meson multiplets lying on the single Regge line.

\begin{figure}
  \includegraphics[width=0.49\textwidth]{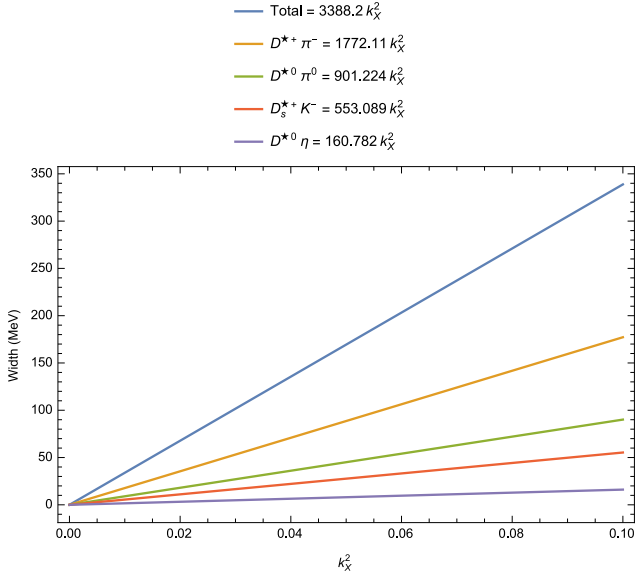}
\caption{Strong decay widths of $1^1D_2$ (in MeV) nonstrange charmed meson state (lying on the Regge line $1^1S_0$ in $(n_r, M^2)$ plane) changing with the square of the coupling $k_X^2$ in HQET.}
\label{fig15}       
\end{figure}

\begin{figure}
  \includegraphics[width=0.49\textwidth]{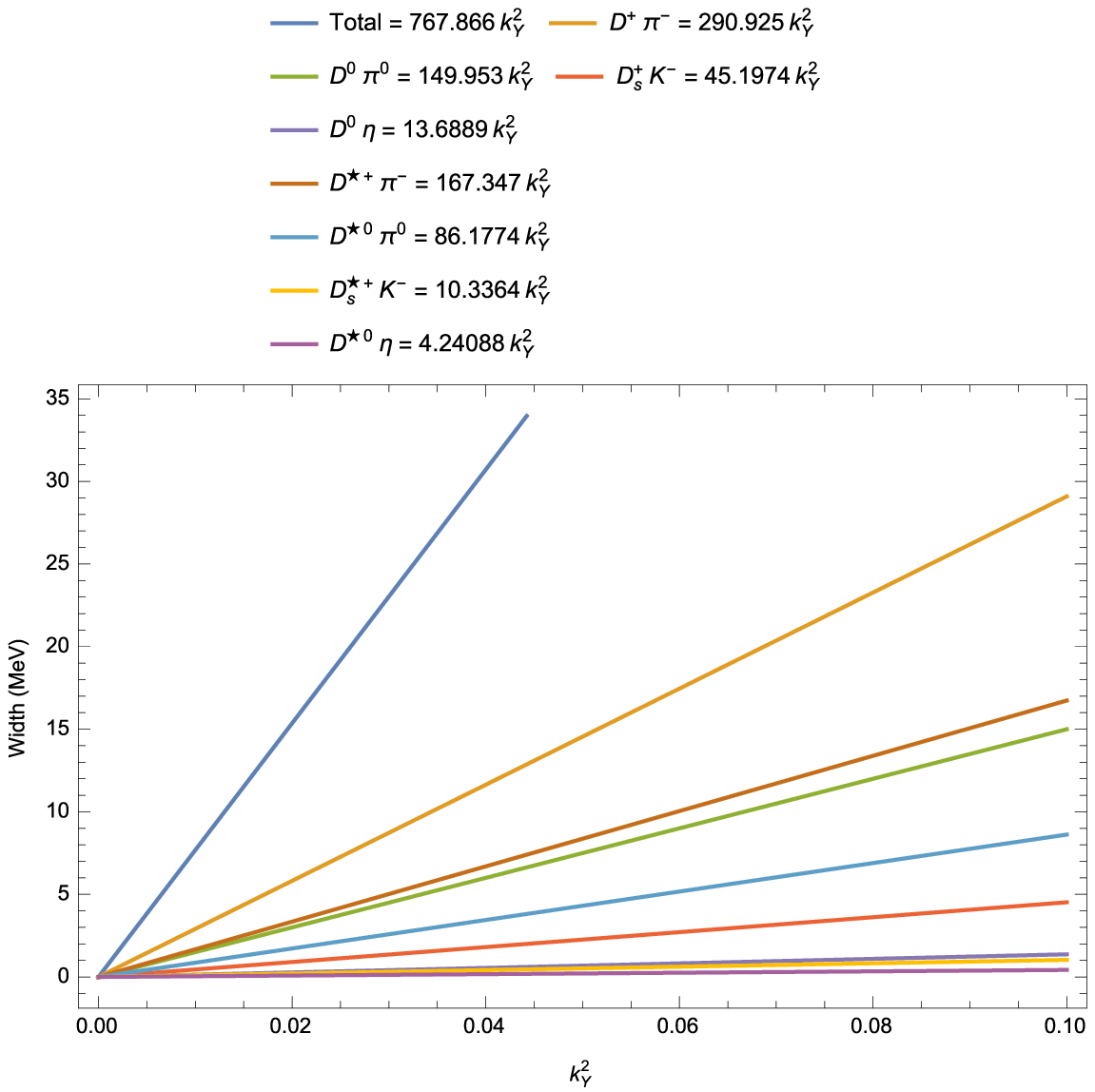}
\caption{Strong decay widths of $1^3D_3$ (in MeV) nonstrange charmed meson state (lying on the Regge line $1^3S_1$ in $(J, M^2)$ plane) changing with the square of the coupling $k_Y^2$ in HQET.}
\label{fig16}       
\end{figure}

\begin{figure}
  \includegraphics[width=0.49\textwidth]{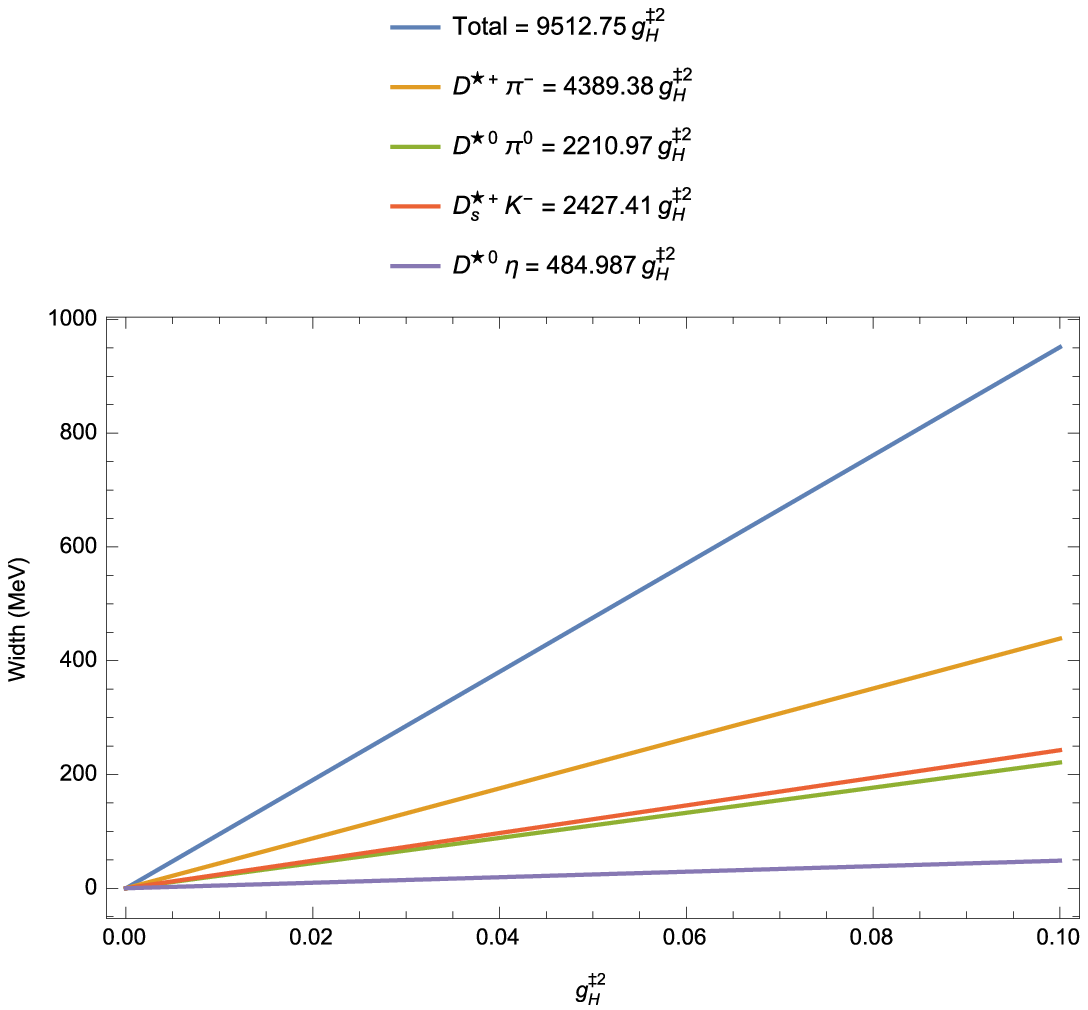}
\caption{Strong decay widths of $3^1S_0$ (in MeV) nonstrange charmed meson state (lying on the Regge line $1^1S_0$ in $(n_r, M^2)$ plane) changing with the square of the coupling $g_H^{\ddag2}$ in HQET.}
\label{fig17}       
\end{figure}

\begin{figure}
  \includegraphics[width=0.49\textwidth]{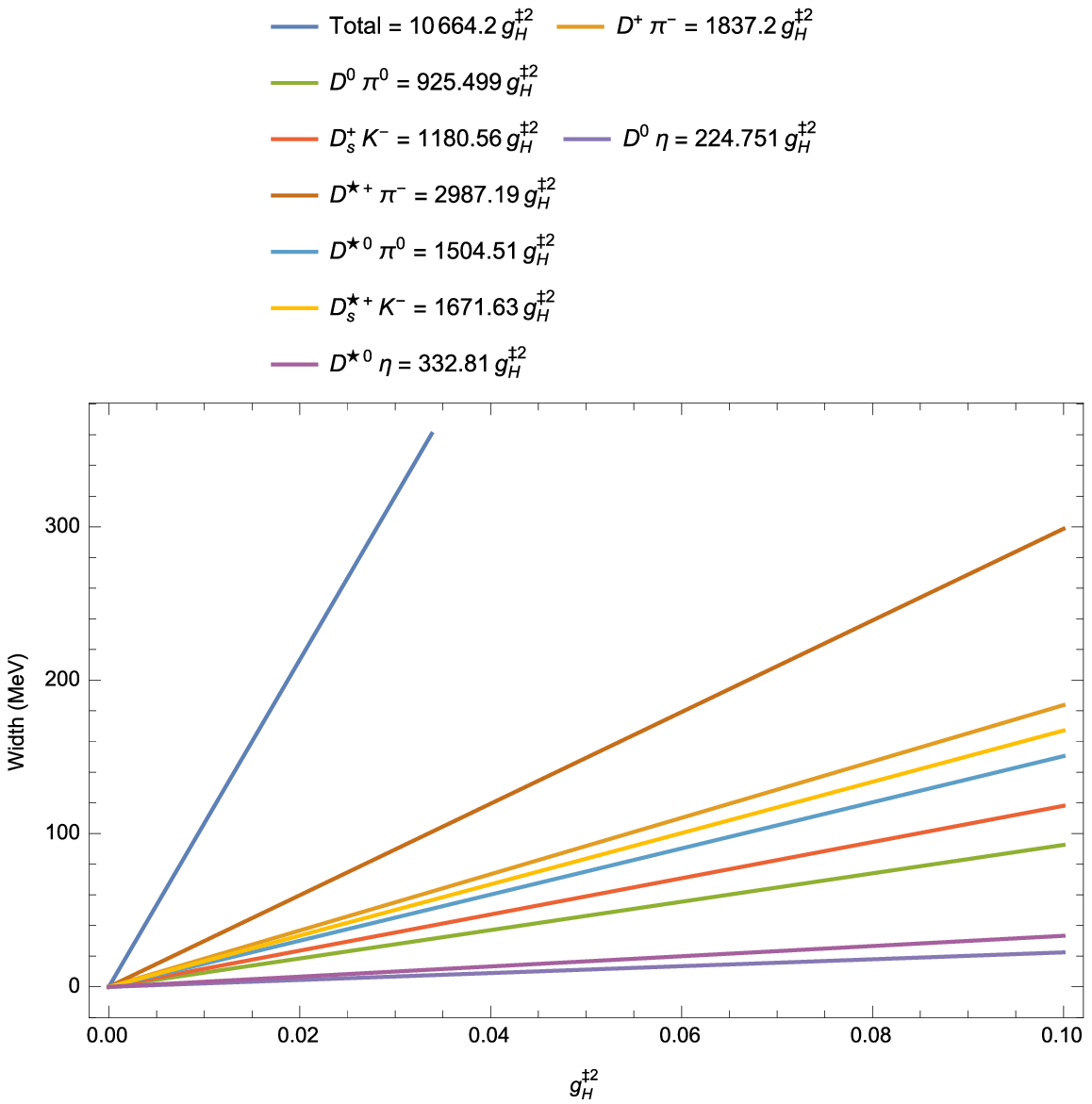}
\caption{Strong decay widths of $3^3S_1$ (in MeV) nonstrange charmed meson state (lying on the Regge line $1^3S_1$ in $(n_r, M^2)$ plane) changing with the square of the coupling $g_H^{\ddag2}$ in HQET.}
\label{fig18}       
\end{figure}

\begin{figure}
  \includegraphics[width=0.49\textwidth]{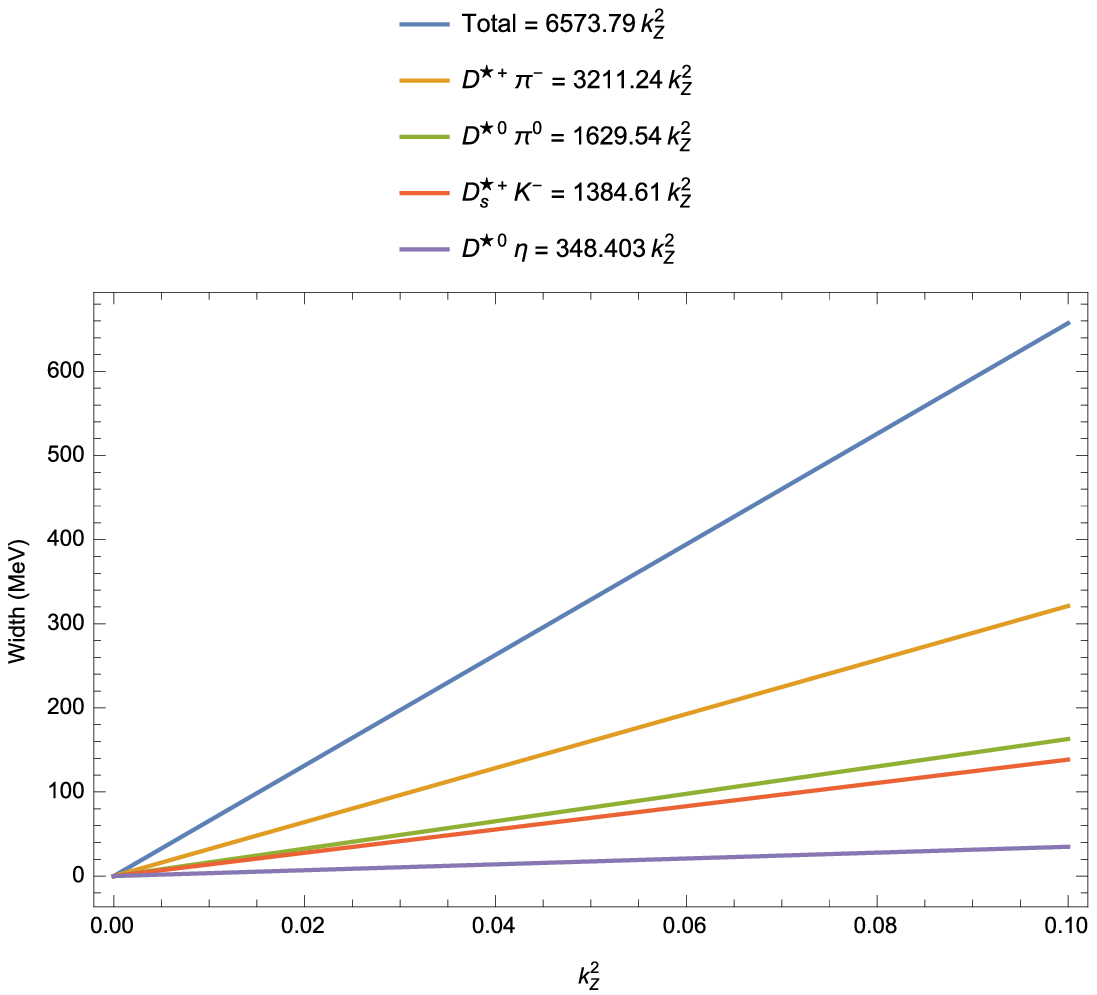}
\caption{Strong decay widths of $1^1F_3$ (in MeV) nonstrange charmed meson state (lying on the Regge line $1^1S_0$ in $(J, M^2)$ plane) changing with the square of the coupling $k_Z^2$ in HQET.}
\label{fig19}       
\end{figure}

\begin{figure}
  \includegraphics[width=0.49\textwidth]{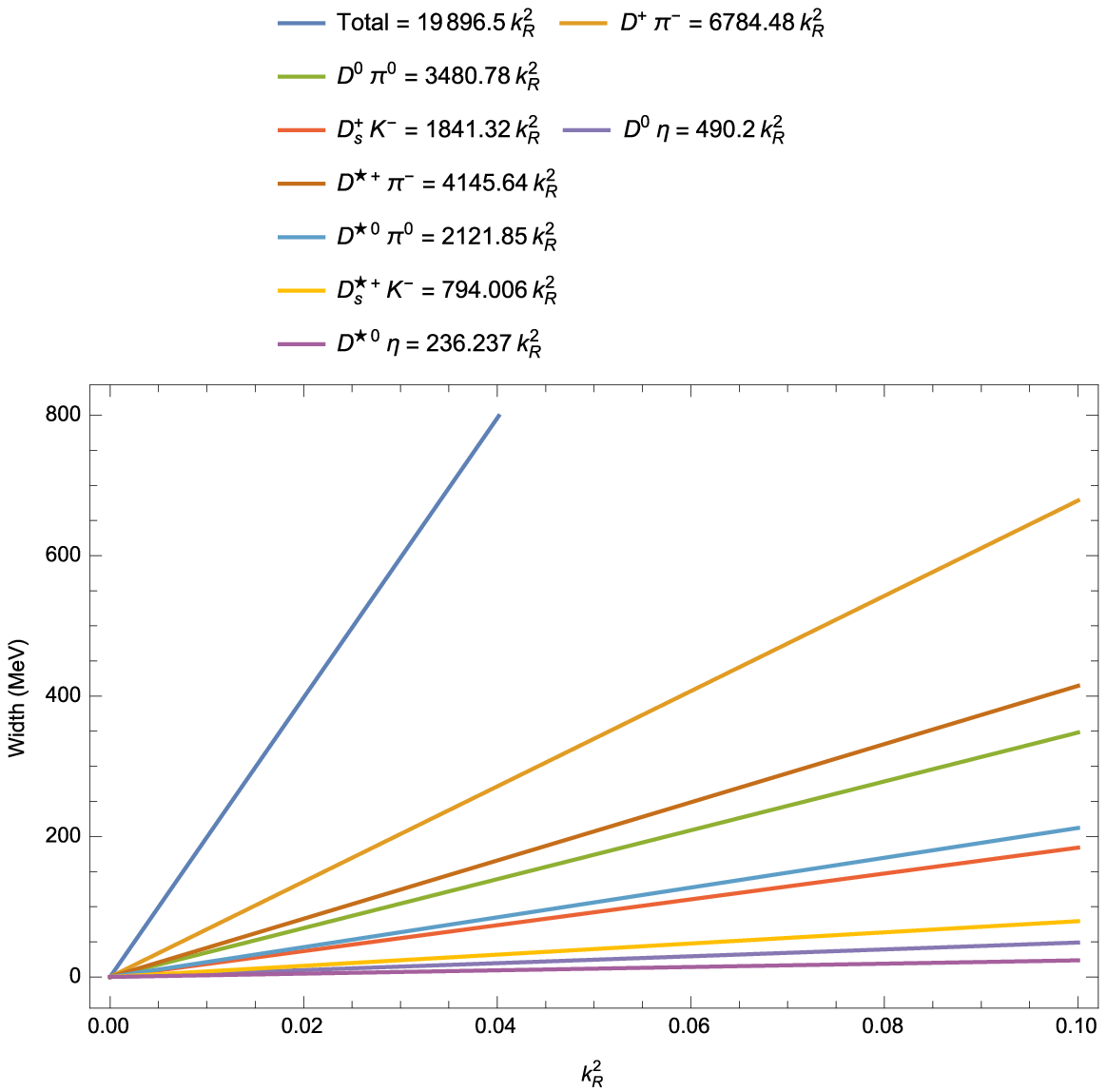}
\caption{Strong decay widths of $1^3F_4$ (in MeV) nonstrange charmed meson state (lying on the Regge line $1^3S_1$ in $(J, M^2)$ plane) changing with the square of the coupling $k_R^2$ in HQET.}
\label{fig20}       
\end{figure}

\begin{figure}
  \includegraphics[width=0.49\textwidth]{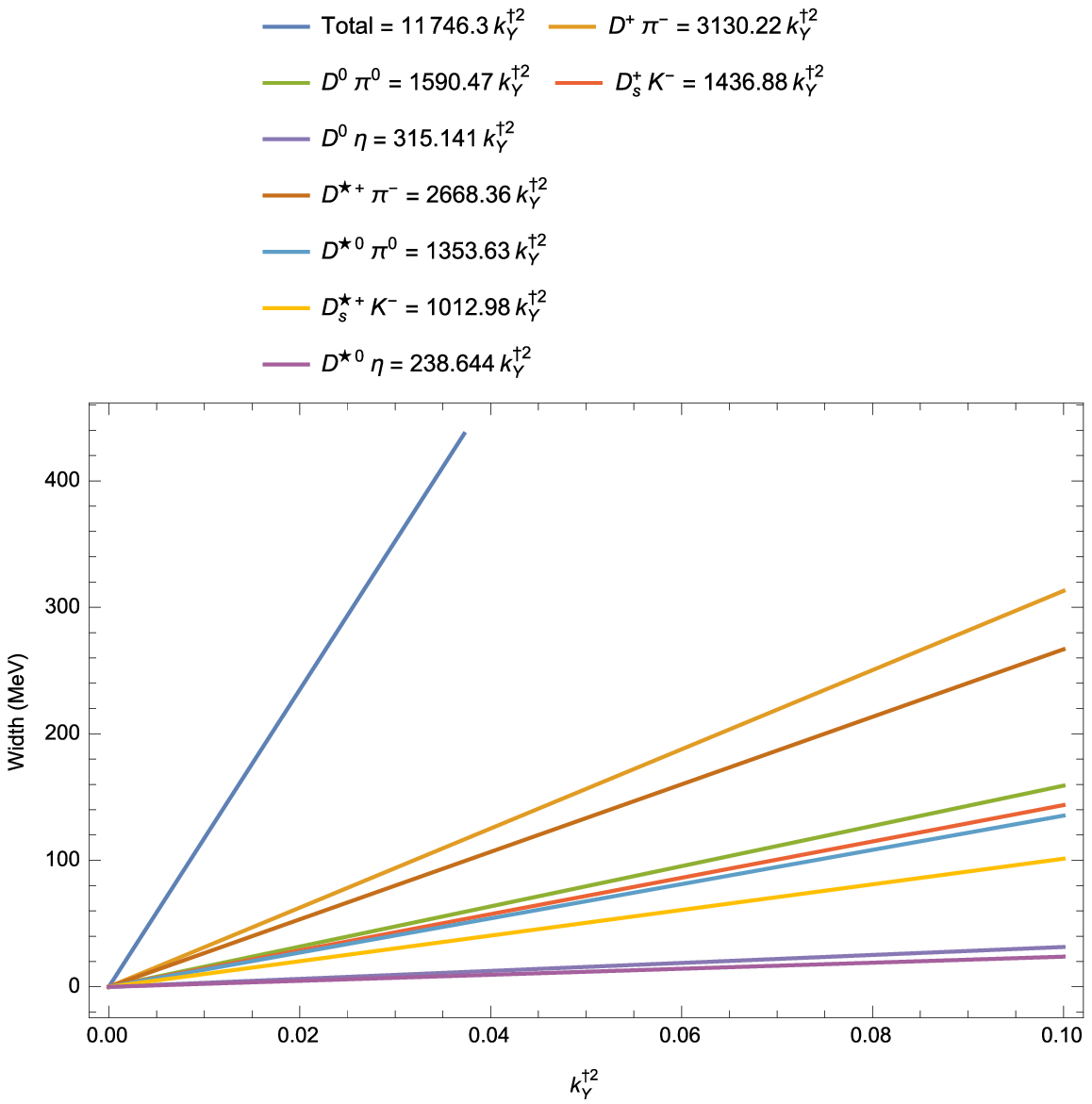}
\caption{Strong decay widths of $2^3D_3$ (in MeV) nonstrange charmed meson state (lying on the Regge line $2^3S_1$ in $(J, M^2)$ plane) changing with the square of the coupling $k_Y^{\dag2}$ in HQET.}
\label{fig21}       
\end{figure}

\begin{figure}
  \includegraphics[width=0.49\textwidth]{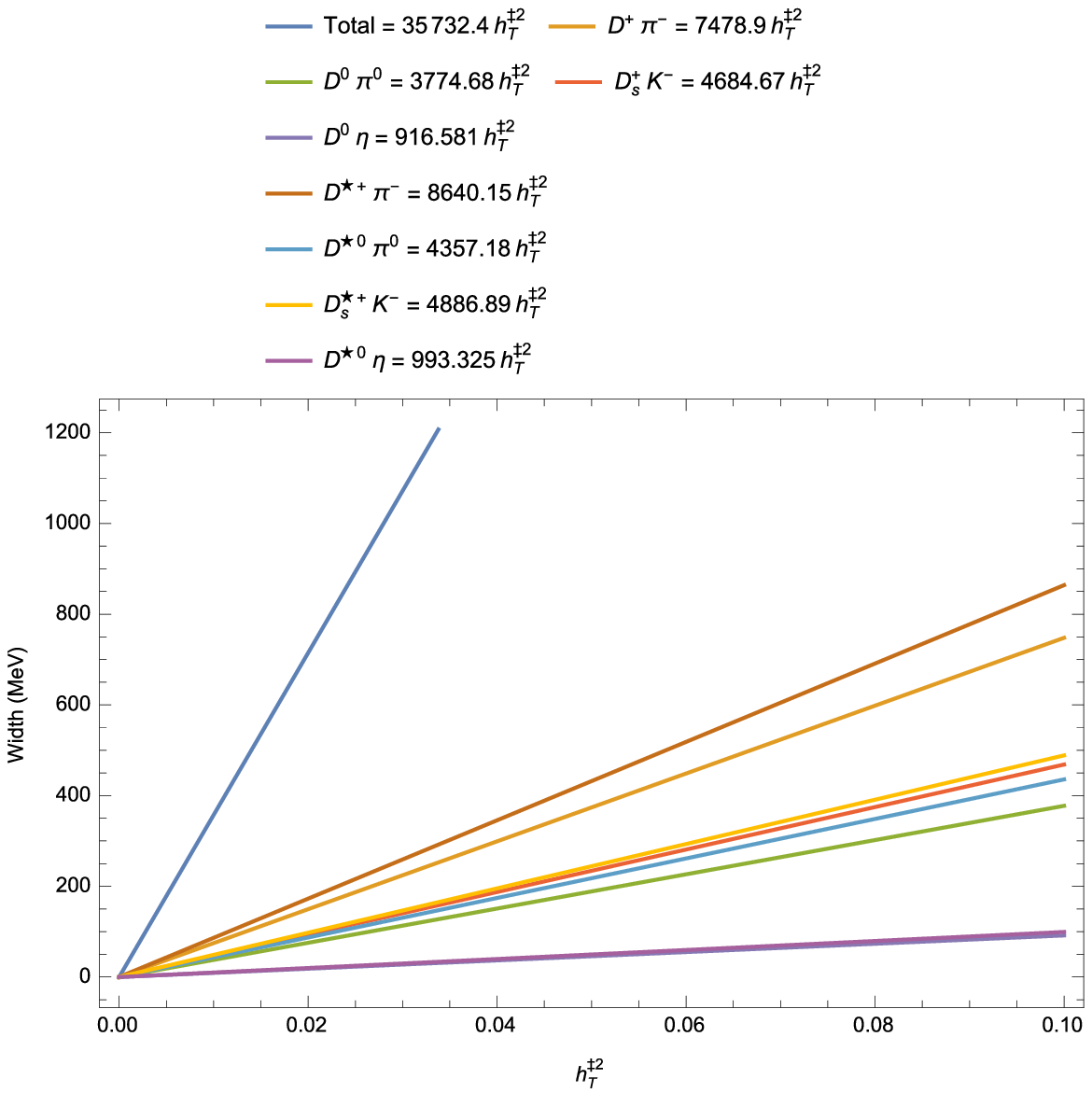}
\caption{Strong decay widths of $3^3P_2$ (in MeV) nonstrange charmed meson state (lying on the Regge line $1^3P_2$ in $(n_r, M^2)$ plane) changing with the square of the coupling $h_T^{\ddag2}$ in HQET.}
\label{fig22}       
\end{figure}

\begin{figure}
  \includegraphics[width=0.49\textwidth]{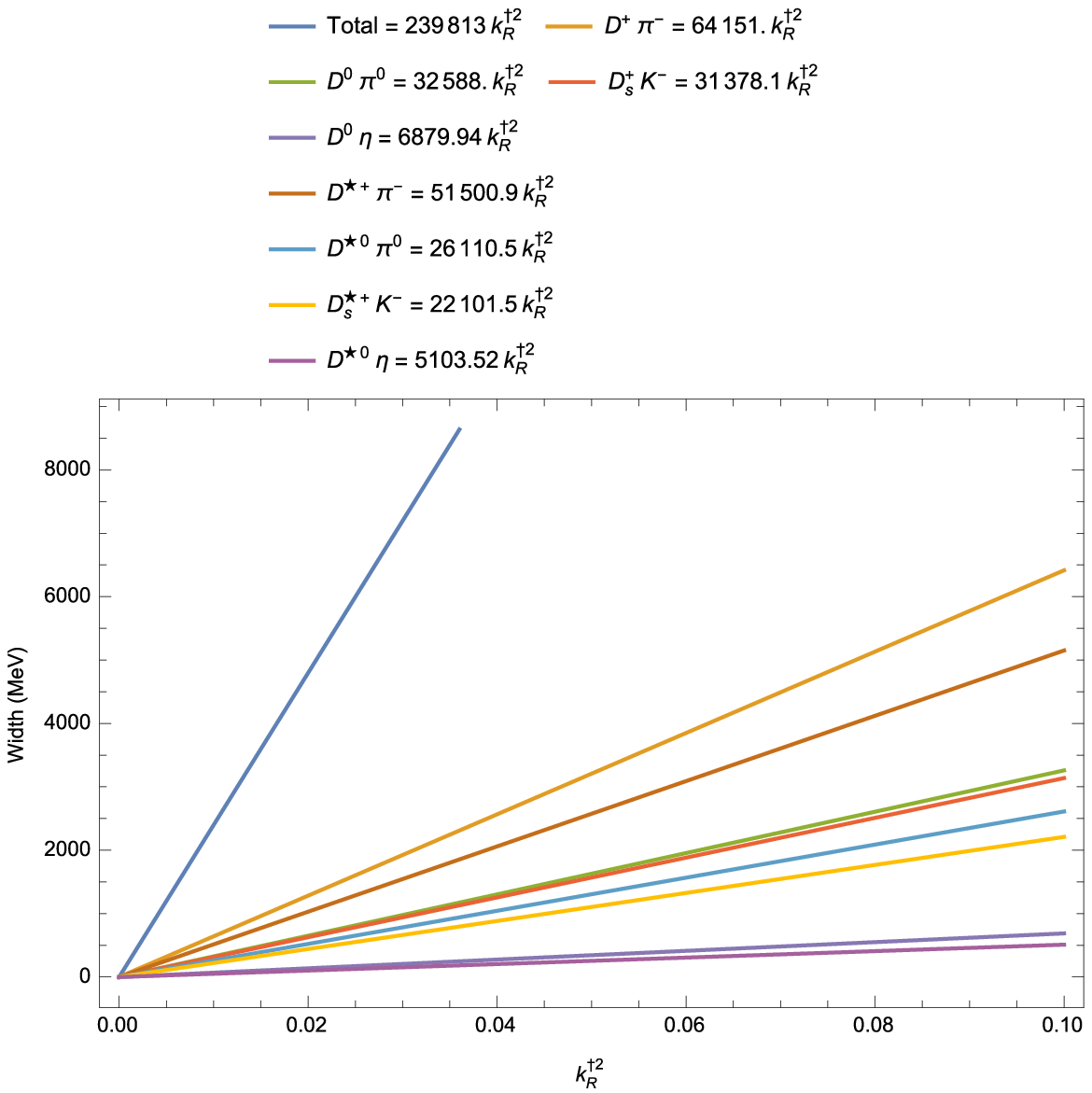}
\caption{Strong decay widths of $2^3F_4$ (in MeV) nonstrange charmed meson state (lying on the Regge line $2^3S_1$ in $(n_r, M^2)$ plane) changing with the square of the coupling $k_R^{\dag2}$ in HQET.}
\label{fig23}       
\end{figure}

The masses of 1$^1D_2$, 1$^3D_3$, 3$^1S_0$, 3$^3S_1$, 1$^1F_3$, 1$^3F_4$, 2$^3D_3$, 3$^3P_2$ and 2$^3F_4$ states are estimated (see in Table \ref{tab7} and \ref{tab8}). The 2.843 GeV of 1$^3D_3$ is overestimated to $D{_{3}^*}(2750)^0$ by a mass difference of 79 MeV. Also, the helicity distribution disfavors the identification of $D{_{3}^*}(2750)$ as a 1$^3D_3$ \cite{Sanchez2010}. But we tentatively identify $(D{_{3}^*}(2750)^0)$ as $(3^-)_{{\frac{5}{2}}^-}$ with $n = 1$. For 1$^3D_3$, 1$^3F_4$, 2$^3D_3$, 2$^3F_4$ and 3$^3S_1$, our results are in agreement with D. Ebert \textit{et al.} \cite{Ebert2010} and are overestimates to the predictions of Refs. \cite{Godfrey2016,Sun2013,DiPierro2001}. Such heavier masses agree with the argument that slopes of Regge trajectories decrease with quark mass increase \cite{Guo2008,Zhang2005,Li2004,Brisudova2000}. The partial strong decay rates of these predicted states are calculated and presented in Table \ref{tab9}. These are also shown in Figures \ref{fig15} to \ref{fig23}, where the strong decay rates change with respect to the square of the couplings. The decay mode $D^{*+}\pi^-$ is dominant in the states 1$^1D_2$, 3$^1S_0$, 3$^3S_1$, 1$^1F_3$ and 3$^3P_2$ with branching fractions 52.30\%, 46.14\%, 28.01\%, 48.85\% and 24.18\% respectively. And, for the 1$^3D_3$, 1$^3F_4$, 2$^3D_3$ and 2$^3F_4$ states the $D^+\pi^-$ decay is dominant with branching fractions 37.87\%, 34.09\%, 26.64\% and 26.74\% respectively. 

\section{Conclusions}
\label{sec5}

In this paper, we have examined the nonstrange charmed mesons $D_1(2420)^0$, $D_2^*(2460)$, $D(2550)^0$, $D_J^*(2600)^0$, $D(2740)^0$, $D_3^*(2750)$, $D_J(3000)^0$, $D{_{J}^*}(3000)^0$ and $D_2^*(3000)^0$ observed by the LHCb \cite{Aaij2016,Aaij2015,Aaij2013} and $BABAR$ \cite{del2010} Collaborations according to their spin, parity and masses. Their strong decays into ground state charmed mesons along with the emission of light pseudoscalar mesons $(\pi, \eta, K)$ are analyzed in the HQET. The branching ratios among the strong decays tentatively identify the quantum numbers of nonstrange charmed mesons. The strong decay widths are retained with the square of the coupling constants $h_T$, $g_H^{\dag}$, $k_Y$, $g_H^{\ddag}$, $h_S^{\dag}$, $h_T^{\dag}$, $k_R$ and $k_Z$, which are determined comparing those with the widths observed by experimental groups given in Table \ref{tab3}. We identify the states $D(2550)^0$, $D_J^*(2600)^0$, $D(2740)^0$ and $D_3^*(2750)$ with spin-parity $0^-$, $1^-$, $2^-$ and $3^-$ respectively. They are in agreement with the strong decays analysis done by Refs. \cite{Wang2011,Wang2012,Wang2013,Batra2015,Gupta2018}. An unclear resonance structures near 3 GeV region motivated our present study. We tentatively assign the quantum states of $D_J(3000)^0$, $D_J^*(3000)^0$ and $D_2^*(3000)^0$ as $2^3P_1$, $2^3P_2$ and $1^3F_2$ respectively. The states $D_J(3000)^0$ and $D_2^*(3000)^0$ are in accordance with the predictions of \cite{Wang2012,Batra2015,Gupta2018}. P. Gupta and A. Upadhhyay \cite{Gupta2018} identified $D_J^*(3000)^0$ as a $2^3P_0$. J.-K. Chen \cite{Chen2018} assigned the states $D_J(3000)^0$, $D_J^*(3000)^0$ and $D_2^*(3000)^0$ as $3^1S_0$, $3^3S_1$ and  $3^3P_2$ respectively. S. Godfrey and K. Moats identified $D_J(3000)^0$ as $3^1S_0$ state and $D_J^*(3000)^0$ as $1^3F_4$. To identify its nature, we expect some more experimental efforts in future.

Using these spin and parity assignments of experimentally observed $D$ mesons, we construct the Regge trajectories in $(J, M^2)$ and $(n_r, M^2)$ planes. By fixing the slopes and intercepts of the Regge lines we estimate the masses of higher excited states 1$^1D_2$, 1$^3D_3$, 3$^1S_0$, 3$^3S_1$, 1$^1F_3$, 1$^3F_4$, 2$^3D_3$, 3$^3P_2$ and 2$^3F_4$ of $D$ mesons. Their strong decays analysis conclude that the $D^{*+}\pi^-$ is dominant decay mode for 1$^1D_2$, 3$^1S_0$, 3$^3S_1$, 1$^1F_3$, 3$^3P_2$ states, and the decay mode $D^+\pi^-$ is dominant for 1$^3D_3$, 1$^3F_4$, 2$^3D_3$, 2$^3F_4$ states. This study can help the experimentalists for searching these higher excited states in such decay modes. We would like to extend this scheme for the study of strong decays of excited strange charmed mesons in future.

\end{document}